\documentclass[useAMS,usegraphicx,usenatbib]{mn2e}

\usepackage{times}
\usepackage{graphicx}
\newcommand{\beq}{\begin{equation}}
\newcommand{\eeq}{\end{equation}}
\def\gs{\mathrel{\lower0.6ex\hbox{$\buildrel {\textstyle >}\over{\scriptstyle \sim}$}}}
\def\ls{\mathrel{\lower0.6ex\hbox{$\buildrel {\textstyle <}\over{\scriptstyle \sim}$}}}

\def\mnras{MNRAS}
\def\aap{A\&A}
\def\aaps{A\&AS}
\def\apj{APJ}
\def\apjs{APJS}
\def\apjl{APJL}
\def\aj{AJ}
\def\pasp{PASP}
\def\pasj{PASJ}
\def\nat{Nature}

\newcommand{\msun}{\mbox{M$_{\odot}$}}
\newcommand{\msol}{\mbox{M$_{\odot}$}}
\newcommand{\lsol}{\mbox{L$_{\odot}$}}

\newcommand{\kms}{\mbox{$\rm{\,km\,s^{-1}}$}}
\newcommand{\logl}{\mbox{$\log \mathrm{L}/ \mathrm{L}_{\odot}$}}

\begin{document}

\title[SN~2008S]{SN 2008S: an electron capture SN from a super-AGB progenitor?}
\author[M.T. Botticella et al. ]{M.T. Botticella$^{1}$
\thanks{E-mail: m.botticella@qub.ac.uk}, A. Pastorello$^{1}$, S.J. Smartt$^{1}$, W. P. S. Meikle$^{2}$, S. Benetti$^{3}$, R. Kotak $^{1}$,  
\newauthor E. Cappellaro$^{3}$, R. M. Crockett$^{1}$,  S. Mattila$^{4}$, M. Sereno$^{5}$, F. Patat $^{6}$, D. Tsvetkov$^{7}$, 
\newauthor  J. Th. Van Loon$^{8}$, D. Abraham, I. Agnoletto$^{3}$, R. Arbour$^{9}$, C. Benn$^{10}$, G. Di Rico$^{11}$,  
\newauthor N. Elias-Rosa$^{12}$,  D.L. Gorshanov$^{13}$,  A. Harutyunyan$^{14}$,  D. Hunter$^{1}$,  V. Lorenzi$^{14}$, 
\newauthor  F. P. Keenan$^{1}$, K.  Maguire$^{1}$,  J. Mendez$^{10}$,  M. Mobberley,  H. Navasardyan$^{3}$,  C. Ries$^{15}$,   
\newauthor V. Stanishev$^{16}$, S. Taubenberger$^{17}$,  C. Trundle$^{1}$,  M. Turatto$^{18}$ and  I.M. Volkov$^{7,19}$\\
$^{1}$Astrophysics Research Centre, School of Mathematics and Physics,  Queen's University Belfast, Belfast BT7 1NN, United Kingdom\\
$^{2}$Astrophysics Group, Blackett Laboratory,  Imperial College London,  Prince Consort Road, London SW7 2BW, United Kingdom\\
$^{3}$INAF- Osservatorio Astronomico di Padova, Vicolo dell'Osservatorio 5, 35122 Padova, Italy \\
$^{4}$Tuorla Observatory, Department of Physics \& Astronomy,  University of Turku,  FI-21500 Piikki\"o, Finland\\
$^{5}$Institut f\"{u}r Theoretische Physik, Universit\"{a}t Z\"{u}rich, Winterthurerstrasse 190, CH-8057 Z\"{u}rich, Switzerland\\
$^{6}$European Southern Observatory (ESO), Karl-Schwarzschild-Str. 2, D-85748,  Garching bei M\"{u}nchen, Germany\\
$^{7}$Sternberg State Astronomical Institute, Universitetskii pr.13, 119992 Moscow, Russia\\
$^{8}$Astrophysics Group, Lennard-Jones Laboratories, Keele University, Staffordshire ST5 5BG, United Kingdom\\
$^{9}$Pennel Observatory, 29 Wrights Way, South Wonston, Hants, S021 3He, United Kingdom\\
$^{10}$Isaac Newton Group of Telescopes,  Apartado 321, E-38700 Santa Cruz de La Palma, Spain\\
$^{11}$INAF - Osservatorio Astronomico di Collurania, Teramo, Italy\\
$^{12}$Spitzer Science Center, California Institute of Technology, 1200 E. California Blvd., Pasadena, CA 91125, USA\\
$^{13}$Central (Pulkovo) Astronomical Observatory, Russian Academy of Sciences, 196140 St. Petersburg, Russia\\
$^{14}$Fundaci\'on Galileo Galilei-INAF, Telescopio Nazionale Galileo, E-38700 Santa Cruz de la Palma, Tenerife, Spain\\
$^{15}$Universit\"{a}ts-Sternwarte M\"{u}nchen, Scheinerstr. 1, 81679 Munchen, Germany\\
$^{16}$Centro Multidisciplinar de Astrofisica, Instituto Superior Tecnico, Av. Rovisco Pais 1, 1049-001 Lisbon, Portugal\\
$^{17}$Max-Planck-Institut f\"{u}r Astrophysik, Karl-Schwarzschild-Str. 1, D-85741 Garching bei M\"{u}nchen, Germany\\
$^{18}$INAF- Osservatorio Astrofisico di Catania, Via S. Sofia 78, I-95123, Catania, Italy\\
$^{19}$Astronomical Institute of the Slovak Academy of Sciences, 059 60 Tatranska Lomnica, Slovak Republic
}

\maketitle
\begin{abstract}
We present comprehensive photometric and spectroscopic observations
of the faint transient SN 2008S discovered in the nearby galaxy NGC
6946. SN 2008S exhibited  slow photometric evolution and almost no
spectral variability during the first nine months, implying a long
photon diffusion time and a high density circumstellar medium.   Its bolometric luminosity ( $\simeq$\,$10^{41}$\,erg\,s$^{-1}$ at peak) is low with respect to most core
collapse supernovae but is comparable to the faintest type II-P
events. Our quasi-bolometric lightcurve extends to 300 days and shows
a tail phase decay rate consistent with that of $^{56}$Co. We propose
that this is evidence for an explosion and formation of $^{56}$Ni ($ 0.0014 \pm 0.0003$\,\msol).  Spectra of SN 2008S show intense emission lines of
H$\alpha$,   [Ca~II] doublet and Ca II NIR triplet, all without obvious
P-Cygni absorption troughs.  
The large mid-infrared (MIR) flux detected shortly after explosion can 
be explained by a light echo from pre-exisiting dust.
The late near-infrared (NIR) flux excess is plausibly due to a combination of warm newly-formed ejecta dust  together with
shock-heated dust in the circumstellar environment. We reassess the progenitor object detected previously in Spitzer
archive images, supplementing this discussion with a model of the
MIR spectral energy distribution.  This supports the idea of a dusty, optically thick shell
around SN 2008S with an inner radius of nearly 90AU and outer radius of
450AU, and an inferred heating source of 3000\,K. The luminosity of
the central star is L $\simeq 10^{4.6}$\,L$_{\odot}$. All the nearby
progenitor dust was likely evaporated in the explosion leaving only
the much older dust lying further out in the circumstellar  environment. The combination of our
long term multi-wavelength monitoring data and the evidence from the
progenitor analysis leads us to support the scenario of a weak
electron capture supernova explosion in a super-asymptotic giant branch (AGB) progenitor star
(of initial mass 6--8\,\msol) embedded within a thick
circumstellar gaseous envelope.  We suggest
that all of main properties of the
electron capture SN phenomenon are observed in SN 2008S and future observations may
allow a definitive answer. 
\end{abstract}



\section{Introduction}\label{Introduction}

In recent years deeper and more frequent searches for transient events
 and stellar explosions in the local and distant Universe have
 provided us with important information
 on the evolution of the most massive
 stars.  However, the simplicity of the emerging picture is compromised
 by the growing number of peculiar events
 \citep[e.g.][]{Kulkarni2007,Smith2007,Quimby2007}. 

From an observational point of view, the challenge is to decide when the 
introduction of 
 new classes is required or if  peculiar or novel transients 
are just variations of an understood scheme. The discovery of some low energy events (in terms of 
their bolometric luminosity and kinetic energies) 
 leads us to investigate in more detail the observational
 differences between explosive (core collapse, pair instability
 explosions) and eruptive (pair instability pulsation, outburst)
 transients. 

From a theoretical point of view, recent observations demonstrate
that the standard scenario of stellar evolution and explosion physics
may not be complete. Both the extremely bright type II SNe 
\citep{Woosley2007,Langer2007}
and the faint type II SNe \citep{Smith2008c,Bond2009,Berger2009}
have been proposed to have physical
origins other than the core-collapse of a degenerate Fe (or O-Ne-Mg)
core. 

SN 2008S is one of the most intriguing  transient events discovered in recent years. 
Although it has been given a supernova designation, (which we will employ in 
this paper) it is not yet certain that it was a supernova of the 
canonical core-collapse type (CCSN). 
The transient
was discovered in NGC 6946 by \cite{Arbour2008CBET1} on February 1.78 UT  with a 30-cm f/6.3 Schmidt-Cassegrain reflector at about 17.6 mag.
 Eight confirming images of SN 2008S were taken on February 2.76 UT, yielding a magnitude of 17.1. 
Furthermore, \cite{Arbour2008CBET2} provided a new image of SN 2008S acquired on 2008 January 24 UT (17.8 mag) and  \cite{Schmeer2008CBET} reported an image obtained on 2008 January 30.529 (16.7 mag).
The transient was classified  as a young reddened Type IIn SN by \cite{Stanishev2008} based on a low resolution spectrum taken at the Nordic Optical Telescope, with narrow H$\beta$ and H$\alpha$ emission lines and strong Na I D doublet.
\cite{Steele2008CBET} reported a new spectrum of SN 2008S obtained on Feb. 29 UT with the 3-m Shane reflector equipped with Kast double spectrograph at the Lick Observatory, and suggested SN 2008S to be a
"SN impostor"  based on peculiar spectral properties and the very faint absolute visual magnitude.

Remarkably, a bright point-like source coincident with SN 2008S was
detected in archival Spitzer MIR images by \citet{Prieto2008}. They
found no optical counterpart to this precursor and suggested this
MIR source was a stellar progenitor with mass of about 10\,M$_{\odot}$
and luminosity of $\sim3.5 \times 10^4$\,L$_{\odot}$, enshrouded in
its own dust. The stellar mass and the total luminosity estimates
result from a blackbody fit to the MIR spectral energy distribution (SED) of the progenitor star.
Shortly afterwards, another transient was discovered in the nearby
galaxy NGC 300 which bears a striking resemblance to SN 2008S
\citep{Berger2009,Bond2009}. 
\cite{Thompson2008} reported the discovery of a similar progenitor star in 
Spitzer MIR prediscovery images and again an optical counterpart was lacking. 
They suggested that both transients share a common evolutionary channel and also that the 
optical transient discovered in M 85 was of similar origin \citep{Kulkarni2007,Pastorello2007}. 
\citet{Thompson2008}  and  \cite{Prieto2008} have proposed that these events
could be low energy electron-capture SNe (ECSNe) from
stars of initial mass around 9\,M$_{\odot}$. 
 The existence of such explosions 
has been theoretically predicted for many years 
\citep{Miyaji1980,Nomoto1984,Miyaji1987,Hashimoto1993,Kitaura2006,Poelarends2008}.
However,  what  exactly the mass range of the 
progenitors would be and how the SN evolution would appear is
far from certain. The nature of these transients has not
yet been firmly established, since recent works on SN 2008S \citep{Smith2008c}
and NGC 300 OT2008-1 \citep{Bond2009,Berger2009} 
 suggest that  these events are the outbursts of a massive star 
and not the cataclysmic stellar deaths of stars after core-collapse.

In this paper we present results from our  extensive photometric
and spectroscopic follow-up of SN 2008S, together with  analysis of supernova and progenitor observations.  The properties of the host galaxy are
described in Sect.~\ref{hostgalsec}.  Photometric 
data reduction and analysis are detailed in Sect.~\ref{Photsec} and  the evolution of the SED is illustrated in Sect.~\ref{SEDsec}.   Spectroscopic data reduction and analysis are detailed in Sect.~\ref{Specsec}.  Sect.~\ref{Progsec}  is
devoted to the analysis of the pre-explosion images and to the discussion
of the progenitor star.  A summary of our observations, a
comparison with other underluminous transients and some 
type II-L SNe, and our conclusions on the nature of SN 2008S are given in
Sect.~\ref{nature}.

\section{Host galaxy, distance and extinction}\label{hostgalsec}
SN 2008S was discovered at R.A.$=20^{h}34^{m}45\fs37$ and Dec.$=60\degr05\arcmin58\farcs3$ (2000), about 53 arcsec West and 196 arcsec South of the nucleus of NGC 6946.
Details of the host galaxy obtained from the NASA/IPAC Extragalactic database\footnote{NASA/IPAC Extragalactic Database, \hspace*{0.16cm} http:/$\!$/nedwww.ipac.caltech.edu/} are summarised in Table~\ref{galprop}.

\begin{table}
\caption{Properties of NGC6946.\label{galprop}}
\begin{footnotesize}
\begin{tabular}{llc}
\hline
$\alpha$ (2000) & $20^{h}34^{m}52\fs3$&1  \\					 
$\delta$ (2000) & $60\degr09\arcmin14\arcsec$&1\\
Galactic longitude & 95.72\degr&1	  \\
Galactic latitude  & +11.67\degr&1	  \\
morphological type &  SAB(rs)cd&1	  \\
Position angle &242\degr &2\\
Inclination angle &$38 \pm 2$\degr &2\\
M$_{B}$&-21.38 mag & 3 \\
L$_{B}$&$5.3\times10^{10}$ \lsol &3\\
redshift & $0.00016 \pm 0.000007$	&1  \\
v$_{Hel}$  & $48 \pm 2$	\kms &1  \\
v$_{galact.}$ & $275 \pm 9 $\kms	&1 \\
v$_{Virgo+GA+Shapley}^a$ \ & $410 \pm 19$ \kms &1	  \\
Galactic reddening & $E(B-V) = 0.342$ mag &4\\
\hline
\end{tabular}
\\[1.5ex]
1 NASA/IPAC Extragalactic Database (NED)\\
2 \cite{Boomsma2008}\\
 3  \cite{Carignan1990}\\
 4 \cite{Schlegel1998}\\
$^a$ based on the local velocity field model given in \cite{Mould2000} using the terms for the influence of the Virgo Cluster, the Great Attractor, and the Shapley Supercluster.
\end{footnotesize}
\end{table}

Optical, far infrared, radio continuum and X-ray observations indicate
vigorous star formation (SF) throughout the NGC 6946 disc, one of the highest among nearby spiral galaxies, a mild starburst at its centre, and an interstellar
medium (ISM) stirred by SNe and stellar winds
\citep{Engargiola1991,Boulanger1992,Kamphuis1993,Schlegel1994,Lacey1997}.
This high level of SF in the disc of NGC 6946 has been attributed both
to its strong spiral density wave \citep{Tacconi1990} and to
stochastic, self-propagating SF \citep{Degioia1984}.
Signs of low level SF such as HII regions and UV bright clusters have been discovered in the far outer regions of galactic disc well beyond the R$_{25}$ radius.

Eight other SNe have been detected in this galaxy
six of which were classified as type II SNe (1917A (II), 1948B (II-P),
1968D (II) 1980K (II-L) 2002hh (II-P) 2004et (II-P)) and two remain
unclassified (SN 1939C and SN 1969P). All these SNe were brighter than
mag 15 except SN 2002hh which was highly reddened. Among these eight
SNe, four have been detected as radio SNe (1968D, 1980K, 2002hh,
2004et) and three as X-ray SNe (1968D, 1980K,  2004et).  
 Many SN
remnants have been detected in NGC 6946 using optical, radio and X-ray
telescopes \citep{Matonick1997,Schlegel2000,Pannuti2007}.

\subsection{Metallicity}
The galactic  metallicity at the position of SN 2008S can be 
estimated in a similar way to that for two other recent SNe
in this galaxy (SN 2002hh and SN 2004et) as shown by 
\citet{Smartt2008}. 
The abundance gradient determined by \citet{Pilyugin2004} ($12+ \log{\rm O/H}  = 8.7 - 0.41(R/R_{25})$) 
and the de-projected galactocentric radius of the SN position 
can be used to determine the likely local metallicity at the 
position of SN 2008S. Using the distance of 5.7\,Mpc as discussed
in Sect.~\ref{Distance}, SN 2008S is at a deprojected galactocentric radius of 
4.9\,kpc, and with R$_{25} = 9.1$\,kpc 
(from HyperLeda\footnote{HyperLeda database, \hspace*{0.16cm} http://leda.univ-lyon1.fr}, \citet{Paturel2003}), the metallicity gradient of \citet{Pilyugin2004}
results in 
an approximate oxygen abundance of 8.5\,dex.  On the \citet{Pilyugin2004} abundance scale, solar
is approximately 8.7\,dex.   Hence the environment of SN 2008S
is mildly sub-solar, although within the uncertainties in this 
method a solar-like composition of the progenitor is still quite
possible. By comparison,  the oxygen abundances estimated for
SN 2002hh and SN 2004et are approximately 8.5 and 8.3\,dex, respectively
\citep{Smartt2008}. 

\subsection{Distance}\label{Distance}
\begin{table*}
\caption{Estimates of the distance to NGC 6946.\label{zgal}}
\begin{tabular}{llll}
\hline
Distance & Distance modulus &	Method & Reference\\
(Mpc) & (mag)                            & & \\					
\hline	
5.5 & 28.70  & HI  Tully-Fisher  relation & \cite{Pierce1994}\\	
5.4 & 28.66 & CO Tully-Fisher relation & \cite{Schoniger1994}	\\
$6.0 \pm 0.5$ & $28.90 \pm 0.18$ & galaxy brightest supergiants & \cite{Sharina1997}\\
$5.9 \pm 0.4$ & $28.85 \pm 0.15$ & brightest supergiants of group & \cite{Karachentsev2000}\\
$5.6 \pm 1.8$& $28.73 \pm 0.68$ & "sosie galaxies" &\cite{Terry2002}\\
$6.1 \pm 0.6$& $28.92 \pm 0.21$ & planetary nebulae
luminosity function  & \cite{Herrmann2008}\\	
$5.7 \pm 0.7$ & $28.78 \pm 0.40$ & EMP SN 1980K &	\cite{Schmidt1994} \\	
$5.7 \pm 0.3$ & $28.78 \pm 0.11$ & SCM SN 2004et & \cite{Sahu2006} \\	
\hline
\end{tabular}
\end{table*}
There are several estimates of the distance to NGC 6946, obtained with
different methods and listed in Table~\ref{zgal}.  
Two SNe hosted in  NGC 6946 have been used as distance indicators.
\citet{Schmidt1994} applied the Expanding Photosphere Method
(EPM) to SN 1980K and found a distance modulus of $28.78 \pm 0.4$\,mag.
\citet{Sahu2006} used a ``standard candle method'' (SCM) for SN 2004et, based on the correlation between the expansion velocities of the SN II-P ejecta and the bolometric luminosities  during the plateau phase \citep{Hamuy2002, Nugent2006}.
They  obtained $\mu= 28.78 \pm 0.11$\,mag, in close agreement with the estimate obtained with EPM for SN 1980K. There is one distance estimate that is significantly
different from the rest:
the radio observations of SN 1980K yield a much 
larger value of $30.5 \pm 0.3$\,Mpc \citep{Weiler1998}.
As this is much larger than the other estimates (which  are 
consistent within the uncertainties) we shall discount this value and 
 use an unweighted mean of $28.78 \pm 0.08$\,mag throughout this paper.

\subsection{Extinction}\label{Extinction}

\begin{figure}
\resizebox{\hsize}{!}{\includegraphics{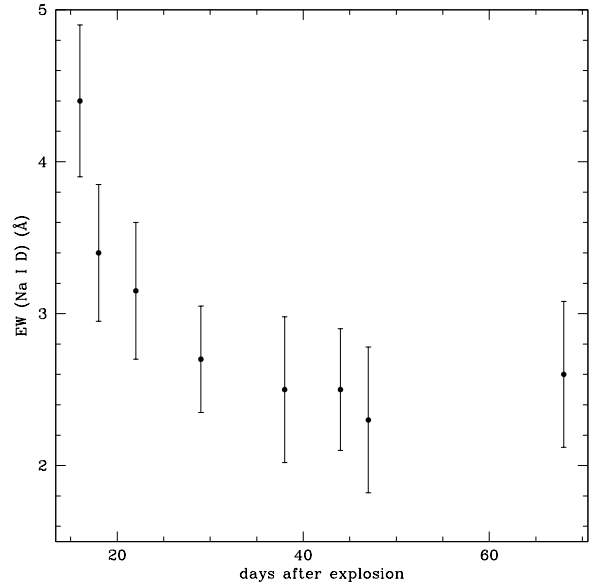}} 
\caption{Temporal evolution of the EW(Na I D). Phase is in days after the explosion epoch (JD $2\,454\,486$).}
\label{NaID}
\end{figure}

NGC 6946 is located close to the Galactic plane (Galactic latitude $\sim$\,12\degr),  with an
 estimated reddening of $E(B-V)=0.342$\,mag \citep{Schlegel1998}.  For
 the SNe which occurred in this galaxy,  different values of 
 extinction have been estimated depending on the SN position. 
In all cases the presence of the Na I D ($\lambda\lambda$5890,5896) lines has been used as an indicator of the presence of
 dust and used to estimate the reddening at the SN position \citep{Zwitter2004,Sahu2006,Meikle2006,Pozzo2006}.
Strong Na I D lines in absorption are also present in the SN 2008S spectra until about
70 days after the explosion. 
The EW(Na I D) appears to show a temporal evolution from 4.4\,\AA\ to 2.5\,\AA\  during this time
(Table~\ref{EWNaID} and Fig.~\ref{NaID}).   

\begin{table}
\caption{Measurements of the EW of Na I D.\label{EWNaID}}
\begin{footnotesize}
\begin{tabular}{ccc}
\hline
 JD$^a$ & ph$^b$ & EW (\AA) \\				
\hline
501 & 16 &$4.4\pm 0.50$   \\
504 & 19 &$3.4\pm 0.45$\\				
508 &22 &$3.2\pm 0.45$ \\
514 &28 &$2.7\pm 0.35$ \\
524 &38 &$2.5\pm 0.48$ \\
530 &44 &$2.5\pm 0.40$ \\
533 &47 &$2.3\pm 0.48$ \\
554 &68 &$2.6\pm 0.48$ \\
\hline
\end{tabular}
\\[1.5ex]
$^a$JD $-$ 2\,454\,000.00\quad\\
$^b$ Phase is in days after  the explosion date JD $2\,454\,486 \pm 4$.\quad
\end{footnotesize}
\end{table}

To estimate the
error in our EW measurements we performed a Monte-Carlo simulation
adding a number of absorption lines with known EW at different
positions in each SN spectrum and re-measured their EW.  We repeated
the simulation for different values of line EW and strength.  These
simulations were performed separately in each spectrum to take account
of the differences in spectral resolution and S/N ratio.  In order to
investigate the reality of the EW changes in the Na I D lines, we
carried out a quantitative statistical test, performing a linear fit
to the data, and found a negative slope at the 2 sigma (95$\%$
confidence) level.  
To test further that the data are better represented
                  by a temporally declining EW, rather than a fixed value,
                  we exploited the  Bayesian information criterion (BIC), which give an
approximation for the Bayes factor \citep[see][and references therein]{lid04}. 
The BIC is defined as 
$\chi^2 + \mathrm{N}_\mathrm{par}\log \mathrm{N}_\mathrm{data}$ where $\chi^2$ is the total
$\chi^2$ for the model, N$_\mathrm{par}$ is the number of parameters of
the model and N$_\mathrm{data}$ is the number of data points used in the
fit. The best model minimizes the BIC. 
A difference of 2 for the BIC is
regarded as positive evidence, and of 6 or more as strong evidence, against
the model with the larger value. 
The BIC corresponding to no evolution is
larger by 6 than the BIC for a straight line with slope ($\Delta \mathrm{BIC} = \mathrm{BIC}_ \mathrm{const} - \mathrm{BIC}_\mathrm{slope} > 6$), clearly
supporting the scenario for a decreasing temporal evolution. The two BIC
values are comparable only if we exclude from the analysis the first two
epochs. We conclude that we find evidence for a change
in EW of the Na I D feature.

 The evolution of the local component EW(Na I D) may be due
to an evolution of the ionization conditions in the CSM and in the
ejecta of SN 2008S since the EW is related to the ionization stage of
Na I.  
The evolution in the EW may also imply that the local extinction
underwent a temporal decline.  However, given (a) the lack of any
well-established EW(Na I D)-exinction correlation at the very large
EWs involved, and (b) the possibility that the Na I D feature includes
saturated components, we make no attempt to use the EW(Na I D) to
determine the extinction or its possible variation. In any case, the
EW(Na I D) variation could be simply due to evolution of the physical
properties of the gas around SN~2008S (see Sect.~\ref{specevol}),
with the extinction taking place at a completely different location.
By days 182 and 256 the Na I D feature has become visible in
emission.  This change is also indicative of the circumstellar origin
of this feature.

The presence of circumstellar Na I D has also been observed in the
type IIn SN 1998S and has been interpreted as a signature of slow
moving outflows originating from the progenitor while its blueshift and
growing intensity between 20--40 days after explosion has been
associated with variable physical conditions in the CSM
\citep{Bowen2000}.   \cite{Chugai2008} studied the
formation of the Na I D and Ca II H$\&$K lines in the RSG wind after a
SN II-P explosion with the goal of using these as a diagnostic of the
wind density. They extrapolated their model to a very high wind
density to reproduce the intensity of these lines observed in SN
1998S.  However, the EW of the absorption depends non-monotonically on
the wind density.  The case of SN 1998S with its very dense wind has
shown that the EW(Na I D) decreases with wind density because of
the ionization of metals in the wind by UV radiation.
Variable EW(NaI D)  have been detected also in a few
Type Ia SNe (2006X,  \cite{Patat2007}; 1999cl,  \cite{Blondin2008};
2007le, Simon et al. in prep).  However, this does not seem to be
a very common phenomenon \citep{Blondin2008} and the interpretation
in terms of evolution of the CSM physical conditions induced
by the SN radiation field \citep{Patat2007}  is still debatable
\citep{Chugai2008a}.

Here we adopt  the Galactic absorption in $V$ band, $A_V = 1.13$ mag,
calculated from the list of $A/E(B-V)$ of  \cite{Schlegel1998} along with their estimate of $E(B-V)$ and an extinction local to the SN, 
$A_V$\,$\sim$\,1\,mag, required by our light echo model
 to fit the observed SED at  17~ days after explosion when the MIR excess is observed, (see Sect.~\ref{MIR excess}).
 If we assume $A_V =2.13$\,mag, the EW of K I ($\lambda$ 7699)  would be
about 0.17\,\AA\ following the calibration by \cite{Munari1997}.  Unfortunately, the K I region of the spectrum  lies close to strong telluric absorption (7570--7750\,\AA\/) and is only ever
covered at low resolution so we did not observe the K I  feature.

\section{Photometric data and analysis}\label{Photsec} 

We commenced monitoring SN 2008S shortly after the discovery epoch and
collected data for the following eight months with a sampling rate among
the highest ever obtained for such a peculiar transient.  Data
obtained before the discovery date by several amateur astronomers are
also included to constrain the explosion epoch of SN 2008S.  The
unfiltered image acquired by D. Abraham on January 16 (JD $2\,454\,482$) shows no
object visible in the SN 2008S location with a limiting magnitude of
19.20 in $V$ band (18.60 in $R$ band), while the first detection of SN
2008S is eight days later on January 24 (JD $2\,454\,490$).  We therefore adopt January
20 (JD $2\,454\,486$) as the explosion epoch, the uncertainty being about 4
d.  The phases in this paper are relative to the explosion date
(when we fix ph=0).

\subsection{Optical data} 
Optical photometry of SN 2008S was obtained  with many telescopes and a summary of their characteristics  is given in Table~\ref{phtel}.
Unfiltered images were obtained for many epochs with a 40~cm telescope with a SXVF H9 camera, a 30~cm telescope with a MX916 camera,  a 35~cm telescope with a ST-9E/9XE camera and with a 25~cm telescope with a ICX424 CCD.

\begin{table*}
\caption{Summary of the characteristics of the telescopes used during the photometric follow up.\label{phtel}}
\begin{footnotesize}
\begin{tabular}{lclllcll}
\hline
Telescope & Primary mirror &Camera & array & CCD & pixel scale & field of view & filters  \\
 &  m& &  & & arcsec/pix & arcmin &   \\
\hline
TNGD &3.6 &DOLORES      & $2048 \times 2048$    &   EEV 42-40    &  0.25      &  $8.6 \times 8.6$ &Johnson $U,B,V$; Cousin $R,I$    \\
 TNGN &3.6& NICS &  $1024 \times 1024$ & HgCdTe Hawaii& 0.25 & $4.2 \times 4.2$ & $JHK $ \\
 NOT & 2.5 & ALFOSC &  $2048 \times 2048$ &EEV 42-40              &0.19  & $6.4 \times 6.4$& Johnson $U,B,V,R$   \\
CAHAT & 2.2  & CAFOS         &  $2048 \times 2048$  & SITe            &  0.53     & $16 \times 16 $&   Johnson $B,V,R,I$   \\
LT &2.0  & RATCam        &  $2048 \times 2048$       &  EEV 42-40     &   0.13       &  4.6 &  Bessel $U,B,V$; Sloan $r',i'$   \\
CAO  &   1.8  & AFOSC&  $1024 \times 1024$  & TK1024AB       &  0.46    & 7.8       &  Bessel $B, V, R$; Gunn $i$     \\
SAO  &1.0 &         &  $2048 \times 2048$ & EEV 42-40       &   0.48    &  8.3      &  Johnson $V$    \\
AZT24 & 1.0 & SWIRCAM         & $256 \times 256$ & HgCdTe PICNIC &  1.03       & 4.4   & $JHK$   \\
MSK & 0.7  &  Apogee AP-7p        & $512 \times 512$    & SITe            &  0.94     &  4      &  Johnson $B,V,I$; Cousin $R$    \\
MSKL &0.7 &  Apogee AP-7p       &   $512 \times 512 $       &    SITe    &   0.64     & 5.5 &  Johnson $B,V,I$; Cousin $R$   \\
CRM &0.6 & Apogee AP-47p        & $1024 \times 1024$ & Marconi47-10   & 0.71      &  6.1      &  Johnson $B,V,I$; Cousin $R$     \\
SLV & 0.5 & SBIG ST-10XME        &  $2184 \times 1472$ & KAF3200ME       &  1.12     &   $20.6 \times 13.9$     &    Johnson $B,V$; Cousin $R,I$   \\
WOT& 0.4 & SBIG ST-10 XME       &$2184 \times 1472$ &   KAF3200ME              &   0.44    &    $16.0  \times $10.8            &      Sloan $r'$\\
\hline
\end{tabular}
\\[1.5ex]
TNGD = the Telescopio Nazionale Galileo (TNG) with the Device Optimized for the LOw RESolution (DOLORES); TNGN = the Telescopio Nazionale Galileo (TNG) with the Near Infrared Camera Spectrometer (NICS);  NOT= the Nordic Optical Telescope (NOT) with the Andalucia Faint Object Spectrograph and Camera (ALFOSC); CAHAT= the 2.2~m telescope at Calar Alto Observatory (CAHA) with the Calar Alto Faint Object Spectrograph (CAFOS);   LT = the Liverpool Telescope (LT) with the optical CCD camera RATCam; CAO = the Copernico telescope at Asiago Observatory with the Asiago Faint Object Spectrograph and Camera (AFOSC);  SAO = the 1~m telescope of Special Astrophysical Observatory of Russian Academy of Sciences; AZT24= the AZT 24 telescope at Campo Imperatore Observatory with SWIRCAM; MSK = the 70~cm  telescope of the Sternberg Astronomical Institute in Moscow; MSKL = the 70~cm  telescope of the Sternberg Astronomical Institute +focal reducing lens; CRM =  the 60~cm reflector of the Sternberg Astronomical Institute Crimean laboratory;   SLV = the  50~cm telescope of the Astronomical Institute of Slovak Academy of Sciences at Tatranska Lomnica; WOT =  the 40~cm telescope at the Wendelstein Observatory
\end{footnotesize}
\end{table*}

Basic data reduction (overscan correction, bias subtraction, flat fielding, trimming) was performed using standard routines in IRAF\footnote{Image Reduction and Analysis Facility (IRAF) is distributed by the National Optical Astronomy Observatories, which are operated by the Association of Universities for Research in Astronomy, Inc., under cooperative agreement with the National Science Foundation.}.
The instrumental magnitudes were obtained with the point spread function (PSF) fitting technique using a custom made DAOPHOT based package (SNOoPY). We did not apply the template subtraction technique, since the host galaxy contamination is negligible around SN 2008S in the optical  and  in the NIR range.

\begin{figure}
\resizebox{\hsize}{!}{\includegraphics{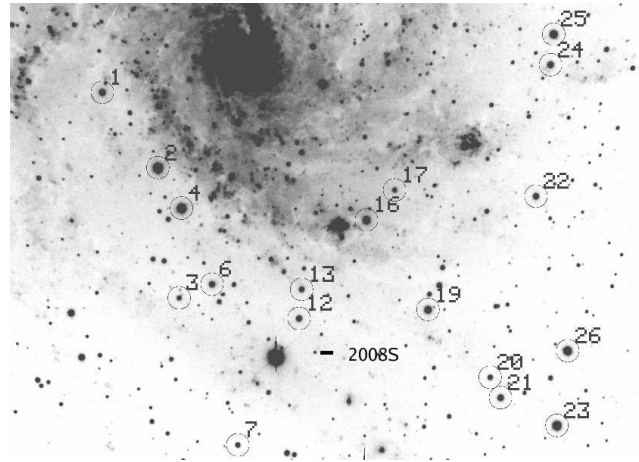}} 
\caption{Finding chart for the local sequence of stars employed for the optical photometric calibration. The numbers are adopted from Pozzo et al. (2006).} \label{chart}
\end{figure}

The photometric calibration was carried out by a comparison with Landolt standard stars observed the same night when possible. 
As our local sequence of stars, we chose a subset of that adopted in \cite{Pozzo2006} for SN  2002hh, shown in Fig.~\ref{chart} and  
calibrated it with respect to a number of Landolt standard fields on several photometric nights. 

\begin{table*}
\caption{Optical, Near Infrared and unfiltered photometry of SN 2008S.\label{phot}}
\begin{scriptsize}
\begin{tabular}{lllllllllll}
\hline
JD$^a$  & ph$^b$  & $U$   & $B$ & $V$  & $R$  & $I$ & $J$ & $H$ & $K$ & Instrument  \\
\hline
490.3 & 4   &                    &                   &17.25 $\pm$ 0.20    & 16.66$\pm$ 0.50  &                  & & & & MX916+ ICX424 unfiltered\\
498.3 & 12  &                    &                   & 16.97 $\pm$ 0.30  &                   &                  & & & &  MX916 unfiltered \\
501.3 & 15  &                    &                   & 16.95 $\pm$ 0.08  & 16.34 $\pm$ 0.05  &                  & & & & SXVF-H9+ST-9E/9XE unfiltered  \\ 
501.4 & 15  &                    &                   &                    & 16.35 $\pm$ 0.01  &                  & & & & NOT  \\
502.2 & 16  &                    & 17.92 $\pm$ 0.16  & 16.87 $\pm$ 0.12   & 16.32 $\pm$ 0.08  & 15.87 $\pm$ 0.07 & & & &  MSK \\
503.3 & 17  &                    &                   & 16.88 $\pm$ 0.10  & 16.39 $\pm$ 0.05 &                  & & & &  SXVF-H9+ST-9E/9XE unfiltered \\
505.3 & 19  &                    &                   & 16.98 $\pm$ 0.14  & 16.33 $\pm$ 0.06 &                  & & & &  SXVF-H9+ST-9E/9XE unfiltered\\
506.3 & 20  &                    &                   &                   & 16.26 $\pm$ 0.06 &                  & & & &  SXVF-H9+ST-9E/9XE unfiltered\\ 
507.8 & 22  & 17.89 $\pm$ 0.04   & 17.77 $\pm$ 0.03  & 16.91 $\pm$ 0.01   & 16.35 $\pm$ 0.02  & 15.79 $\pm$ 0.02 & & & & TNGD   \\ 
508.3 & 22  &                    &                   & 17.10 $\pm$ 0.14  & 16.36 $\pm$ 0.06 &                  & & & &  SXVF-H9+ST-9E/9XE unfiltered\\
508.7 & 23  &                    & 17.81$\pm$  0.30  & 16.90 $\pm$ 0.50   & 16.37 $\pm$ 0.04  & 15.86 $\pm$ 0.07 & & & & CAHAT \\ 
509.3 & 23  &                    &                   & 17.12 $\pm$ 0.14  & 16.44 $\pm$ 0.11&                  & & & &  SXVF-H9+ST-9E/9XE unfiltered \\
510.3 & 24  &                    &                   & 17.10 $\pm$ 0.14  &                  &                  & & & &  SXVF-H9 unfiltered\\
512.3 & 26  &                    &                   & 17.08 $\pm$ 0.12  & 16.42 $\pm$ 0.09 &                  & & & & SXVF-H9+ST-9E/9XE unfiltered\\
514.3 & 28  &                    &                   & 16.99 $\pm$ 0.13  &                  &                  & & & &  SXVF-H9 unfiltered\\
514.7 & 29  &                    & 17.91 $\pm$ 0.03  & 17.00 $\pm$ 0.01   & 16.48 $\pm$ 0.01  & 15.91 $\pm$ 0.01 & & & & CAO  \\
515.3 & 29  &                    &                   & 17.06 $\pm$ 0.40  &                    &                 & & & &  SXVF-H9 unfiltered\\
515.8 & 30  & 18.26 $\pm$ 0.10   & 17.89 $\pm$ 0.03  & 17.01 $\pm$ 0.02   & 16.47 $\pm$ 0.01  & 15.91 $\pm$ 0.01 & & & & LT\\
517.3 & 32  &                    &                   &                    & 16.52 $\pm$ 0.08  & 16.03 $\pm$ 0.13 & & & &  MKS \\
522.7 & 37  &                    & 17.98 $\pm$ 0.01  & 17.10 $\pm$ 0.01   & 16.57 $\pm$ 0.01  & 16.11 $\pm$ 0.02 & & & &  CAHAT \   \\
523.2 & 37  &                    & 18.15 $\pm$ 0.14  & 17.08 $\pm$ 0.07   & 16.61 $\pm$ 0.04  & 16.20 $\pm$ 0.06 & & & &   MSK \\
523.3 & 37  &                    &                   & 17.05 $\pm$ 0.08  & 16.54 $\pm$ 0.07   &                  & & & &SXVF-H9+ST-9E/9XE  unfiltered\\
524.7 & 39  & $>18.70$           & 18.04 $\pm$ 0.06  & 17.18 $\pm$ 0.02   & 16.67 $\pm$ 0.02  & 16.10 $\pm$ 0.01 & & & & LT  \\
525.7 & 40  &                    &                   & 17.13 $\pm$ 0.06   &                    &                  & & & & SXVF-H9 unfiltered\\
526.8 & 41  &                    &                   &                    &                   &                  &15.20   $\pm$ 0.09 &15.16  $\pm$ 0.09 &14.95  $\pm$ 0.1  & TNGN \\
528.7 & 43  &                    &                   & 17.22 $\pm$ 0.11  &                    &                  & & & &  SXVF-H9 unfiltered\\
530.3 & 44  &                    &                   & 17.25 $\pm$ 0.12  &                     &                  & & & &  SXVF-H9 unfiltered\\
530.8 & 45  & 18.90 $\pm$ 0.15   & 18.22 $\pm$ 0.05  & 17.37 $\pm$ 0.03   & 16.75 $\pm$ 0.03  & 16.20 $\pm$ 0.02 & & & & LT \\
532.7 & 47  &                    & 18.39 $\pm$ 0.02  & 17.38 $\pm$ 0.01   & 16.83 $\pm$ 0.01  & 16.26 $\pm$ 0.02 & & & & CAHAT \\
533.3 & 47  &                    &                                  &                                     & 16.65 $\pm$ 0.11  &                     &                  & & &  ST-9E/9XE unfiltered\\
537.6 & 52  &                    & 18.46 $\pm$ 0.11  & 17.46 $\pm$ 0.05   & 16.91 $\pm$ 0.03  & 16.45 $\pm$ 0.04 & & & &SLV   \\
539.7 & 54  &                    & 18.59 $\pm$ 0.01  & 17.57 $\pm$ 0.02   & 16.99 $\pm$ 0.02  & 16.41 $\pm$ 0.01 & & & & LT \\
543.6 & 57  &                    &                   &                    & 17.15 $\pm$ 0.05  & 16.55 $\pm$ 0.05 & & & &  SLV \\
544.7 & 59  &                    &                   &                    & 17.14 $\pm$ 0.04  &                  & & & & NOT  \\
546.6 & 61  &                    &                   & 17.88 $\pm$ 0.14   & 17.27 $\pm$ 0.10  & 16.70 $\pm$ 0.07 & & & & SLV    \\
546.7 & 61  &                    &                   &                    &                   &                 & 15.75  $\pm$ 0.10  &  & & AZT24  \\
554.2 & 68  &                    & 19.25 $\pm$ 0.08  & 17.96 $\pm$ 0.01   & 17.35 $\pm$ 0.02  & 16.70 $\pm$ 0.01 & & & & LT   \\
555.5 & 69  &                    &                   &                    & 17.31 $\pm$ 0.21  &                  & & & & SLV  \\
556.6 & 71  &                    &                   & 18.18 $\pm$ 0.09   & 17.44 $\pm$ 0.04  & 16.81 $\pm$ 0.05 & & & & SLV \\
556.6 & 71  &                    &                   &                    &                   &                 & 16.00   $\pm$0.09 &15.62   $\pm$ 0.09 &15.39  $\pm$ 0.15 & AZT24  \\
556.7 & 71  &                    & 19.30 $\pm$  0.09 & 18.16 $\pm$ 0.02   & 17.43 $\pm$ 0.01  & 16.78 $\pm$ 0.01 & & & & CAO  \\
557.7 & 72  &                    &                   & 18.15 $\pm$ 0.04   & 17.43 $\pm$ 0.04  & 16.77 $\pm$ 0.03 & & & & LT   \\
570.7 & 84  & 20.92 $\pm$ 0.10   & 20.16 $\pm$  0.07 & 18.63 $\pm$ 0.03   & 17.82 $\pm$ 0.02  & 17.11 $\pm$ 0.01 & & & & LT  \\
572.7 & 86  & 20.93 $\pm$ 0.14   & 20.17 $\pm$  0.04 & 18.78 $\pm$ 0.02   & 17.87 $\pm$ 0.01  & 17.16 $\pm$ 0.01 & & & & LT   \\
573.6 & 87  & 20.93 $\pm$ 0.30   & 20.18 $\pm$  0.09 & 18.79 $\pm$ 0.03   & 17.90 $\pm$ 0.01  & 17.14 $\pm$ 0.01 &16.32  $\pm$0.08 &  15.96  $\pm$ 0.08&15.65  $\pm$ 0.09 & TNGD,TNGN  \\
574.6 & 88  &                    & 20.13 $\pm$  0.07 & 18.83 $\pm$ 0.03   & 18.03 $\pm$ 0.02  & 17.20 $\pm$ 0.02 & & & & LT   \\
576.6 & 90  &                    & 20.34 $\pm$  0.14 & 18.82 $\pm$ 0.04   & 18.10 $\pm$ 0.03  & 17.32 $\pm$ 0.01 & & & & LT  \\
577.6 & 91  & 21.01 $\pm$ 0.30   & 20.36 $\pm$  0.01 & 18.97 $\pm$ 0.03   & 18.12 $\pm$ 0.02  & 17.33 $\pm$ 0.02 & & & & LT   \\
579.5 & 93  &                    &                   & 19.10 $\pm$ 0.10   & 18.18 $\pm$ 0.10  & 17.33 $\pm$ 0.07 & & & &  MSKL\\
582.6 & 96   &                   &                   & 19.15 $\pm$ 0.05   &                   &                  & & & & CAHAT   \\
585.4 & 99  &                    &                   & 19.24 $\pm$ 0.30   & 18.35 $\pm$ 0.09  & 17.54 $\pm$ 0.08 & & & & MSKL  \\
590.4 & 105  &                   &                   &                    & 18.45 $\pm$ 0.11  & 17.67 $\pm$ 0.10 & & & &  MSKL \\
590.7 & 105  &                   &                   & 19.51 $\pm$ 0.01   & 18.55 $\pm$ 0.01  &                  & & & & LT  \\
592.6 & 107  &                   & 20.88 $\pm$  0.09 & 19.57 $\pm$ 0.06   & 18.61 $\pm$ 0.03  & 17.80 $\pm$ 0.02 & & & & CAO   \\
593.6 & 108  &  21.68 $\pm$ 0.20 & 21.07 $\pm$ 0.13  & 19.65 $\pm$ 0.05   & 18.67 $\pm$ 0.02  & 17.82 $\pm$ 0.02 & & & & TNGD  \\
601.5 & 115  &                   &                   &                    & 19.10 $\pm$ 0.2  &                  & & & &  WOT       \\
602.2 & 116  &                   &                   & 20.06 $\pm$ 0.09   & 19.15 $\pm$ 0.04  &                  & & & & LT   \\
602.5 & 117  &                   &                   &                    & 19.17 $\pm$ 0.47  &                  & & & &  MSKL   \\
604.5 & 118  &                   &                   &                    & 19.28 $\pm$ 0.37  & 18.10 $\pm$ 0.26 & & & & CRM \\
607.5 & 122  &                   &                   & 20.26 $\pm$ 0.26   &                   &                  & & & & CRM   \\
607.7 & 122  &                   &                   & 20.27 $\pm$ 0.06   & 19.26 $\pm$ 0.04  &                  & & & & LT    \\
608.1 & 123  &                   &                   & 20.25 $\pm$ 0.11   & 19.35 $\pm$ 0.05  &                  & & & & LT   \\
613.4 & 127  &                   &                   &                    & 19.39 $\pm$ 0.19  & 18.45 $\pm$ 0.15 & & & &  MSKL \\
618.5 & 132  &                   &                   & 20.58 $\pm$ 0.21   &                   &                  & & & &   SAO \\
619.5 & 133  &                   &                   &                    & 19.45 $\pm$ 0.20  &                  & & & &  WOT  \\
621.5 & 135  &                   &                   & 20.65 $\pm$ 0.27   & 19.49 $\pm$ 0.20  & 18.55 $\pm$ 0.08 & & & & CAO   \\
627.6 & 142  &                   &                   & 20.93 $\pm$ 0.13   & 19.67 $\pm$ 0.07  &                  & & & & LT   \\
629.6  & 144 &                    &                   &                    &                   &                 & 17.68  $\pm$0.06 & 16.72  $\pm$ 0.05 &16.12  $\pm$ 0.05 & TNGN \\
637.4  & 151  &                  &                   & 20.94 $\pm$ 0.14   & 19.87 $\pm$ 0.06  & 18.95 $\pm$ 0.03 & & & & LT \\
642.6  & 157  &                  &                   & 21.01 $\pm$ 0.21   & 19.94 $\pm$ 0.07  & 19.05 $\pm$ 0.04 & & & & LT\\
646.6  & 159  &                  & 22.47 $\pm$ 0.22  & 21.08 $\pm$ 0.14   & 20.05 $\pm$ 0.09  &                  & & & & CAHAT\\
648.4  & 162  &                  &                   & 21.18 $\pm$ 0.30   & 20.02 $\pm$ 0.07  & 19.07 $\pm$ 0.07 & & & & LT \\
651.5  & 165  &                  &                   & 21.30 $\pm$ 0.30   & 20.06 $\pm$ 0.05  & 19.07 $\pm$ 0.05 & & & & LT\\
658.5  & 172  &                  &                   & 21.34 $\pm$ 0.28   & 20.15 $\pm$ 0.08  & 19.13 $\pm$ 0.05 & & & & LT\\ 
659.5  & 174  &                    &                   &                    &                   &                & 17.77  $\pm$0.06 & 16.75  $\pm$ 0.08 &16.05  $\pm$ 0.08 & TNGN \\
667.5  & 181  &                  &                   &                    & 20.21 $\pm$ 0.26  & 19.24 $\pm$ 0.32 & & & & LT \\
672.4  & 186  &                  &                   & 21.52 $\pm$ 0.15   & 20.27 $\pm$ 0.05  & 19.34 $\pm$ 0.05 & & & & LT\\
677.4  & 191  &                  &                   & 21.53 $\pm$ 0.23   & 20.28 $\pm$ 0.06  & 19.33 $\pm$ 0.04 & & & & LT\\
697.5  & 212  &                    &                   &                    &                   &                & 17.99  $\pm$0.10  & 16.78  $\pm$ 0.09 &16.00  $\pm$ 0.08 & TNGN \\
714.4  & 228  &                  &                   & 22.09 $\pm$ 0.30   & 20.75 $\pm$ 0.11  & 19.65 $\pm$ 0.10 &18.14  $\pm$0.10 &16.89  $\pm$ 0.1  &15.90  $\pm$ 0.1 & TNGN,TNGD \\
782.4 & 296   &                  &                   & 22.85 $\pm$0.30    & 21.30$\pm$ 0.20   & 20.30 $\pm$ 0.10 & & & & LT  \\
790.4  & 304  &                    &                   &                    &                   &                & 18.85 $\pm$ 0.10  &17.20 $\pm$0.05  &15.82 $\pm$0.05 &TNGN \\
\hline
\end{tabular}
\\[1.4ex]
$^a$JD $-$ 2\,454\,000.00\quad\\
$^b$ Phase is in days after the explosion epoch JD $2\,454\,486 \pm 4$.\quad
\end{scriptsize}
\end{table*}

Our calibration is in agreement with that of \cite{Pozzo2006} with an average difference of 0.01\,mag in the $V$  and $I$ bands and of 0.005\,mag in the $R$ band.
The magnitudes of the calibrated local sequence,  listed in Table~\ref{seqstar}, were subsequently  used to measure the relative SN 2008S magnitude for each observation.
The $UBVRI$ magnitudes of SN  2008S are reported with their uncertainties estimated by combining in quadrature the error
of the photometric calibration and the error in the PSF fitting in Table~\ref{phot}.  The responses of the SXVF H9 camera used by
     A. Arbour, and the ST-9E/9XE camera used by M. Mobberly peak,
     respectively, in the $V$ and $R$ bands.  Therefore, although strictly
     these cameras were unfiltered, we nevertheless list the magnitudes
     obtained in the $V$ or $R$ columns of Table~\ref{phot}.

\subsection{Near Infrared data} 

The near infrared (NIR) photometry was obtained with the TNG with NICS and with the AZT 24 telescope  with SWIRCAM, in the $JHK$ filters (Table~\ref{phtel}). 
The NIR images were reduced using standard IRAF routines, with the jittered exposures first median-combined to obtain sky images in each band. Jittered images were then sky subtracted, registered and finally combined. The instrumental magnitudes were measured on the combined images with SNOoPY package.
Photometric calibration was carried out via relative 2MASS photometry of the same local sequence stars used for optical data. 
The NIR magnitudes of SN 2008S are listed in Table~\ref{phot}.

\subsection{Photometric evolution and bolometric lightcurve}\label{bolsec}

\begin{figure*}
\resizebox{\hsize}{!}{\includegraphics{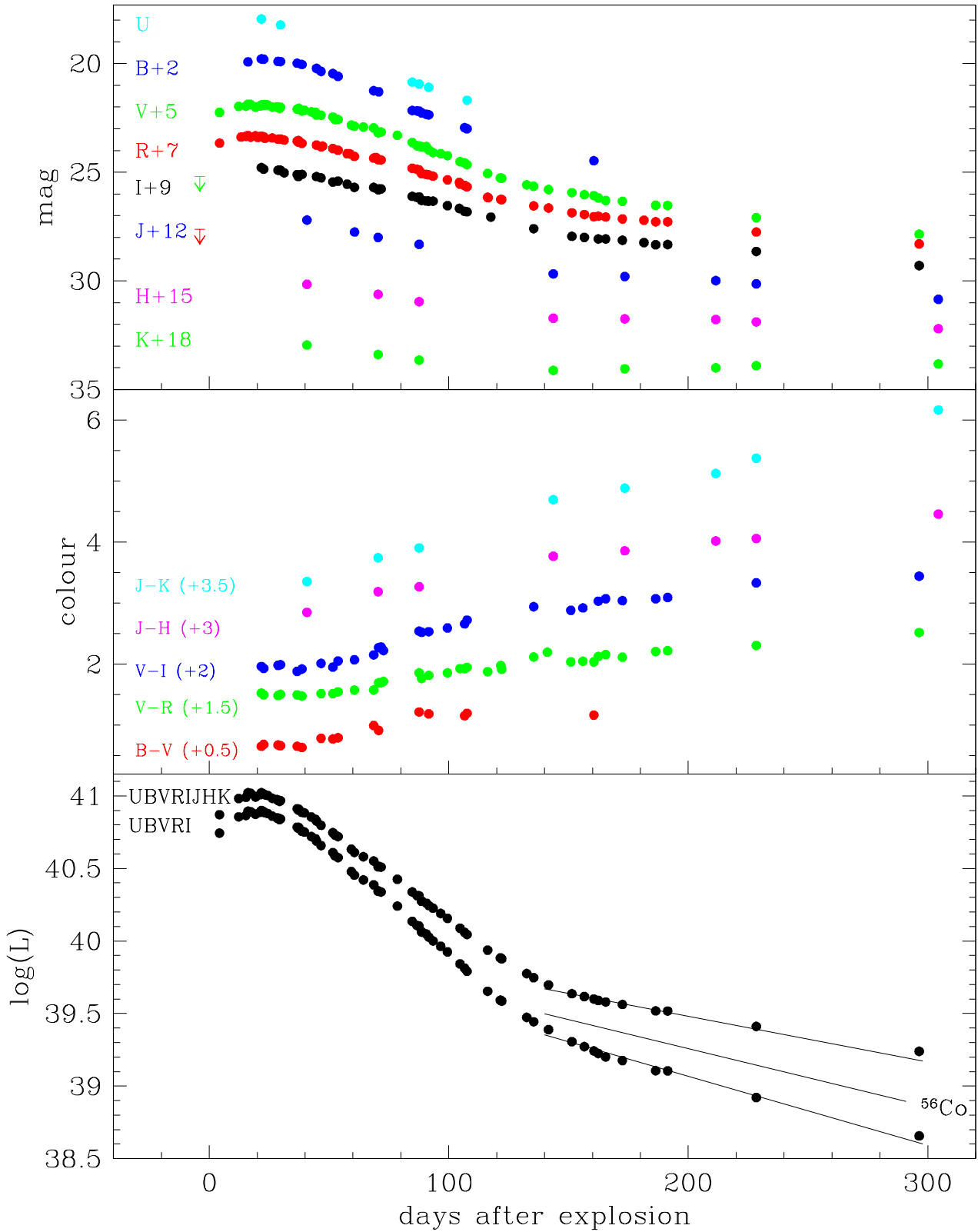}}
\caption{Top panel: $UBVRIJHK$ light curves of SN2008S. Middle panel: evolution of the $B-V$, $V-R$, $V-I$, $J-H$, $J-K$ colours. The magnitudes have been corrected for the adopted extinction according to the Cardelli extinction law. Bottom panel:  the $UBVRIJHK$ and $UBVRI$ quasi-bolometric light curves. The $^{56}$Co decay slope and the slope of $UBVRIJHK$ and $UBVRI$ curves are shown. Phase is in days after the explosion epoch (JD $2\,454\,486$).}
\label{lc}
\end{figure*}

In Fig.~\ref{lc} (top panel) the $UBVRIJHK$
light curves of SN 2008S are illustrated. These are
characterised by a broad peak and a subsequent slow decline.  In the $R$ and
$V$ bands there is clearly a fast rise to peak  shown by early 
 observations from a range of amateur telescopes.
 The peak occurs progressively earlier from the blue to the red bands (Table~\ref{maxdata}).
The absolute magnitudes at maximum (Table~\ref{maxdata}),
calculated adopting $\mu =28.78 \pm 0.08$\,mag and correcting for
Galactic ($A_V = 1.13$\,mag) and  internal extinction ($A_{V}=1$\,mag), reveal
that SN 2008S is brighter by 1-1.5 mag than the 
NGC 300 OT2008-1 and M 85 OT2006-1 transients with which it has been compared 
(see Sect.~\ref{NGC300-M85-comp}). 

\begin{table}
\caption{Epochs, apparent and absolute magnitudes of light curve maximum in the $BVRI$ bands.\label{maxdata}}
\begin{footnotesize}
\begin{tabular}{lcccc}
\hline         
Filter &     JD$^a$    & ph$^b$ &     m$_\mathrm{max}$  &     M$_\mathrm{max}$ \\
\hline 
$B$  &       $509 \pm 2$  & 23 &  $17.83 \pm 0.05$ & $-13.76 \pm 0.16$\\
$V$  &       $505 \pm 2$  & 19 &  $16.95 \pm 0.05$ &  $-13.97 \pm 0.16$ \\
$R$  &       $503 \pm 2$  & 17 &  $16.33 \pm 0.05$ &  $-14.17 \pm 0.16$\\
$I $ &        $502 \pm 3$  & 16 &  $15.85 \pm 0.05$ & $-14.20 \pm 0.16$\\
\hline
\end{tabular}
\\[1.5ex]
$^a$JD $-$ 2\,454\,000.00\quad\\
$^b$ Phase is in days after the explosion epoch JD $2\,454\,486 \pm 4$.\quad
\end{footnotesize}
\end{table}

The $BVRI$ light curves show very similar temporal evolutions with three phases characterized by a
different decline rate: a
broad peak (about two weeks), a  phase of steeper decline ($\gamma_1$ in
Table~\ref{gammaopt}) starting about 60 days after  explosion, more pronounced at shorter
wavelengths,  and a flattening  ($\gamma_2$) after 140 days, more evident at longer wavelengths.
Due to the faintness of SN 2008S we do not have $U$ and $B$ band observations at
the late phases to obtain an accurate estimate of the decline rate.
In the NIR bands the decline rates are very slow until 120 days and show further a flattening (or a slight increase in $K$ band) after this epoch.

The time evolution of the $B-V$,  $V-R$, $V-I$, $J-H$ and $J-K$ colours of SN 2008S is shown in Fig.~\ref{lc}.  All optical colours become
progressively redder until about  200 days after the explosion. Subsequently
the colours do not show significant evolution.  This trend is most
evident for the $V-I$ colour.
The NIR colours show a different evolution after about 200 days after the explosion: $J-H$ is slightly increasing while $J-K$ is steepening.
\begin{table}
\caption{Decline rates in the $BVRIJHK$ bands.\label{gammaopt}}
\begin{footnotesize}
\begin{tabular}{lcccr}
\hline   
 Filter &  ph$_1$$^a$ & $\gamma_1$(mag/100d)   & ph$_2$$^a$ &  $\gamma_2$ (mag/100d) \\  
  \hline
$B$ & 50-100 & $4.5 \pm 0.10$ & 100-160 &$ 2.8 \pm 0.10$\\
$V$ & 50-120 & $4.0 \pm  0.05$ &140-300 & $1.3  \pm 0.06$  \\ 
$R$ & 60-120 & $3.4 \pm 0.05$  & 140-300 & $1.0 \pm  0.05$ \\
$I$ & 60-120 & $2.8 \pm 0.05$ & 140-300 & $0.9 \pm  0.06$ \\
$J$ &40-120 &$2.4 \pm 0.10$ & 140-310 &$0.7 \pm  0.05$\\
$H$ &40-120 &$1.5 \pm 0.10 $&140-310  &$0.4 \pm 0.10$\\
$K$ &40-120 &$1.5 \pm 0.09$& 140-310 &$-0.2\pm 0.06$ \\
\hline
\end{tabular}
\\[1.5ex]
$^a$ Phase is in days after the explosion epoch JD $2\,454\,486 \pm 4$.\quad
\end{footnotesize}
\end{table}

A ``bolometric''  light curve (Fig.~\ref{lc}, bottom panel) was obtained by first converting $UBVRIJHK$ magnitudes
 into monochromatic fluxes per unit wavelength, then correcting these fluxes for the adopted extinction ($A_{V}=2.13$\,mag) according to the extinction law from \cite{Cardelli1989}, and finally  integrating the resulting SED over  
wavelength, assuming zero flux at the integration limits (the blue end  
of the $U$ band and the red edge of the $K$ band). 
We estimated the flux only for the phases in which $V$ band observations were available. The photometric data in the other bands were estimated at these phases by interpolating magnitudes in subsequent nights.  

During the period 140--290 days,
          the $UBVRIJHK$ ``bolometric'' light curve tail shows a decay rate of $0.88 \pm 0.05$\,mag/100d, very similar to that of $^{56}$Co  \citep[1.023\,mag/100d from][] {Huo1987},  while 
the $UBVRI$  ``bolometric''  lightcurve is slightly steeper ($1.3 \pm 0.05$\,mag/100d).
Assuming that
 radioactive material was powering the late time photometric evolution
 of SN 2008S, we tried to estimate the $^{56}$Ni mass synthesized by
 SN 2008S by comparing its $UBVRIJHK$ quasi-bolometric light curve with
 that of SN 1987A from 140 to 300 days after explosion and assuming similar $\gamma-$ray deposition fraction:
\beq
\mathrm{M}(^{56}\mathrm{Ni})_\mathrm{08S}=\mathrm{M}(^{56}\mathrm{Ni})_\mathrm{87A} \times \frac{\mathrm{L}_\mathrm{08S}}{\mathrm{L}_\mathrm{87A}} 
\eeq
where M($^{56}$Ni)$_\mathrm{87A}$ is the mass of $^{56}$Ni produced by SN
1987A, L$_\mathrm{08S}$ is the luminosity of SN 2008S and L$_\mathrm{87A}$ is the
luminosity of SN 1987A (at a similar epoch)
also obtained from $UBVRIJHK$ data. We
adopt M($^{56}$Ni)$_\mathrm{87A}=0.073 \pm 0.012$\,M$_{\odot}$ which is the
weighted mean of values given by \cite{Arnett1989} and by
\cite{Bouchet1991}.  For SN 2008S we obtain a  M($^{56}$Ni) of
$0.0014\pm 0.0003$\,M$_{\odot}$ where the error includes both the
uncertainties in the assumed distance of SN 2008S and in the $^{56}$Ni
mass of SN 1987A.   If we considered only optical data for both SN  2008S and SN 1987A  we would obtain a M($^{56}$Ni) of $0.0011\pm 0.0002$\,M$_{\odot}$.
We estimated the $^{56}$Ni mass also using the
method of \cite{Hamuy2003} assuming that all the $\gamma-$rays
resulting from the decay of $^{56}$Co into $^{56}$Fe are fully
thermalized:
\beq
\mathrm{M}(^{56}\mathrm{Ni})_\mathrm{08S}=7.866\times 10^{-44}\mathrm{L}  exp[\frac{(t-t_{0})/(1+z)-\tau_\mathrm{Ni}}{\tau_\mathrm{Co}}]  \mathrm{M}_{\odot}
\eeq
where t$_{0}$ is the explosion epoch, $\tau_\mathrm{Ni}=8.76$ days the life time of $^{56}$Ni and $\tau_\mathrm{Co}=111.26$ days is the life time of $^{56}$Co. Using this method we estimated M($^{56}$Ni) for each point of the radioactive tail and the average of these estimates gives an ejected $^{56}$Ni mass of $\sim$\,0.0012\,M$_{\odot}$.
We remark that the estimated value of $^{56}$Ni  mass has to be considered as an upper limit if the ejecta-CSM interaction or an IR echo contributes to the tail of the bolometric light curve.

\subsection{Data at other wavelengths}\label{Other data}
SN 2008S was serendipitously observed
   with Spitzer on 2008 Feb 6.8 UT  during scheduled
   observations of the nearby SN~2002hh  showing a strong MIR emission \citep{Wesson2008}.  Images were obtained with IRAC (3.6\,$\mu$m,
4.5\,$\mu$m, 5.8\,$\mu$m and 8.0\,$\mu$m) on day~17.3 and MIPS (24\,$\mu$m) on day~18.0,  uniquely early epochs for coverage of
this wavelength region.
 The Swift satellite also observed SN 2008S in the optical, UV and X-ray bands on Feb. 4.8, 6.0, and 10.5  UT and detected it only in the optical images \citep{Smith2008c}.
SN 2008S was not detected on February 10.62 UT at radio frequencies with the Very Large Array by \cite{Chandra2008ATel}.
The contribution to the overall energy budget of SN 2008S at about 20 days after the explosion seems to be negligible at UV wavelengths but it is considerable at MIR wavelengths. 
In Sect.~\ref{MIR excess}, we shall
        demonstrate that the large MIR luminosity must have been caused
        by an IR echo from pre-existing circumstellar dust.

\section{Spectral energy distribution evolution}\label{SEDsec}
\subsection{Up to day~120}\label{MIR excess}
\begin{figure}
\resizebox{\hsize}{!}{\includegraphics{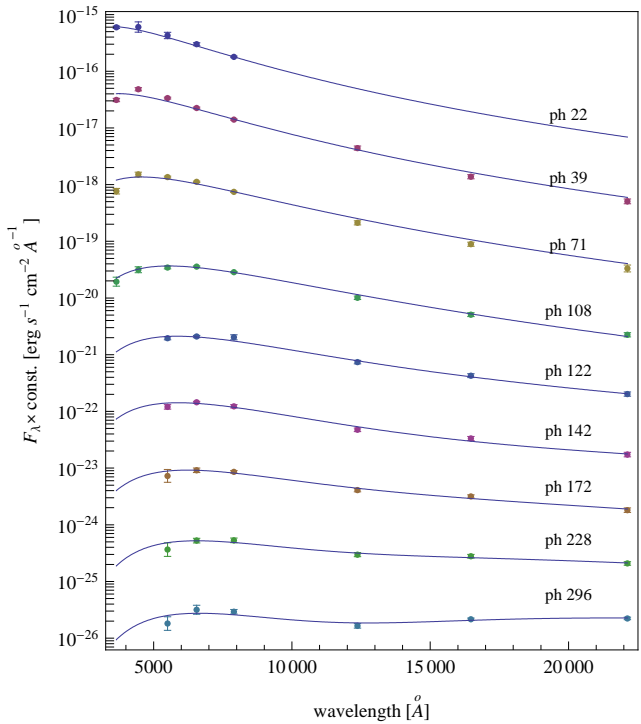}} 
\caption{Temporal evolution of the observed SED of SN 2008S.  Lines show the blackbody fits to the SED.  Observed fluxes are corrected for the adopted extinction ($A_{V}=2.13$\,mag) according to the extinction law from Cardelli et al (1989). Phase is in days after the explosion epoch (JD $2\,454\,486$). }
\label{sedev}
\end{figure}

\begin{figure}
\resizebox{\hsize}{!}{\includegraphics{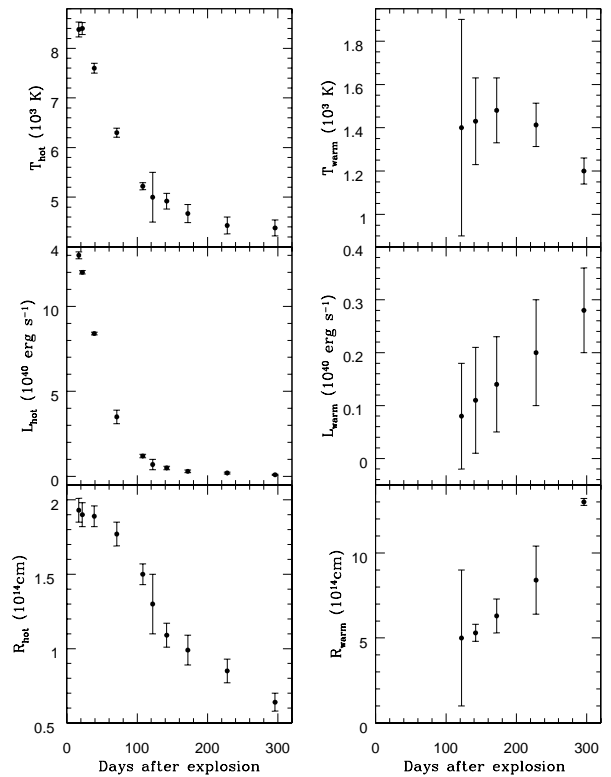}}
\caption{Temporal evolution of the parameters  of the hot (left) and warm (right) blackbodies fitting the SED of SN 2008S.  Phase is in days after  the explosion epoch (JD $2\,454\,486$).}
\label{bb_evol}
\end{figure}

The SED evolution of SN 2008S  is shown in Fig.~\ref{sedev}.  During
the first $\sim$\,120~days, the optical-NIR fluxes can be well
reproduced by a single hot blackbody. The blackbody temperatures,
radii  and luminosities  for some epochs are shown in Table~\ref{bbfit} and plotted in Fig.~\ref{bb_evol}.

\begin{table*}
\caption{Parameters for a single blackbody fit (until 108 days after explosion) and for a two-component (hot and warm blackbodies) fit to the $UBVRIJHK$ fluxes of SN 2008S.\label{bbfit}}
\begin{footnotesize}
\begin{tabular}{llllllll}
\hline
ph$^a$  & T$_\mathrm{hot}$ & R$_\mathrm{hot}$ & L$_\mathrm{hot}$ & T$_\mathrm{warm}$ & R$_\mathrm{warm}$ &  L$_\mathrm{warm}$ & L$_\mathrm{warm}$/L$_\mathrm{tot}$   \\
            & K &  $10^{14}$ cm & $10^{40}$ erg s$^{-1}$ & K & $10^{14}$ cm &$10^{40}$ erg s$^{-1}$ & per cent \\
\hline
17  &   $8380 \pm      150$   & $1.93 \pm 0.08$ & $13 \pm 0.2$ &                    &               &           \\  
22  &	$8400 \pm	120$  &	$1.90 \pm 0.08$ &	$12 \pm	0.1$&                     &                    &               &           \\       
39 &	$7600 \pm	100$  &	$1.89 \pm 0.07$ &	 $8.4 \pm 0.1$&                     &                    &               &         \\        
71 &	$6300 \pm	 90$  &  $1.77 \pm  0.08$ & $3.5 \pm 0.4$&                     &                    &               &          \\
108 &	$5220 \pm 70$   &	$1.50 \pm 0.07$ &	 $1.2 \pm 0.1$&                    &                    &               &          \\
122 &	$5000 \pm	500 $ &  $1.30 \pm 0.20$ &	 $0.7 \pm 0.3$&   $1400 \pm 500$ &	$5.0  \pm 4.0$  & $0.08 \pm 0.1$ & 10 \\ 
142 &       $4920 \pm 160$ & $1.09 \pm 0.08$ & $0.5 \pm0.1$& $1430 \pm 200$& $5.3 \pm 0.5$ & $0.11 \pm 0.1$& 25\\
172 &	$4670 \pm      180 $ &  $0.99 \pm  0.10$&	 $0.3 \pm 0.08$&   $1480    \pm 150$ &      $6.3  \pm 1.0 $& $0.14 \pm 0.09$ & 32  \\
228 &	$4430 \pm	170 $ &  $0.85 \pm  0.08$& $0.2 \pm 0.05$&   $1413	\pm 100$ &$8.4  \pm 2.0$  &$ 0.20 \pm 0.1$ & 50\\
296 &   $4380 \pm    160 $ &  $0.64 \pm 0.06$ &  $0.1\pm 0.03$&               $1200    \pm 60$ &   $ 13   \pm 0.2 $ & $0.28  \pm 0.08$ & 74 \\ 
\hline
\end{tabular}
\\[1.5ex]
$^a$Phase is in days after the explosion epoch JD $2\,454\,486 \pm 4$.\quad
\end{footnotesize}
\end{table*}

The temperature fell monotonically from $\sim$\,8300\,K to $\sim$\,5000\,K
during this time. The radius declined slowly from about
$1.9\times10^{14}$\,cm to $1.3\times10^{14}$\,cm,  suggesting that the
blackbody surface was defined by the photosphere through which the
ejecta flowed. 

Remarkably, a strong MIR excess was observed at $\sim$\,18 days after explosion by \cite{Wesson2008}. 
The SED obtained with the Spitzer data  was fitted by a 500\,K blackbody modified by a dust
emissivity that is inversely proportional to the wavelength, giving a luminosity
of $21 \times 10^{6}$\,L$_{\odot}$. 
We
have re-measured the Spitzer images, using the post-basic
calibrated data (PBCD) products provided by the  Spitzer
pipeline. The 8\,$\mu$m and 24\,$\mu$m images showed strong
irregular background at or near the location of the SN.  We
therefore subtracted serendipitously obtained pre-explosion
``templates'' from the post-explosion images before proceeding with
our flux measurements.  The IRAC templates were taken from 
Spitzer programme 30292 (P.I. Meikle)  and the
MIPS template from programme 0230 (P.I. Barlow).  The image matching
and subtraction was performed as implemented in the ISIS v2.2 image
subtraction package \citep{Alard2000}, and modified in a
manner analogous to that described in \citet{Meikle2006}. Aperture
photometry was performed on the background-subtracted IRAC and MIPS
images using the Starlink package GAIA \citep{Draper2000}.  A circular
aperture of radius 5\arcsec was used for the photometry. This was chosen
as a compromise between minimising the effects of the residual
irregular background emission at the SN location
and minimising the the size of aperture correction needed in the final
flux determination.  The aperture radius corresponds to a distance of
$\sim$\,140\,pc at SN~2008S.  Residual background in the
template-subtracted IRAC and MIPS images was measured and subtracted
by using a clipped mean sky estimator, and a concentric sky annulus
having inner and outer radii of 1.5 and 2 times the aperture radius,
respectively.  Aperture corrections were derived from the IRAC and
MIPS point response function frames available from the  Spitzer
Science Center, and ranged from $\times1.04$ at 3.6\,$\mu$m to
$\times 2.12$ at 24\,$\mu$m.  
For IRAC, the aperture was centred by
centroiding on the SN image. For MIPS, the aperture was centred
visually on the SN, checking against the WCS coordinates.  
We found fluxes of 
$1.60\pm 0.02$\,mJy at 3.6\,$\mu$m, $1.97 \pm 0.03$\,mJy
at 4.5\,$\mu$m, $3.06 \pm 0.07$ mJy at 5.8\,$\mu$m and $4.15\pm 0.05$\,mJy
at 8\,$\mu$m.  
The errors are statistical only and do not include
calibration uncertainties which may amount to an additional
$\sim$\,$\pm5\%$. The image subtraction at 24\,$\mu$m left a substantial
residual. We therefore also carried out aperture photometry on the
unsubtracted 24\,$\mu$m image. Consistent results were obtained for the
subtraction and non-subtraction procedures.  The mean flux derived
from the two methods is $0.7 \pm 0.1$\,mJy.  Our IRAC and MIPS fluxes
are consistent with those reported by \citet{Wesson2008}.  However,
their errors are at least several times larger.

\begin{figure}
\resizebox{\hsize}{!}{\includegraphics{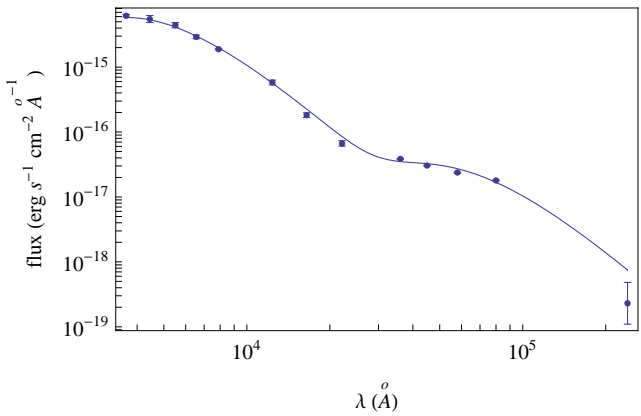}}
\caption{ Fit of two blackbodies to the SN 2008S SED from optical to MIR at 17.3 days after explosion. The NIR data are obtained extrapolating data from  phase 41. Observed fluxes are corrected for the adopted extinction ($A_{V}=2.13$\,mag) according to the extinction law from Cardelli et al (1989).}
\label{bb17}
\end{figure}

To investigate the nature of the MIR source, we carried out a
simultaneous fit of two blackbodies to the SED at 17.3 days, and this is
shown in Fig.~\ref{bb17}.
It can be seen that the blackbodies provide a plausible representation
of the data. The hot blackbody has a temperature of  $8076 \pm 150$\,K and
a radius of $(2.1\pm0.1)\times10^{14}$\,cm, corresponding to an
expansion velocity of 1430\,\kms. These estimates are consistent with
emission from the hot photosphere.  The warm blackbody has a
temperature of  $585 \pm 5$\,K. This temperature together with the
reasonable match to the blackbody leads us to propose that the MIR
emission must be due to warm dust.  The radius of the warm blackbody
is $(9.9\pm0.4)\times10^{15}$\,cm. For the SN to produce a surface
of this radius in just 17.3~days would take a velocity of
$66,200 \pm 2700$\,\kms. Such an enormous velocity immediately rules out
newly-formed ejecta dust as the source of the MIR emission. The high
velocity also rules out collision of the fastest moving ejecta with a
dusty CSM. The only viable alternative appears to be an IR echo from
CSM dust. To explore this possibility we have matched an IR echo model
to the MIR SED. Details of the model are given in \cite{Meikle2006}.
This assumes a spherically symmetric cloud of carbon grains
centred on the SN, with a concentric dust-free cavity at the centre.
For simplicity, a single grain radius, $a$, is adopted.  For ease of
computation, we assumed that the grain material was amorphous carbon
where, for wavelengths longer than $2 \pi a$, the grain
absorptivity/emissivity can be well-approximated as being proportional
to $\lambda^{-1.15}$ \citep{Rouleau91}.  For shorter wavelengths, an
absorptivity/emissivity of unity was used.  The material density is
1.85\,g\,cm$^{-3}$ \citep{Rouleau91}.  Free parameters are the grain
size, grain number density, the CSM radial density law, the CSM extent
and the size of the concentric dust-free cavity.  The outer limit of
the CSM was set at 10 times that of the cavity radius, although this
parameter is not critical.  The input luminosity is a parametrized
description of the $UBVRIJHK$ ``bolometric''  light curve 
(Fig.~\ref{lc}). Given the apparently low temperature of the dust, it is
reasonable to assume that the IR echo made only a small contribution
to the NIR flux implying a negligible overestimate of the  
bolometric light curve. 

\begin{figure}
\resizebox{\hsize}{!}{\includegraphics{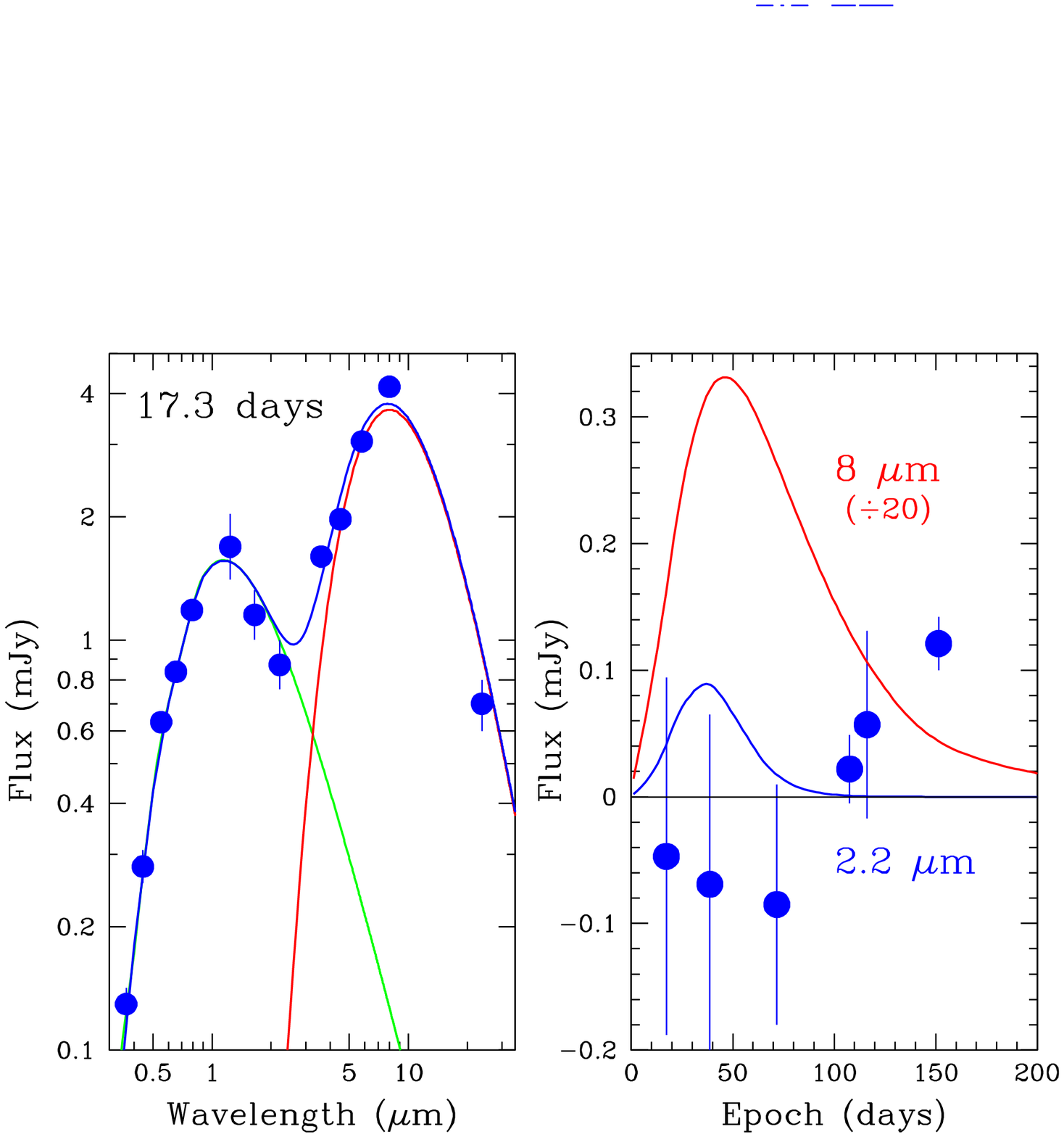}}
\caption{The left hand panel shows the day~17.3 hot blackbody spectrum
(green), the IR~echo spectrum (red) and the combined spectrum
(blue). These are compared with the day~17.3 optical-NIR points and
the day~17.3 MIR points. The hot blackbody has a temperature of
$8076 \pm 150$\,K and a radius of $(2.1\pm0.1)\times10^{14}$\,cm,
reddened using the extinction law of Cardelli et al. (1989) with
$A_V=2.13$\,mag.  The total optical depth through the CSM in the optical
region is about 0.20, or $A_V=0.22$\,mag.  The reddening effects of the
CSM dust are included in the model.  Consequently the final IR~echo
model was reddened with $A_V=2.13-0.22=1.91$\,mag.  For the IR~echo model,
a single grain size was used with radius $a = 0.5$\,$\mu$m.  The grain
material is amorphous carbon where, for wavelengths longer than
$2\pi/a$, the grain absorptivity/emissivity is proportional to
$\lambda^{-1.15}$ \citep{Rouleau91}.  For shorter wavelengths, an
absorptivity/emissivity of unity was used.  The material density is
1.85\,g\,cm$^{-3}$ \citep{Rouleau91}.  The CSM extended from
$3\times10^{16}$\,cm to $3\times10^{17}$\,cm, with the dust density
declining as $r^{-2}$ (steady wind).  The dust mass is
$1.4\times10^{-3}$\,M$_{\odot}$.  The right hand panel shows the residual
fluxes at 2.2\,$\mu$m (blue points) at a number of epochs together with
the 2.2\,$\mu$m light curve from the IR~echo model (blue).  For the
first five points it can be seen that the model flux never exceeds the
residual by much more than 1$\sigma$.  The latest three residual
points show the gradual emergence of the NIR excess (see text).  Also
shown is the IR~echo light curve at 8\,$\mu$m (red).}
\label{echomodel}
\end{figure}

It was found that a range of parameters were able plausibly to
reproduce the MIR flux. Matches were found for grain radii of
0.001--0.5\,$\mu$m.  The corresponding cavity range was
$1\times10^{17}$\,cm to $3\times10^{16}$\,cm and the dust mass range
was $10^{-2}-10^{-3}$\,M$_{\odot}$. Such masses of dust are entirely
plausible within the CSM of a red (super)giant. 
For a typical wind velocity of 20\,\kms
and a dust-gas mass ratio of 0.01, the CSM dust masses we require in
the IR echo model would be produced by a mass loss rate of a few times
10$^{-5}$\,\msol\,yr$^{-1}$.
All the
progenitor circumstellar dust within a particular radius would have
been evaporated in the SN explosion.  Given the peak luminosity of the
SN, the evaporation radius for carbon grains is $3.5 \times 10^{15}$\,cm. Silicate grains would allow an evaporation cavity several
times larger. However, there is little sign of an excess in the
8~$\mu$m band which we would expect from optically thin silicate
grains. Nevertheless, even to account for a cavity radius of $3 \times
10^{16}$\,cm we must invoke some sort of episodic mass-loss history.
The echo model spectrum is shown in Fig.~\ref{echomodel}, left-hand
panel. Also shown are the hot blackbody spectrum obtained in the
two-blackbody fits described above, and the combined hot blackbody and
IR echo spectrum.  This is compared with the optical-NIR-MIR
observations at this epoch.  The earliest NIR data were obtained at
40~days and so the 17.3~days fluxes were estimated by extrapolation of
the NIR light curves. The extrapolation was performed by assuming
constant $V-J$, $V-H$, $V-K$ colours during the first month. (If, instead,
we assume a linear evolution of these colours, then the fluxes in the
NIR bands would be smaller by $6\%$, $22\%$ and $27\%$ in $J, H, K$
respectively.)  The hot blackbody was reddened using the extinction law
of  \cite{Cardelli1989} with $A_V=2.13$\,mag.  The total optical depth in
the optical region through the CSM in the IR~echo model is about 0.20,
or $A_V=0.22$\,mag.  The reddening effects of the CSM dust are included in
the model.  Consequently the final IR~echo model was reddened with
$A_V=2.13-0.22=1.91$\,mag.  Inspection of Fig.~\ref{echomodel} shows that a
satisfactory match to the data is obtained.

While the echo model easily reproduced the MIR flux, it also produced
a certain level of NIR emission, with $K$ being the brightest of the
observed NIR bands.  This, therefore, provides an additional
constraint on the echo model i.e. the model has to be consistent with
the NIR flux and its evolution. As explained above, satisfactory fits
to the early-time optical-NIR SEDs were obtained using a single
blackbody. There was little evidence of a NIR excess up to
around 120~days. However, the blackbody fits are subject to error.  We
therefore investigated whether or not the predicted $K$~band fluxes of
the echo model could be consistent with the uncertainties in the
observed values and in the hot blackbody fit.  For a number of epochs
(17.3, 38.7, 71.7, 107.6, 116.2, and 151.4~days) we compared the echo
model with the residual $K$~band fluxes derived by subtracting the hot blackbody flux to the observed one.  The first five epochs were all consistent with a zero
residual, but with uncertainties of $\sim$\,$\pm 0.1$\,mJy. We then selected
the echo model which yielded the smallest $K$~band fluxes and still
produced a satisfactory match to the MIR fluxes. This was achieved
with a 0.5\,$\mu$m grain radius and a dust-free cavity of
$3\times10^{16}$\,cm. The dust mass is $1.4\times10^{-3}$\,M$_{\odot}$.
The right hand panel of Fig.~\ref{echomodel} shows the residual fluxes
at 2.2\,$\mu$m together with the 2.2\,$\mu$m light curve from the echo
model (coloured blue). The error bars on the residual points were
derived as a combination of the uncertainty in the
observed/extrapolated points and the uncertainties in the hot
blackbody fits.  It can be seen that the model flux never exceeds the
residual by much more than 1\,$\sigma$. The marginal negative shift of
the first three residual points may be due to slightly imperfect
blackbody fits such as might be caused by the actual extinction law
being different from the one adopted.  We deduce that it is quite
plausible that the $K$~band flux from an IR~echo would have been
undetected, hence accounting for the near-zero residuals.  The later
three residual points show the gradual emergence of the NIR excess,
discussed below.  We conclude that the IR echo model accounts for the
MIR flux as well as being consistent with all the shorter wavelength
fluxes.

Also shown in Fig.~\ref{echomodel} (right hand panel) is the 8\,$\mu$m light
curve predicted by the IR~echo model.  We note that neither the
2.2\,$\mu$m nor 8\,$\mu$m light curves exhibit the flat top or plateau
characteristic of the IR~echo phenomenon. This is because, owing to
the exceptionally low bolometric luminosity of SN~2008S, the radius of
the dust free cavity had to be small in order to attain a sufficiently
high dust temperature. Indeed, the radius of the cavity in the model
illustrated here is only 12~light days ($3 \times 10^{16}$\,cm).
Consequently the decline of the light curves are actually dominated by
the characteristic decline timescale of the input bolometric light
curve, which is about 70~days per e-fold, together with the $r^{-2}$
dust density decline.

The final MIR observation of SN~2008S was at about 180~days, but at
the time of writing, the data are not yet out of embargo (due 2009
August).  Nevertheless, we can use the IR~echo model to predict the
fluxes at this epoch.  We find that the 8\,$\mu$m flux at 180~days is
1.3\,mJy for the model discussed above, rising to around 5\,mJy for
models having a grain radius $\sim$ 0.01\,$\mu$m. Thus, the observed
180~day MIR fluxes should provide constraints for the grain size.

\subsection{After day~120: dust formation?}\label{NIR excess}
To fit the SED at later epochs we need the sum of a hot and a warm
blackbody.  We find statistical evidence for the emergence of a warm
component as early as 120 days, viz. $\Delta \mathrm{BIC} =
\mathrm{BIC}_\mathrm{hot}-\mathrm{BIC}_\mathrm{hot+warm} \simeq$\,1.5. This
becomes stronger with time,  reaching $\Delta \mathrm{BIC} \simeq$\, 5.7
by 135 days.  The hot component was still dominant at 172 days but by 296 days it
contributed only $26\%$ of the total luminosity.  The temperature of
the warm component stayed roughly constant at 1400-1500\,K until at
least 228 days, but by 296~days had cooled to 1200\,K.  The luminosity and the
radius monotonically increased during the 122--296~days period
(Fig.~\ref{bb_evol}).



A clear NIR excess was also observed  at late phases in several
other SNe  (1997ab, 1979C, 1980K, 1982E, 1982L, 1982R, 1985L, 1993J,
1994Y, 1995N, 1998S, 2005ip, 2006jc) \citep[e.g.] [] {Gerardy2002}.  The
simplest explanation for the IR excess in both SN~2008S and these
other SNe is thermal emission from warm dust heated by the SN.
However, different locations, origins and heating mechanisms are
possible.  Specifically, IR excesses may be due to (a) newly-formed
dust in the ejecta, heated by radioactive decay or a reverse shock,
(b) newly-formed dust in a cool dense shell (CDS) formed by the SN
shock/CSM collision and heated directly by the shock, by absorption of
X-rays from the interaction region or by the bolometric light curve of the SN (a type of
IR~echo), (c) ejecta collision with pre-existing circumstellar dust or
(d) an IR~echo from pre-existing circumstellar dust heated by the SN
bolometric light curve.  These scenarios do not necssarily exclude one another.  For
example, in the case of SN~2006jc the NIR excess was explained by
\cite{Mattila2008} as a combination of IR echoes from newly-formed CDS
dust and from pre-existing dust in the CSM.

In the case of SN~2008S, we can immediately rule out an IR~echo,
whether from CDS or CSM dust, as being the cause of the late-time NIR
excess.  The bolometric light curve is too weak by a factor of $\sim$\,100 to account for
the high dust temperatures and luminosities seen during the
200--300~days period. Moreover, for an IR echo scenario we would expect
the phase and rise-time of the echo flux to be determined by the
temporal behaviour of the bolometric light curve, which is about 35~days per e-fold
beginning at 0~days (cf. Fig.~\ref{echomodel} right-hand panel). Yet the NIR
excess does not appear until at least 120~days and has a
characteristic risetime of about 100~days.

This leaves dust heated by radioactivity or by energy from shock
interaction with CSM.  We first estimated the mass
of such dust required to reproduce the NIR flux.  We employed the
escape probability model described in \cite{Meikle2007}. Setting
the expansion velocity of the dust cloud at $1000 \pm 50$\,\kms, we found
that for silicate dust grains in the ejecta the NIR excess was
reproduced with a dust mass rising monotonically from about
$0.2\times10^{-5}$\,M$_{\odot}$ on day~122 to
$1.2\times10^{-5}$\,M$_{\odot}$ on day~296.  During this time, the dust
temperature fell from about 1300\,K to 1100\,K and the optical depth to
the centre at 2.2\,$\mu$m rose from 0.2 to 0.4. Thus, optically thin
dust lying well within the velocity limit of refractory elements (say
up to $\sim$ 2000\,\kms) was able to reproduce the NIR excess. It
follows that the derived masses indicate the minimum dust mass
required.  Reducing the size of the dust cloud caused it to quickly
become optically thick, with the blackbody limit being approached at a
radius corresponding to about 500\,\kms. Clearly, in this optically
thick case the NIR observations do not restrict the mass of dust that
could be concealed in the ejecta. If instead of ejecta dust, we assume
the dust lay in the cloud which gave rise to the early time MIR echo,
and was heated by ejecta collision, then the mass of dust required to
reproduce the NIR flux remains much the same, but the optical depth
falls to around 0.001. Finally, for optically thin dust, replacing
silicate grains with amorphous carbon grains reduces the derived dust
mass by a factor of about 10, due to the higher emissivity of carbon
in the NIR region.

We conclude that a very modest mass of optically thin dust can account
for the NIR excess flux. Moreover, it appears that the dust mass
involved grew quite rapidly between 122 and 296~days. This dust may
have condensed in the ejecta, heated by radioactive decay or a reverse
shock.  Alternatively, its location may have been much farther out in
the CSM. To account for the early-time MIR flux we had to invoke a
``dust-free cavity'' of radius at least $3\times10^{16}$\,cm, with a
dust mass beyond this limit of at least
$10^{-3}$\,M$_{\odot}$. However, to account for the later NIR excess
flux, we have shown that the required dust mass would be barely $1\%$ of
this. Thus, it is quite feasible that sufficient dust to account for
the NIR flux lay within the MIR~echo cavity, with an inner limit fixed
only by a dust evaporation radius which could be as small as
$\sim$\,3$ \times 10^{15}$\,cm (see above). For the ejecta to reach this
distance in, say, 150~days, would take a velocity of just 2300\,\kms.
The apparently increasing mass of dust might have been due to dust
newly forming within a CDS. Alternatively, it simply may be that the
shock swept up an increasing mass of pre-existing CSM dust.

Is it plausible that the NIR excess was due to new,
radioactively-heated ejecta dust?  We have already shown that this can
be achieved with a small mass of optically-thin dust lying comfortably
within the limits of refractory elements. However, examination of the
energy budget reveals problems.  In subsection \ref{comp98S79C}  we estimate that
SN~2008S produced only about $\times 0.02$ of the mass of $^{56}$Ni
that was released by SN~1987A. If we scale the radioactive energy
deposition law for SN~1987A \citep{Li1993} to SN~2008S we find that the deposited radioactive energy
exceeded that of the dust luminosity by $\times 8$ on day~122 falling
to $\times 1.25$ by day~228 i.e. there was sufficient radioactive
energy to power the dust luminosity up to about day~228. However, by
day~296, even if we assume the entire radioactive luminosity was
deposited in the ejecta, it could only supply half of the dust
luminosity i.e. for radioactivity to account for the warm component
flux at 296~days we would have to double the derived mass of
$^{56}$Ni.  Another possible difficulty is that, even if we focus on
just the period 122--228~days, the fraction of radioactive energy
deposited in the dust increases from $12\sim \%$ to $80\sim \%$ and yet the
average number density of ejecta dust stayed approximately
constant. Why then did the dust-deposition fraction show such a large
increase?  It may be related to the detailed distribution of the grain
growth and radioactive material. But, given the day~296 energy
deficit, perhaps a more plausible explanation is that at least part of
the NIR excess had another cause. This might be reverse-shock heating
of new ejecta dust. Alternatively it might be IR radiation from
shock-heated newly-formed CDS dust or pre-existing circumstellar
grains.  We conclude that, while the earlier NIR excess could have been
entirely due to new, radioactively-heated dust in the ejecta, as time
went by, an increasing proportion of the flux must have been powered
by other mechanisms.

NIR/MIR-based evidence for newly-formed dust in the ejecta or CDS of
SNe has been reported in the cases of SN~1987A
\citep{Moseley1989,Suntzeff1990,Meikle1993,Roche1993}), SN~1998S \citep{Pozzo2004},
SN~2004gd \citep{Sugerman2006,Meikle2007}, SN~2004et  \citep{Kotak2009}
and SN~2006jc \citep{Smith2008d,DiCarlo2008,Mattila2008}. Dust masses of typically $10^{-4}$--$10^{-3}$\,M$_{\odot}$ are directly observed, although larger masses may exist in
optically thick clumps.  Dust condensation in the ejecta or in a CDS
can be also be demonstrated via the effects of the dust on optical
radiation.  In particular, it can attenuate the red wings of spectral
lines, causing an apparent blue shift of the line profiles
\citep{Lucy1991,Danziger1991,Elmhamdi2003,Pozzo2004}. However, the spectra of SN~2008S are of
insufficient resolution or wavelength precision to enable us to make
this test.

\section{Spectroscopic data reduction and analysis}\label{Specsec}

We spectroscopically monitored SN 2008S for eight months with several telescopes and details of the spectroscopic observations are reported in Table~\ref{logspec}.
\begin{table*}
\caption{Journal of spectroscopic observations of SN 2008S.\label{logspec}}
\begin{footnotesize}
\begin{tabular}{rrcccrr}
\hline
JD$^a$     & ph$^b$  & Telescope   & Grism or Grating & Range     & Resolution  & exp. time \\
                  &       &                        &                               & \AA               & (FWHM) \AA            & s\\
\hline
500.8 &  15     & WHT         & R1200R         & 6200--6880           &   0.7          & 1800        \\
501.3 &  16     & NOT        & Gr 4           & 3600--8700           &   19           &  800        \\
504.3 &  19     & WHT        & R158R          & 5500--10000          &   10           &  800         \\
507.7 &  22     & TNGD        & LR-R           & 5150--10225          &   11           & 1800               \\
513.7 &  28     & CAO  & Gr 4           & 3900--7800           &   24          & 3600            \\
514.8 &  29     & WHT         & R1200R         & 7025--7560           &   0.8          &  900              \\
514.8 &  29     & WHT        & R300B          & 3170--5350           &   3.5          &  900            \\
523.7 &  38     & CAHAT       & G-200          & 4950--10260          &   13           & 2700                 \\
529.8 &  44     & INT        & R300V          & 3750--9340           &    5           & 1000              \\
532.7 &  47     & CAHAT & G-200          & 4950--9750           &   14           & 3000               \\
539.8 &  54     & TNGD         & VHR-V          & 4650--6600           &    5           & 2400                \\
552.8 &  67     & CAO  & Gr 4           & 3790--7790           &   24           & 3600            \\
553.7 &  68     & CAHAT        & r-100          & 5800--9600           &    9           & 3600                \\
557.8 &  72     & CAO   & Gr 4           & 3500--8450           &   24           & 3600            \\
561.7 &  76     & TNGD      & VHR-R          & 7110--7560           &   4.1          & 2700                \\   
582.6 &  97     & CAHAT       & r-200          & 6178--10600          &   11           & 3600               \\
588.7 & 103     & CAHAT        & b-200          & 3900--8850           &   11           & 1700               \\
606.7 & 121     & WHT          & R316           & 5520--8260           &    6           & 2700                \\
609.7 & 124     & WHT         & R316           & 7190--10400          &   4.7          & 1800                \\
667.7 & 182     & WHT          & R158           & 5400--9900           &   10           & 2700                \\
668.7 & 183     & WHT         & R158           & 5400--10200          &    6           & 2700                \\
741.7 & 256     & HP200          & 300            & 3300--5600           &    8           & 1800                        \\
741.7 & 256     & HP200          & 158            & 5800--10300          &   11           & 1800                   \\
\hline
\end{tabular}
\\[1.5ex]
$^a$JD $-$ 2\,454\,000.00\quad\\
$^b$ Phase in days with respect to the explosion date JD $2\,454\,486 \pm 4$.\quad\\
WHT = the 4.2~m William Herschel Telescope (WHT) with the Intermediate dispersion Spectrograph and Imaging System (ISIS);  NOT= the NOT with ALFOSC;  TNGD = the TNG with DOLORES;  CAO = the Copernico telescope with AFOSC;  CAHAT= the 2.2~m telescope at CAHA with CAFOS; 
INT= the 2.5~m Isaac Newton Telescope (INT) with the Intermediate Dispersion Spectrograph (IDS); HP200= the 5.08~m Hale Telescope with the Double Spectrograph Specs (DBSP) at Palomar Observatory.
\end{footnotesize}
\end{table*}
Spectra were reduced (trimmed, overscan and bias corrected,
flat-fielded) using standard routines within IRAF. An optimal
variance weighted extraction of the spectra was carried out using the
IRAF routine APALL.  Wavelength calibration was performed using the 
spectra of comparison lamps acquired with the same instrumental
configuration as the SN observation. Flux calibration was done
using spectrophotometric standard stars observed with the same
instrumental set-up.  Approximate spectral resolutions were estimated
from the full-width-at-half-maximun (FWHM) of the night sky lines. The
wavelength calibration is accurate to $\pm 1$~\AA\ for ISIS spectra,  $\pm 2$~\AA\ for DOLORES and DBSP spectra and $\pm 3$~\AA\ for CAFOS, ALFOSC and AFOSC
spectra.  Atmospheric extinction corrections were applied using
tabulated extinction coefficients for each telescope site. The spectra
of standard stars have also been used  to identify telluric features
and to remove these from the SN spectra.  Spectra of similar quality
obtained during the same night were combined to increase the signal to noise ratio ($S/N$)
ratio.
 To check the flux calibration, $BVRI$ magnitudes were estimated by integrating the spectra convolved with standard filter functions using  the task CALCPHOT within the IRAF
 package STSDAS. The spectro-photometric magnitudes were compared to the photometric observations and, whenever necessary, a  scale factor was applied to match the photometric observations.
The flux calibration is accurate to within approximately 10\%.
\subsection{Line identification}
\begin{figure*}
\resizebox{\hsize}{!}{\includegraphics{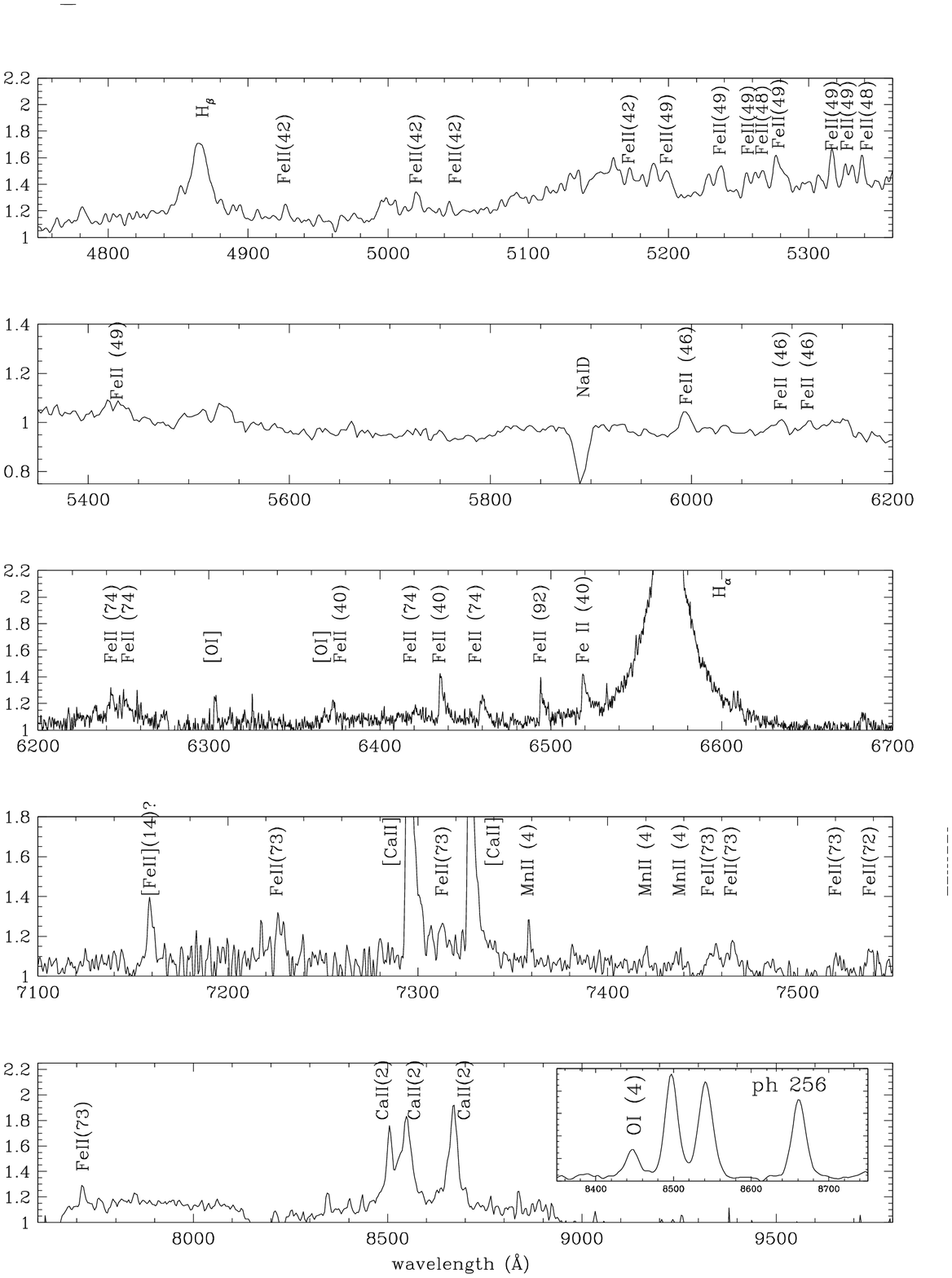}}
\caption{Spectral lines observed in SN 2008S spectra. From the top to the bottom panel:  spectra obtained at  WHT  (phase 29), at TNG (phase 22), at WHT (phase 15),  at WHT (phase 29), at WHT (phase 19).  Inset: spectrum obtained with HP200 (phase 256). Phase is in days after the explosion epoch (JD $2\,454\,486$) and wavelength is in the observer frame.}
\label{lineidfig}
\end{figure*}

We identified H$\alpha$, H$\beta$, H$\delta$, [Ca II] doublet ($\lambda$$\lambda$7292,7324) and Ca II triplet ($\lambda$$\lambda$$\lambda$8498,8542,8662), Fe II (multiplets 27, 28, 37, 38, 40, 42, 46, 48, 49, 72, 73, 74, 92, 186, 199),  [O I] and Mn II (multiplet 4) in emission and Na I doublet ($\lambda$$\lambda$5890,5896) in absorption in  early spectra (Fig.~\ref{lineidfig} and Table~\ref{lineid}). 
In the more recent spectra O I (multiplet 4, $\lambda$8446) and Na I D  are visible in emission, while the O I triplet in absorption seems to disappear. 
 These emission lines are likely produced in different regions with different ionization states and velocities. The low ionization is likely due to high column density material with high optical depth.  
\begin{figure}
\resizebox{\hsize}{!}{\includegraphics{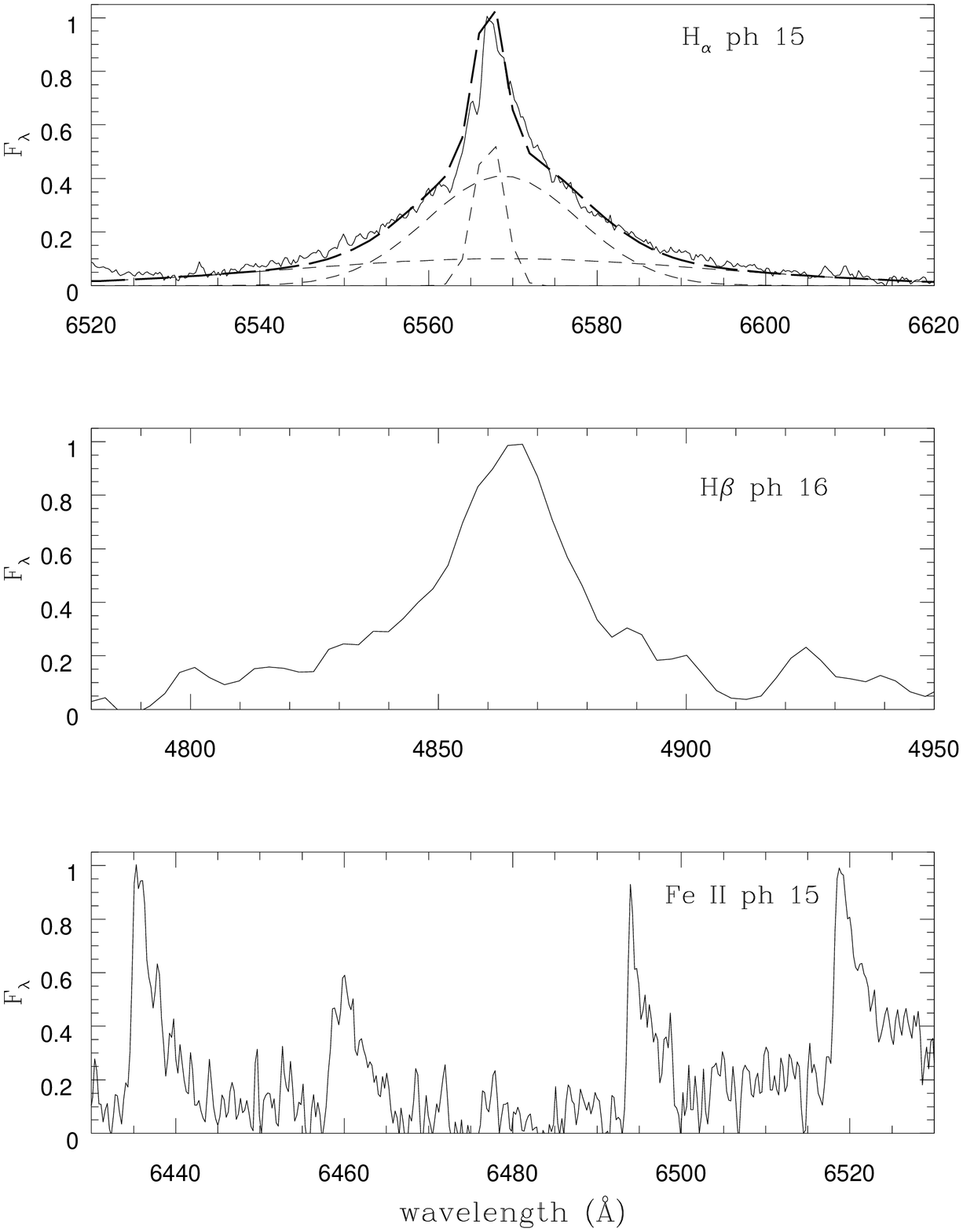}} 
\caption{Line profiles at about two weeks after  the explosion epoch (JD $2\,454\,486$):  in the top panel the multicomponent profile of H$\alpha$ in  the high resolution ISIS spectrum (the three Gaussian component are shown as thin dotted lines and their combination is shown as thick dotted line),  in the middle panel H$\beta$ profile in the NOT spectrum, in the bottom panel the profiles of  Fe II lines in the ISIS spectrum. Wavelength is in the observer frame. The flux densities have been normalised to the line peaks.}
\label{profFeII}
\end{figure}

\subsection{Spectroscopic evolution} 
\label{specevol}
\begin{figure*}
\resizebox{\hsize}{!}{\includegraphics{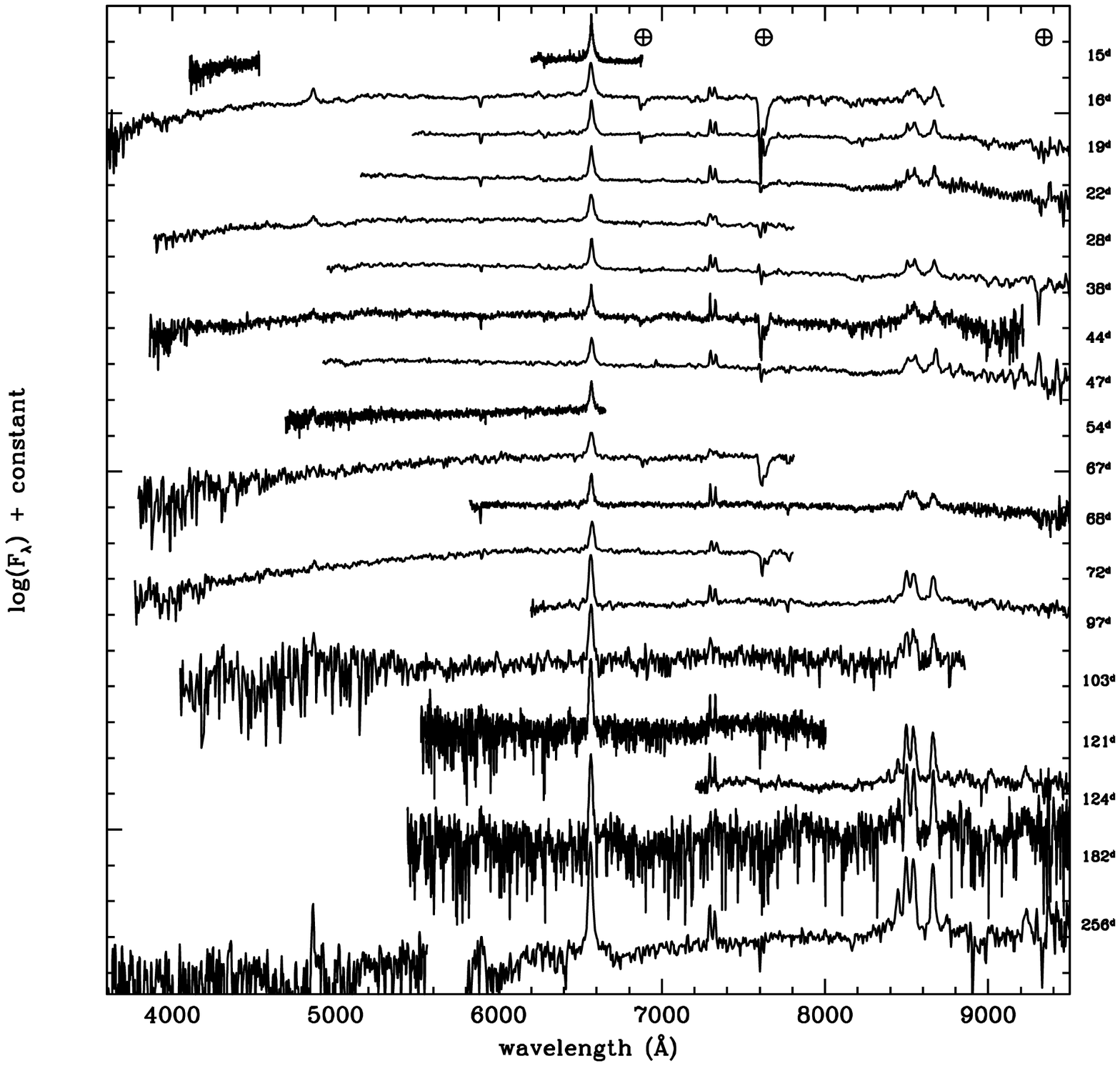}}
\caption{Time sequence of SN 2008S spectra. The spectra are corrected for Galactic extinction.  The phase reported to the right of each spectrum is relative to the explosion epoch (JD $2\,454\,486$) and  wavelength is in the observer frame. The $\oplus$ symbols mark the positions of the
most important telluric absorptions.}
\label{specev}
\end{figure*}

The spectra of SN 2008S do not show significant evolution during the
temporal interval from 15 days to eight months after the
explosion (Fig.~\ref{specev}), the continuum becoming progressively
redder and fainter with time.  All spectra consist of a nearly
featureless continuum with superimposed strong Balmer emission lines,
[Ca II] doublet and Ca II near-infrared triplet.   The most
remarkable change in the latest spectra is the presence of Na I D and
O I ($\lambda$8446) in emission.
 We do not detect high velocity absorption lines at any 
phase, suggesting that the ejecta are not directly  visible. 
The lines do not show a P-Cygni
profile typical of a SN explosion at very early phase but exhibit two
different kinds of profiles.  [Ca II] and Fe II lines show only a narrow
asymmetric component with a red wing (Fig.~\ref{profFeII}  and Fig.~\ref{profCaII}).  The H$\alpha$, H$\beta$  and Ca II NIR triplet lines 
show evidence for a multicomponent profile (Fig.~\ref{profFeII}, top and middle panels)
We focused our analysis on H$\alpha$,  the [Ca II] doublet and the Ca II triplet since they are the most prominent lines and visible in
 all spectra estimating line parameters  with the IRAF task SPLOT.
  The velocities are FWHM and those of
 H$\alpha$ and Ca II ($\lambda$8662)  have been measured by
 deblending the multiple-component profile and assuming a Gaussian
 profile for each component.  To measure the peak position at
 different phases we selected the spectra acquired with a slit width
 smaller than the seeing, hence considering only spectra with accurate wavelength
 calibrations.  

\begin{figure}
\resizebox{\hsize}{!}{\includegraphics{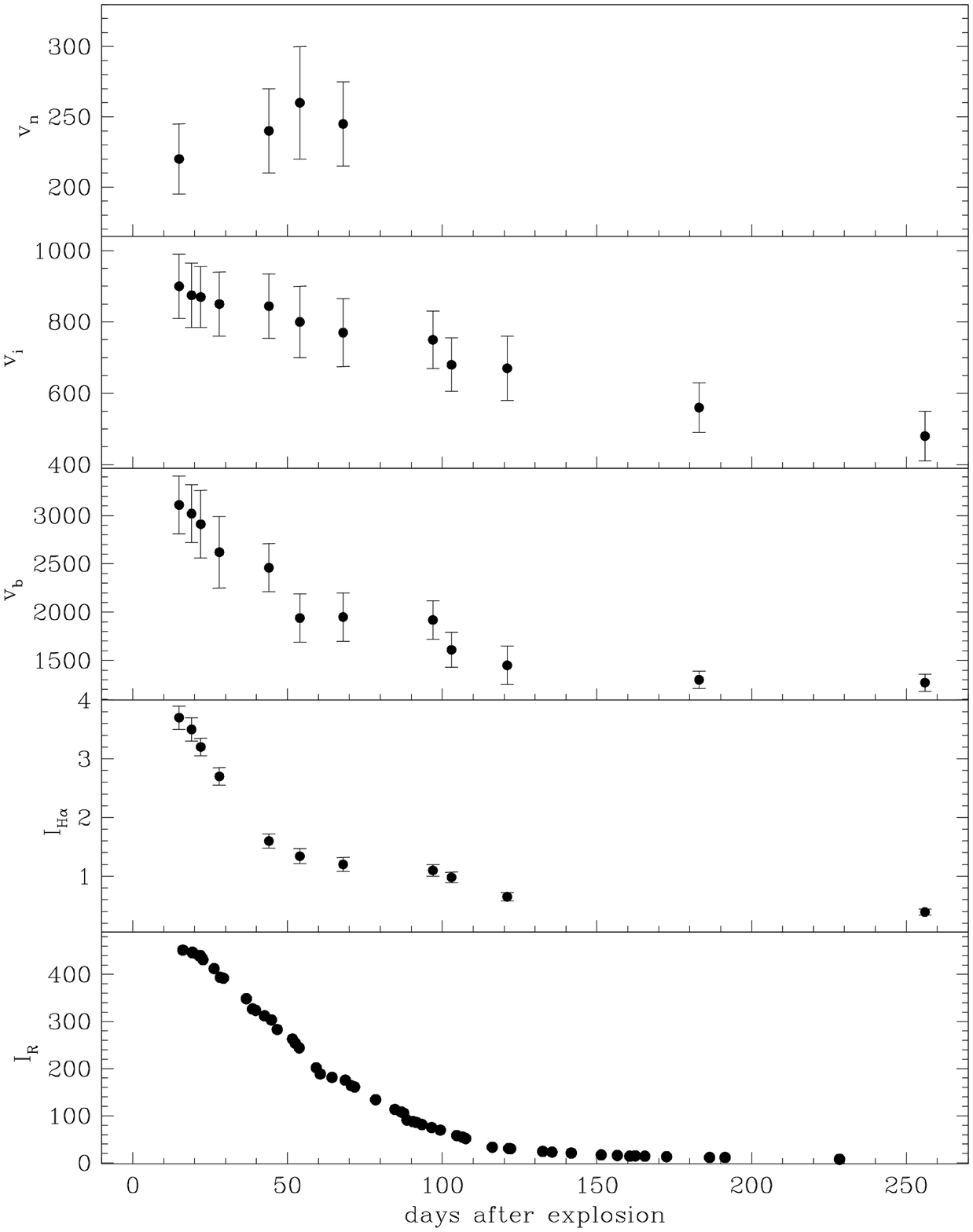}}
\caption{Temporal evolution of  H$\alpha$ kinematic components and intensity.  From the top to the bottom panel:   evolution of the velocity of the narrow  (v$_\mathrm{n}$),  intermediate  (v$_\mathrm{i}$), broad (v$_\mathrm{b}$)  component,  the integrated flux of H$\alpha$ and the $R$ band.  Velocity widths (FWHM) are in \kms  and integrated fluxes are in units of $10^{-14}$\,erg\,cm$^{-2}$\,s$^{-1}$. }
\label{velHa}
\end{figure}

{\em Hydrogen}\\
The intensity of H$\alpha$ decreases by a factor of
approximately ten from the earliest spectrum to the latest (Table~\ref{fluxcompCatr}). 
It follows the flux in $R$ band (see Fig.~\ref{velHa})
during the first 60 days, then shows a flattening
(the $R$ band light curve changes the decline slope shortly after).  
\begin{table*}
\caption{Evolution of the intensity of H$\alpha$, Ca II ($\lambda$8662) and [Ca II] doublet. Intensities are in units of $10^{-14}$\,erg\,cm$^{-2}$\,s$^{-1}$.\label{fluxcompCatr} }
\begin{footnotesize}
\begin{tabular}{llllll}
\hline
JD$^a$ & ph$^b$  & I (H$\alpha$) & I (Ca II $\lambda8662$) & I ([Ca II] $\lambda7291$) & I ([Ca II] $\lambda7323$)  \\
\hline
501  &15  & $3.7 \pm 0.20$  & $1.40 \pm 0.08$  &$0.42 \pm 0.02$ &$0.41\pm 0.02$\\
505  &19  & $3.5 \pm 0.20$  & $1.26 \pm 0.07 $&               &  \\
508  &22  & $3.2 \pm 0.15 $ & $1.18 \pm 0.07$ &                &  \\
514  &28 & $2.7 \pm 0.15$   &                &                &  \\
524  &38  & $2.1 \pm 0.12$  & $0.86 \pm 0.06$ &$0.37 \pm 0.01$ &$ 0.35\pm 0.01$ \\
530  &44 & $1.6 \pm 0.12$   &                &              &  \\
540  &54 & $1.3  \pm 0.10$  &                &              &  \\
554  &68 & $1.2 \pm 0.10$   &                &              &   \\
562  &76 &                 &                &$0.17 \pm 0.009$ & $0.15 \pm 0.009 $\\
583  &97  & $1.1 \pm 0.10$  & $0.45 \pm 0.05$  &$0.12\pm 0.008$ & $0.10 \pm 0.009$\\
589  &103 &$0.98 \pm 0.09$  &                &              &  \\
607  &121 &$ 0.65 \pm 0.07$ &$ 0.37 \pm 0.04$  &$0.051 \pm 0.003$    & $0.053\pm 0.004$\\ 
742  &256 & $0.39 \pm 0.05$ & $0.22 \pm 0.03$  &$0.022 \pm 0.001$    & $0.021\pm 0.001$\\
 \hline
\end{tabular}
\\[1.5ex]
$^a$JD $-$ 2\,454\,000.00\quad\\
$^b$ Phase is in days with respect to the explosion date JD $2\,454\,486 \pm 4$.\quad
\end{footnotesize}
\end{table*}
The H$\alpha$ profile in the first high resolution ISIS spectrum
exhibits three different kinetic components as shown in Fig.~\ref{profFeII}.  The narrow, intermediate and broad components
correspond to velocity widths (FWHM) of v$_\mathrm{n} \sim 250$\,\kms,  v$_\mathrm{i} \sim 1000$\,\kms and  v$_\mathrm{b} \sim 3000$\,\kms, respectively, in this
early spectrum.  
The H$\alpha$ intensity and the broad component velocity
            width, v$_\mathrm{b}$, show a similar temporal behaviour, decreasing quickly until about 60 days after explosion and
more slowly after this epoch (Fig.~\ref{velHa} and Table~\ref{Havel}).   The
            intermediate component velocity width, v$_\mathrm{i}$,  shows a less
            abrupt decline.
The narrow component, v$_\mathrm{n}$,  is not resolved in several spectra at intermediate
phases but appears to stay constant until 60 days after the explosion
and is not visible after this epoch (Fig.~\ref{velHa} and
Table~\ref{Havel}).  
This  component of  H$\alpha$ seems also  to be asymmetric,  but
the resolution of our spectra is not adequate to analyse  its
profile in detail.  

\begin{figure}
\resizebox{\hsize}{!}{\includegraphics{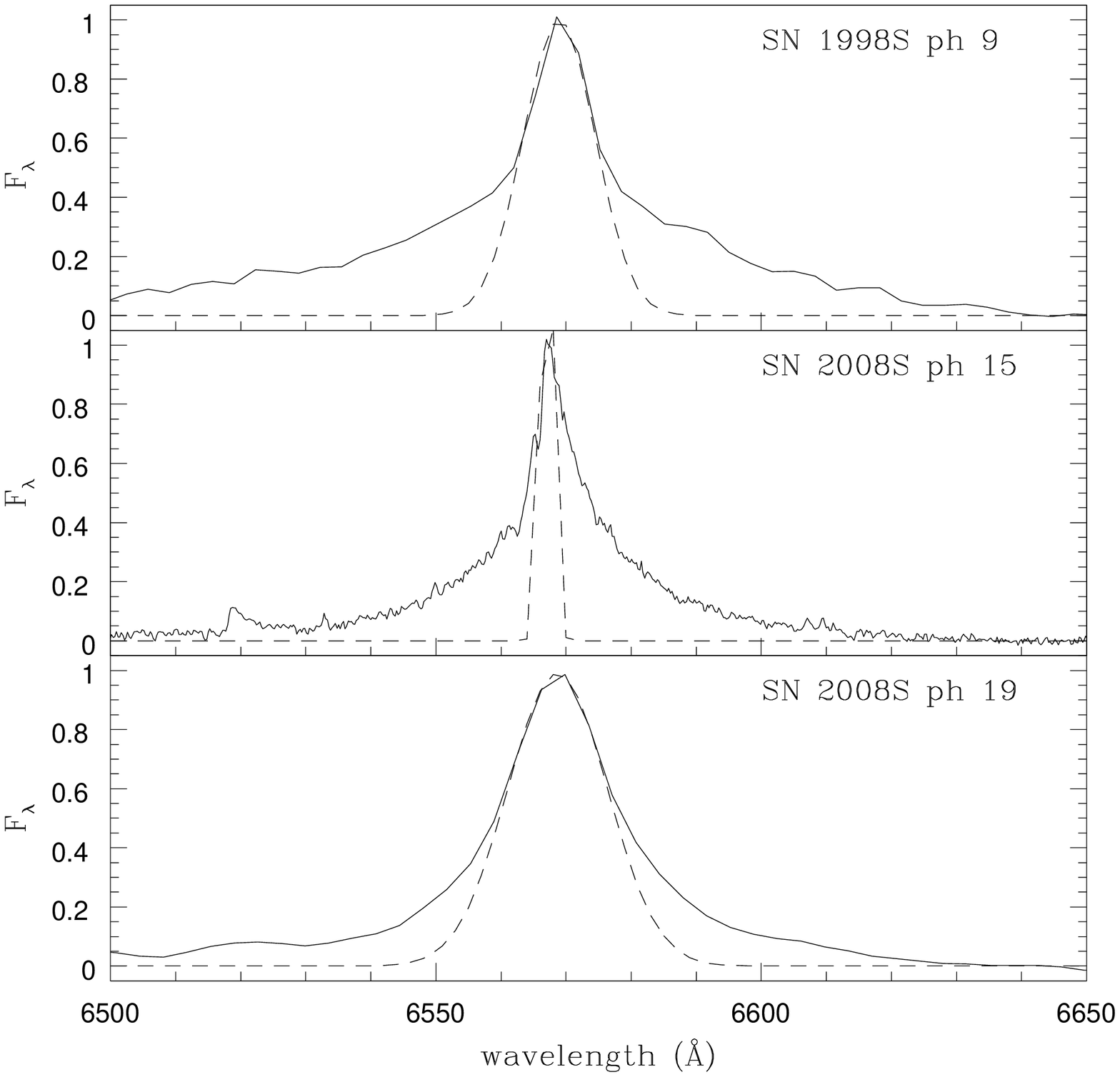}}
\caption{Comparison between H$\alpha$ profiles in SN 1998S (phase 9 days) and SN 2008S at two different phases (15 and 19 days).  Phase is in days after the explosion epoch (JD $2\,450\,869$ for SN 1998S and JD $2\,454\,486$ for SN 2008S). Wavelength is in the observer frame. A Gaussian profile that matches the line peak is shown as a dashed line.}
\label{compprof98S}
\end{figure}

\begin{table*}
\caption{Evolution of the three kinematic components (broad, v$_\mathrm{b}$; intermediate, v$_\mathrm{i}$; narrow, v$_\mathrm{n}$) in H$\alpha$ and Ca II ($\lambda$8662) features.  The velocity widths (FWHM) are in \kms. \label{Havel}}
\begin{footnotesize}
\begin{tabular}{llllllll}
\hline
JD$^a$ & ph$^b$ & v$_\mathrm{b}$ (H$\alpha$)   & v$_\mathrm{i}$ (H$\alpha$) &  v$_\mathrm{n}$(H$\alpha$) & v$_\mathrm{b}$  (Ca II $\lambda8662$)   &  v$_\mathrm{i}$(Ca II $\lambda8662$) & v$_\mathrm{n}$ (Ca II $\lambda8662$) \\
\hline
501  &15 & $3110 \pm 300$  &  $ 900 \pm 90$  & $ 220 \pm 30$  &   &  \\
504  &19 & $3020 \pm 300 $  &  $ 875 \pm 90$ &             & $ 3000 \pm 350$       & $ 850\pm 90$   &$ 264\pm 30 $    \\
508  &22 & $2910 \pm 350$   & $  870 \pm 85 $ &            &        &         \\
514  &28 & $2620 \pm 370$   &  $ 850 \pm 90$ &             &        &   \\
524  &38 &                                  &                             &            & $ 2280 \pm 300$    & $ 753 \pm 60$ \\
530  &44 & $2460 \pm 250 $  &  $ 844 \pm 90$  & $ 240 \pm 30$ &        & \\
540  &54 & $1940 \pm 200$   &  $ 790 \pm 90$  & $ 260 \pm 40$ &        & \\
554  &68 & $1950 \pm  250 $  &  $ 770 \pm 95 $ & $ 245 \pm 30$ &        & \\
583  &97 & $1920 \pm  200$   &  $ 750 \pm 80$  &            & $ 2040 \pm 200$    &$  720 \pm 80 $          \\
589  &103& $1610 \pm 180 $  &  $ 680 \pm 75$ &             &        & \\
607  &121 &$1450 \pm 100$   & $  670 \pm 90$   &             &                 &\\
610  &124 &                                 &                            &             & $ 1340 \pm 200$    & $ 660 \pm 70$ \\
669  &183 &$1300 \pm  90 $   &  $  560 \pm 70$  &            & $ 1290 \pm 200$    & $ 550 \pm 70$ \\
742  &256 & $1270 \pm 100 $   & $ 480 \pm 70$  &            & $ 1260 \pm 300$    &$ 540 \pm 80$ \\
\hline.
\end{tabular}
\\[1.5ex]
$^a$JD $-$ 2\,454\,000.00\quad\\
$^b$ Phase is in days with respect to the explosion date JD $2\,454\,486 \pm 4$.\quad
\end{footnotesize}
\end{table*}
The H$\alpha$  profile could be 
interpreted as broad underlying emission due to the ejecta,  an
intermediate component resulting by shocked material behind the
interaction front between ejecta and CSM and a narrow component from
unshocked CSM. However,  the lack of any high velocity absorption
component, or P-Cygni profile, is a fairly strong argument that the broad
emission is not a direct measurement of the ejecta velocities. 
This profile may be also interpreted as a narrow core with broad wings resulting from 
multiple scattering events  with thermal electrons, an escape mechanism  for line photons favoured by the small photospheric velocity observed for SN 2008S at all phases \citep{Dessart2008}. 
The electron scattering in a dense (Thomson optical depth of $\sim$ 3--4) CS shell  lying
beyond the SN photosphere was suggested by  \cite{Chugai2001} as an explanation for the very broad wings of H$\alpha$ in 
SN 1998S.
 We compare the H$\alpha$
profiles of SN 1998S and SN 2008S in Fig.~\ref{compprof98S}.  
The redshift in the H$\alpha$  peak decreases after about
            100~days (Fig.~\ref{shift}), although the uncertainties in the
            wavelength calibration mean that this shift has only a
            modest significance.  The EW(H$\alpha$) increases with time,
            becoming very large at late phases (about 900\AA\/).
\begin{figure}
\resizebox{\hsize}{!}{\includegraphics{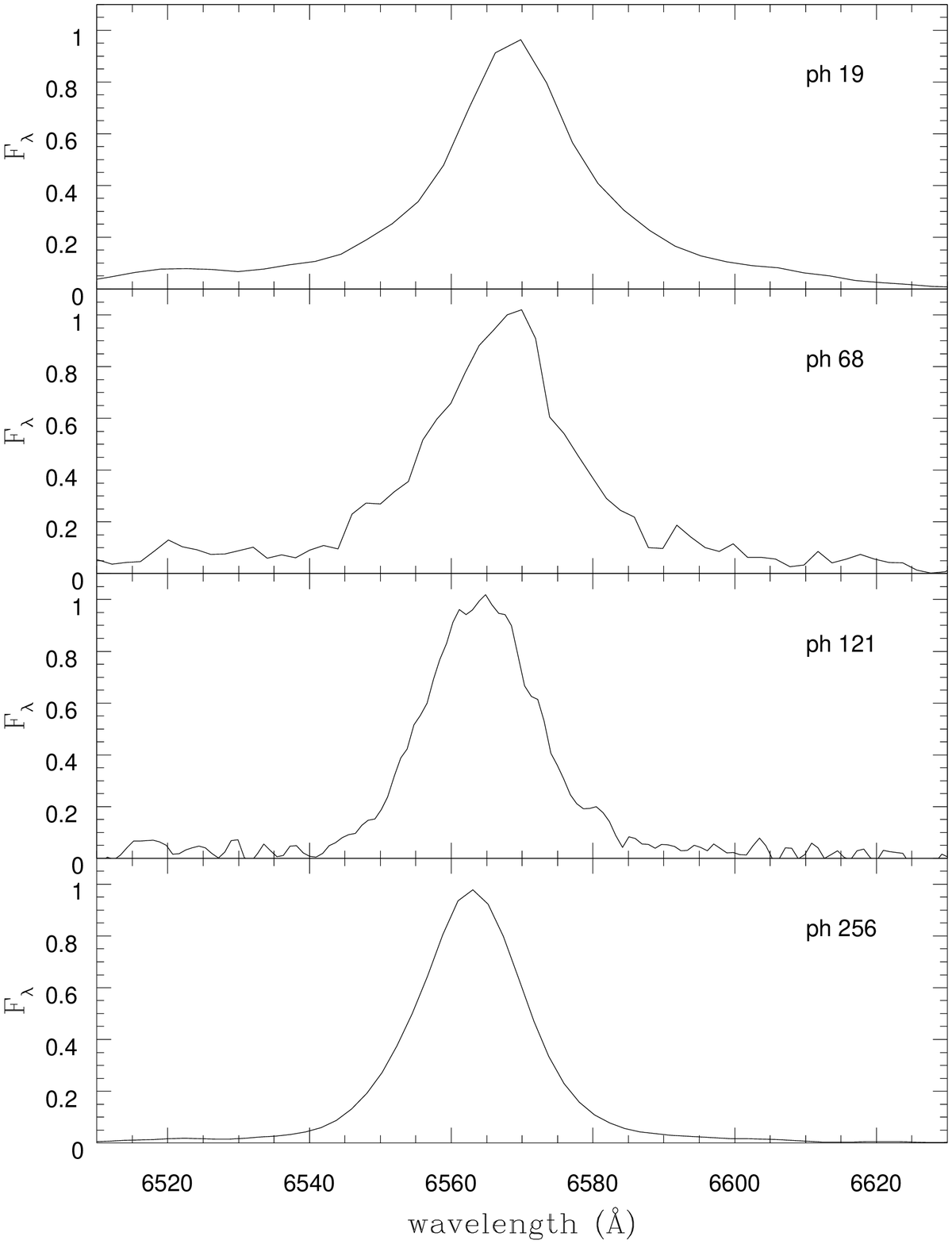}}
\caption{The peak position and profile of H$\alpha$ at  at 19 days, about 2 months, 4 months and 9 months after the explosion epoch (JD $2\,454\,486$). Wavelength is in the observer frame.  Flux density has been normalised to the line peak at each phase.}
\label{Hashift}
\end{figure}
There are
two processes proposed for the origin of H emission lines:
photoionisation with subsequent emission by recombination or
collisional excitation \citep{Drake1980}.  The first can not explain the increase of
the EW(H$\alpha$) with time so collisional excitation of
H$\alpha$ from the $n=2$ level seems to be the more likely mechanism to
explain Balmer emission, especially at the latest phases.
In the first spectra H$\beta$ shows a multicomponent profile
similar to H$\alpha$ (Fig.~\ref{profFeII}), but  this feature is too
weak to analyse  its
profile in detail.  
The H$\alpha$/H$\beta$ ratio is larger than the recombination value of about 3 and
increased from 4 at 15 days to 10 at 256 days after the explosion.  
This ratio depends critically on several parameters:
electron density (N$_{e}$), electron temperature, external radiation
field strength and optical depth in $\tau_\mathrm{H\alpha}$
\citep{Drake1980}.  The  three most important processes which
can change the line intensity with respect to the recombination value are:
Balmer self absorption (important at low density), collisional
excitation and de-excitation (important at high density)
\citep{Drake1980}. 

Collisional excitation
processes  cause the H$\alpha/$H$\beta$ intensity ratio to increase,  and
the N$_{e}$ value at which these processes become important inversely
depends on $\tau_\mathrm{H\alpha}$.  At the first epochs the
H$\alpha$/H$\beta$  intensity ratio indicates of $10^{10}$\,cm$^{-3}\le$\,N$_{e}\le10^{12}$\,cm$^{-3}$ given the observed temperatures in SN 2008S (see Sect.~\ref{SEDsec}) so the
collisional processes may have an important role.  The Balmer
decrement, observed in SN 2008S, may be a sign of the high optical
depth and the interaction with a high density CSM. In particular, its
evolution is likely led by the increase of $\tau_\mathrm{H\alpha}$ with
 time.

{\em Calcium}\\
The strength of the [Ca II] doublet decreases by a factor of $\sim$\,10
over 100 days and  line ratio H$\alpha$/[Ca II] ranges
from 9 to 18 over about 260 days (Table~\ref{fluxcompCatr}).  We did not
see any temporal evolution of the intensity ratio for the [Ca II]
doublet (always around the value of 1$\pm 0.2$) and of the position of
the peak.  The [Ca II]
doublet  very likely originates
in a lower density region with respect to the
Ca IR triplet. 
As discussed by \cite{ChevalierFransson1994}, the [Ca II]
doublet originates from radiative de-excitation from the
metastable 3$d^{2}$D level, which is highly populated. The
probabilities for the two [Ca II] transitions originating within the
same multiplet are approximately equal and the threshold to ionize
Ca$^{+}$ to Ca$^{++}$ from this metastable level is at 1218.8\,\AA. 
The gap between this threshold and the Ly$\alpha$ line is 3.2\,\AA\
corresponding to about 800\,\kms. Only gas with higher velocity
can ionize Ca II by absorbing Ly$\alpha$ photons suppressing  [Ca II] emission. The
presence of strong [Ca II] doublet in the first SN 2008S spectra may be
indicative of little Ca II being ionized, and hence a narrow Ly$\alpha$ line.
 The [Ca II] doublet shows a velocity width (FWHM) of about 250$\pm60$\,\kms
that is comparable to that of the narrow component of both H$\alpha$ and
Ca II triplet. This  velocity width  remains constant while the line profile shows an evolution: it is
asymmetric during the first three months after the explosion and starts
to be more symmetric after this epoch (Fig.~\ref{profCaII}).   
In Sect. \ref{summaryev} we propose an explanation for this
   behaviour based on a toroidal geometry in the CSM.\\

\begin{figure}
\resizebox{\hsize}{!}{\includegraphics{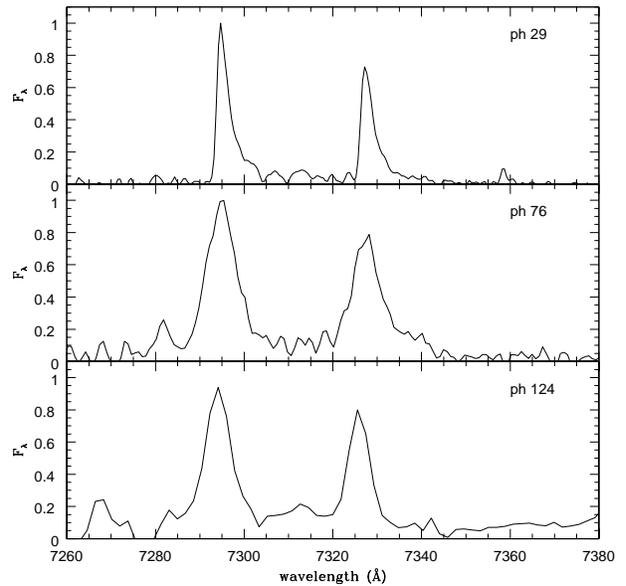}} 
\caption{Profile of [Ca II] doublet  at about one month, 3 months and 4 months after the explosion epoch (JD $2\,454\,486$).  Wavelength is in the observer frame. Flux density has been normalised to the line peak at each phase.}
\label{profCaII}
\end{figure}

\begin{figure}
\resizebox{\hsize}{!}{\includegraphics{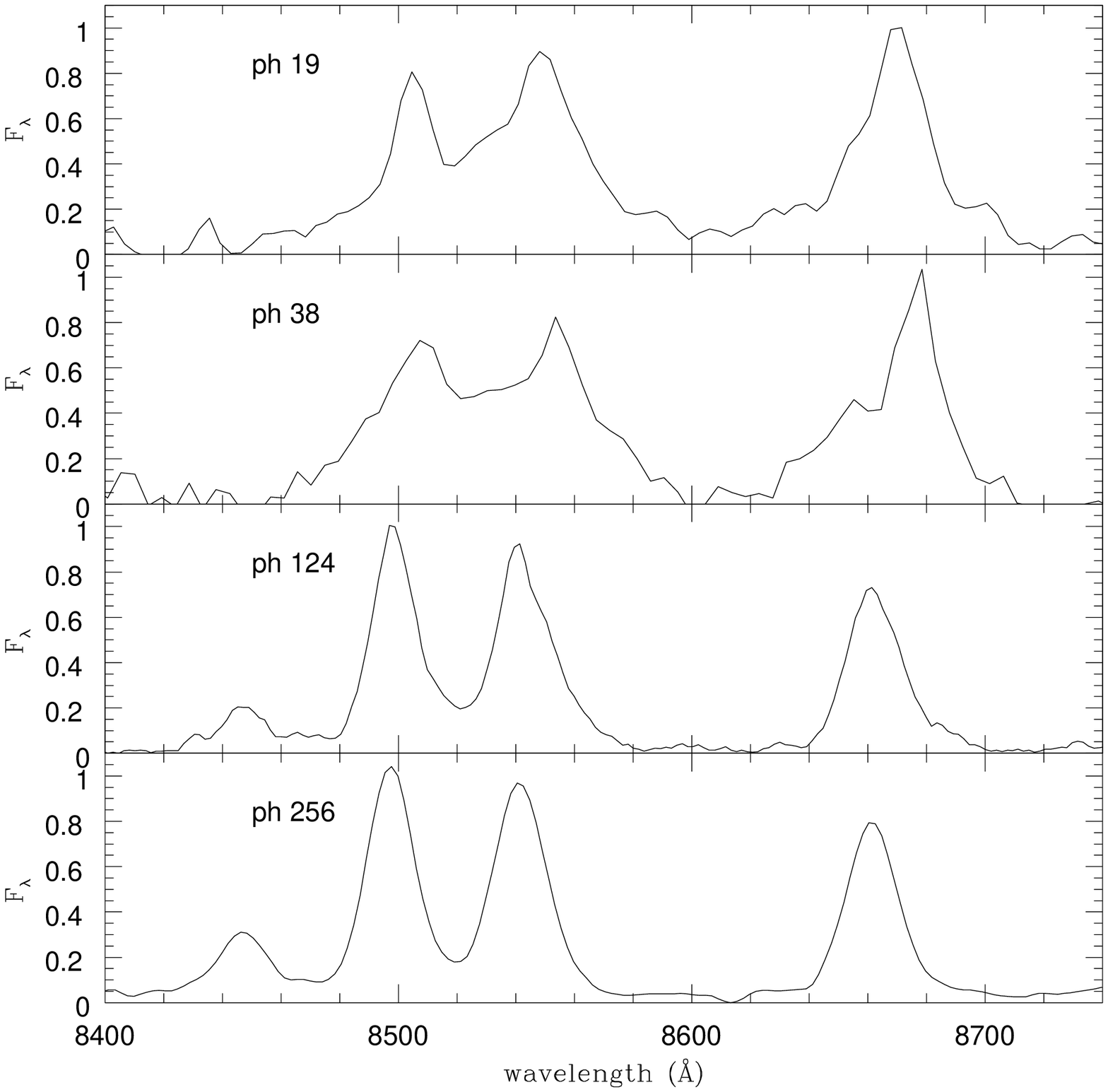}}
\caption{Profile of Ca II NIR triplet at about 20 days, one month, 3 months, 4 months and 9 months after the explosion epoch (JD $2\,454\,486$). Wavelength is in the observer frame. Flux density has been normalised to the line peak at each phase.}
\label{CaIINIRprof}
\end{figure}

\begin{figure}
\resizebox{\hsize}{!}{\includegraphics{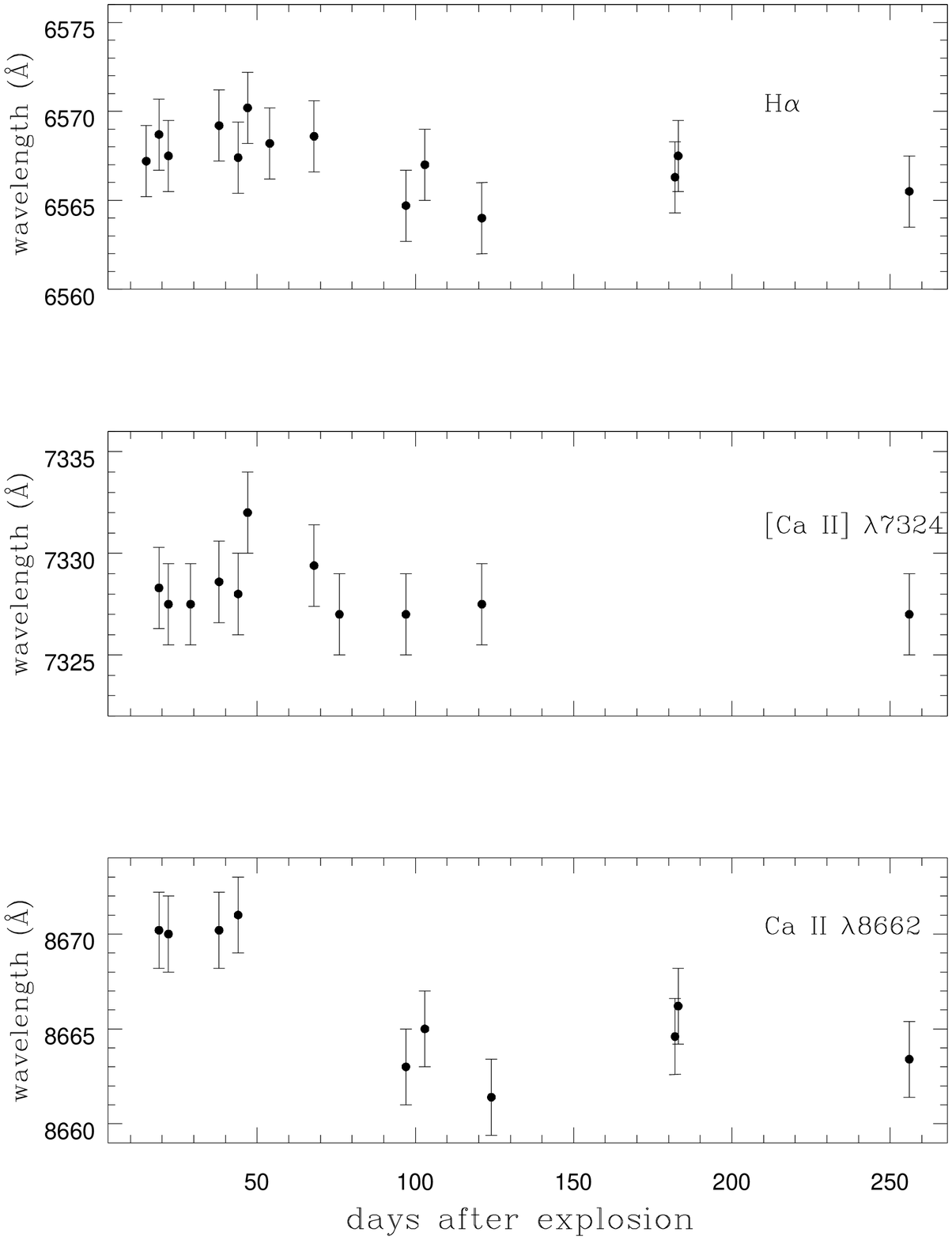}}
\caption{The peak position of H$\alpha$, [Ca II] ($\lambda$7324) and Ca II ($\lambda$8662)  as a function of phase.  Phase is in days after the explosion epoch (JD $2\,454\,486$) and wavelength is in the observer frame.}
\label{shift}
\end{figure}

The Ca II NIR triplet must originate in a different region 
from that which produces the [Ca II] doublet 
since it shows both a multicomponent
profile (Fig.~\ref{CaIINIRprof}) and a decreasing redshift with time
(Fig.~\ref{shift}).  Velocities for the different
components are reported in Table~\ref{Havel} and are consistent with
those of H$\alpha$.  The intensity ratios of the infrared triplet
to H$\alpha$ and to [Ca II] increase with the time
(Table~\ref{fluxcompCatr}). 
To investigate if the peak position of  H$\alpha$, the
[Ca II] doublet and the CaII NIR triplet show any temporal evolution,
we again exploited the BIC factor.
 We compare the BIC values obtained by fitting the spectral
time series with either a position constant in time or a temporal
evolution (parameterized as a straight line with a non null slope),
$\Delta \mathrm{BIC} = \mathrm{BIC}_\mathrm{slope} - \mathrm{BIC}_\mathrm{const}$.
For $H\alpha$ there is only a marginal evidence for an evolution,
$\Delta \mathrm{BIC} \simeq 0.3$. For the Ca II doublet, evolution is
not favored ($\Delta \mathrm{BIC}\simeq-1.0$ and $-0.8$ for the first
and the second line, respectively). Finally, there is strong evidence
for a decreasing redshift for the Ca II NIR triplet, $\Delta \mathrm{BIC} \simeq 13.9$.\\

{\em Oxygen, Iron and Sodium}\\
The [O I] ($\lambda$$\lambda$6300,6364) doublet is visible in the first high
 resolution spectrum with a very low velocity width (FWHM), about
 80\,\kms. These are collisionally suppressed at high
 density so likely originate in a slow moving and low density
 region.    O I($\lambda$8446)
 appears at the latest phases with a velocity width of
 about 530\,\kms similar to those of  the H$\alpha$ and Ca II
 intermediate components at the same phase.  
 The Fe II lines, visible in the first
 high resolution spectrum, appear to have only a narrow component with
 velocity width around 200\,\kms and the same profile as [Ca II]. 
The Na I D feature, visible in absorption in the early spectra, seems to show two blended lines:  the
Galactic contribution and the host galaxy doublet.
 In the spectra acquired at
182 and 256 days  this feature clearly appears in emission. 
This evolution is indicative of the circumstellar origin of the Na I D and  is a clear sign of the
high density of CSM. 
The evolution of EW(Na I D), illustrated in Sect.~\ref{Extinction},  may
due to an 
evolution of the ionization conditions in the CSM and in the ejecta of
SN 2008S since the EW is related to the ionization stage of Na I.

\section{The progenitor of SN2008S}\label{Progsec}

\subsection{Optical and NIR pre explosion images}
We carried out an independent analysis 
on the deepest 
pre-explosion optical and NIR images that we could locate, 
similar to that presented in \cite{Prieto2008}. We 
added some further data and recalculated
all the limits as some values in that paper were
taken from heterogeneous references rather than original data. 
We used optical images from the  
Gemini North and Large Binocular Telescope and NIR images from the
Bok 2.3\,m telescope. 
To ensure an accurate positioning of SN 2008S on all the pre-explosion
images we used an image of  
SN  2008S obtained with the Auxiliary Port Imager (AUX) on the WHT on 2008 February 08.  The camera has a pixel scale
of 0\farcs11/pixel, and a 1024 pixel TEK2 CCD.  A single $R$ band
image of 300\,s was taken, and although it was at high airmass (2.47)
the image quality was 1\farcs03. The position of SN 2008S was
determined on all the pre-explosion images to an accuracy of 
0\farcs1 and we confirm the results of \cite{Prieto2008} that there is
no optical or NIR counterpart detected. 
We calculated the 3$\sigma$ detection limits for each of the optical
images from the LBT and Gemini images and these are reported in
Table~\ref{proglim}. 

The $K'$ band image was obtained  from \cite{Knapen2003}
with the Bok 2.3\,m.  telescope of the Steward Observatory on 1999
October 17 with the PISCES camera, a HAWAII HgCdTe array of
1024$\times$1024 pixels of $0\farcs5$ on the sky. The dithered exposures
resulted in co-added total exposure time of between 1000-2000 seconds,
depending on field location, and the image quality was $2\farcs3$. The WHT
AUX  $R$ band image was used to locate the position of SN 2008S on
the frame by matching the positions of 12 stars located within $1.7\arcmin$
of the SN in both $R$ and $K$  images. The geometric transformation resulted in an RMS to the fit
of 108 milli-arcsec and at this transformed position there is no
detection of a source in the $K'$ band image. The zero-point for this
frame was estimated using five 2MASS stars located close to the SN and
the calculated 3$\sigma$ limiting magnitude was estimated to be
$K=18.0$\,mag.  The mean magnitudes of several of the faintest point
sources visible in the vicinity of the SN was $K=18.3 \pm 0.3$\,mag. Hence
we adopt $K=18$\,mag as the sensitivity limit of this frame.

\subsection{Spitzer MIR pre explosion images}
As discussed in \cite{Prieto2008}, several epochs of archival MIR imaging from the 
Spitzer Space Telescope are available. We analaysed  IRAC
(3.6--8.0\,$\mu$m) images  with 
the longest integration times from programmes 3249, 20256,
and 30292 (P.I. Meikle). Aperture photometry was carried
out on the post-BCD images using GAIA. A circular aperture of
2\farcs0 was used. Aperture corrections were derived using the
point response function frames available from the Spitzer
Science Center. The residual background level was measured 
using a clipped-mean sky estimator, and a concentric sky
annulus having inner and outer radii of 1.5 and 2 times the
aperture radius, respectively. The resulting flux densities
are listed in Table~\ref{proglim} and are consistent with the values 
reported in \cite{Prieto2008}.
We used the AUX image of SN 2008S and the deep, wide field
Gemini $i$ band image to determine the position of SN 2008S on the 
Spitzer  4.5, 5.8 and 8.0\,$\mu$m
images using differential astrometry. In each case the 
uncertainty in the positioning of SN 2008S on the Spitzer images
was 0\farcs3. This uncertainty is a combination in quadrature of the 
geometric transformation RMS and the uncertainty in the 
measurement of the centroid of the 
progenitor object detected by \cite{Prieto2008}. The difference
between SN 2008S and the progenitor object is 
0\farcs16$\pm$0\farcs30 and 
0\farcs14$\pm$0\farcs30 for the 5.8 and 8.0\,$\mu$m images respectively. 
However,  there is a small difference in the position of the progenitor source
in the 4.5\,$\mu$m image. This is separated from the 
SN 2008S position by 0\farcs51$\pm$0\farcs33. 
On close inspection of the images it appears that the source in 
the 4.5\,$\mu$m  image may well be 
extended compared to its 5.6 and 8.0\,$\mu$m counterparts (see also Fig.~1 in \cite {Prieto2008}).  This might suggest 
that the 4.5\,$\mu$m source is actually a blend of two or more sources. 
It seems fairly secure that SN 2008S is coincident with the 5.8 and 
8.0\,$\mu$m sources and we have no evidence to suggest that they 
are extended or non-stellar. The suggestion that the 4.5\,$\mu$m 
source may be slightly extended could just be due to contamination
from nearby, but unrelated flux at this wavelength. But the 
possibility remains that the source is a blend and this 
needs to be clarified with late, deep imaging with Spitzer and 
ground-based high resolution studies. 

\begin{figure*}
\resizebox{\hsize}{!}{\includegraphics{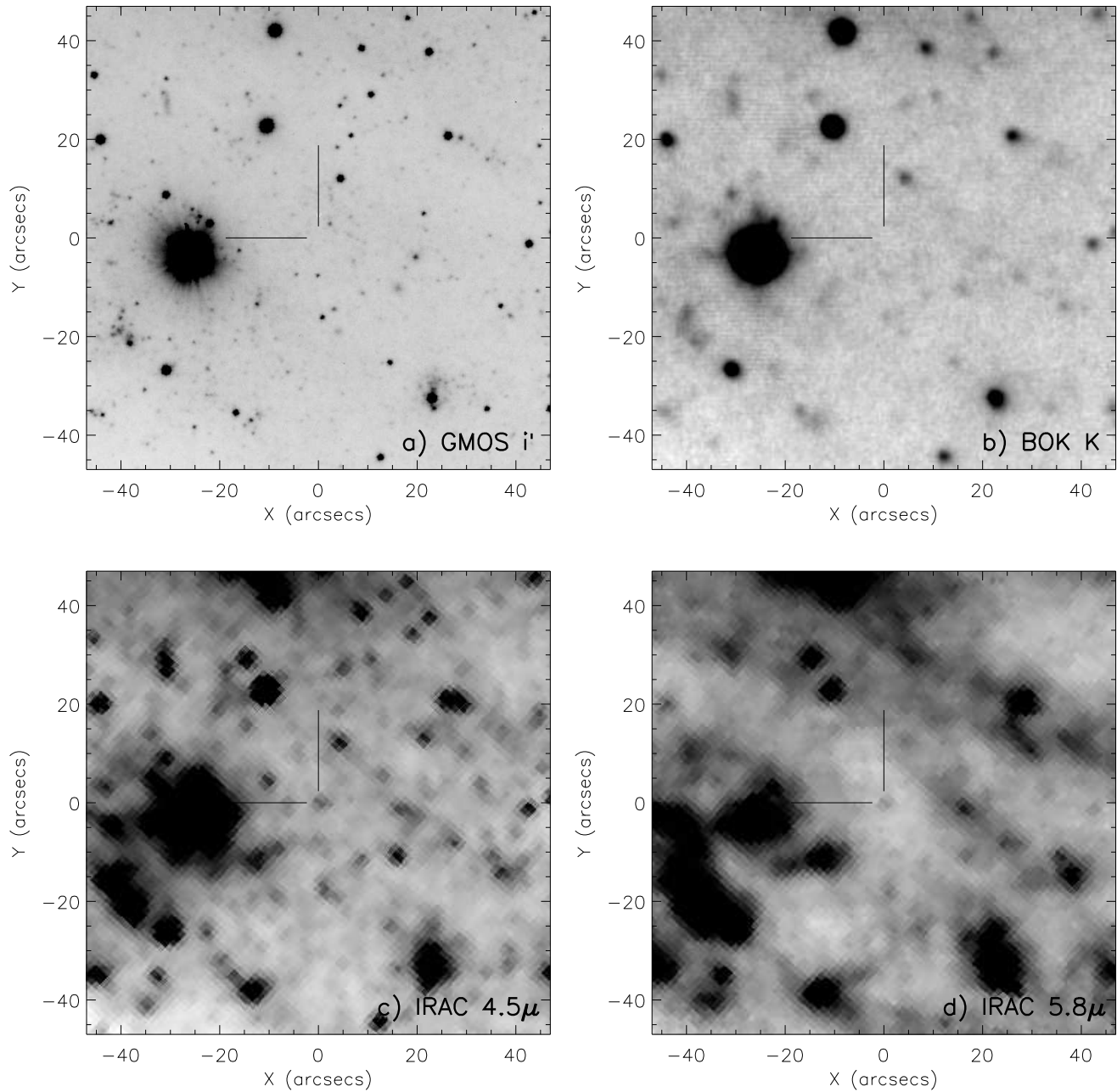}}
\caption{Pre-explosion images of SN 2008S. {\bf (a)} The Gemini GMOS $i'$ band image showing no 
progenitor detection to deep limits.
{\bf (b)} $K$ band image from the Bok telescope, again showing
no detection in the NIR. {\bf (c) \& (d)} : Spitzer
images at 4.5 and 5.8\,$\mu$m, originally presented by 
Prieto et al (2008). We find some evidence that the 4.5\,$\mu$m
may be slightly extended. This can only be tested with  higher 
spatial resolution images once the SN fades.}
\label{Progenitor-images}
\end{figure*}

\begin{table}
\caption{Fluxes and magnitudes measured in the position of SN 2008S on pre-explosion images.\label{proglim}}
\begin{tabular}{lllr}
Waveband & Instrument & Flux or magnitude & Units  \\
\hline
$U$   & LBT           & $>25.1$     & Vega Mag\\
$B$   & LBT           & $>24.5$     & Vega Mag\\
$V$   & LBT           & $>24.5$     & Vega Mag\\
$i$   & GMOS-N        & $>24.4$     & Vega mag  \\
$K'$  & PISCES        & $>18$       & Vega Mag \\
3.6$\mu$m    & IRAC   & $<3.6$     & $\mu$Jy \\
4.5$\mu$m    & IRAC   & $21.3 \pm 1.5$     & $\mu$Jy \\
5.8$\mu$m    & IRAC   & $45.6 \pm 2.4$     & $\mu$Jy \\
8.0$\mu$m    & IRAC   & $59.5 \pm 4.3$    & $\mu$Jy \\
\hline
\end{tabular}
\end{table}

\cite{Prieto2008} also reported aperture photometry for the individual
pre-explosion Spitzer epochs. They found no evidence for variability
in the flux of the pre-explosion source over the $\sim$\,1000 days covered.
Our photometry is based on the combined Spitzer images. To detect
possible variability on the source flux we compared the Post-BCD
4.5\,$\mu$m images from 2004-11-25 and 2006-08-12 where the pre-explosion
source was well detected. We first aligned the images using 
the centroid coordinates of 13 point-like sources around the pre-explosion
source position. The aligned images were matched and subtracted using 
the ISIS 2.2 \citep{Alard2000} image subtraction package (for details see
also \cite{Meikle2006}. No residual above the noise was apparent in
the subtracted image at the pre-explosion source position. This confirms
the findings of  \cite{Prieto2008} on the lack of variability of the 
pre-explosion source.
\subsection{Analysis of the pre-explosion images}\label{proganalysis}
\begin{figure}
\resizebox{\hsize}{!}{\includegraphics{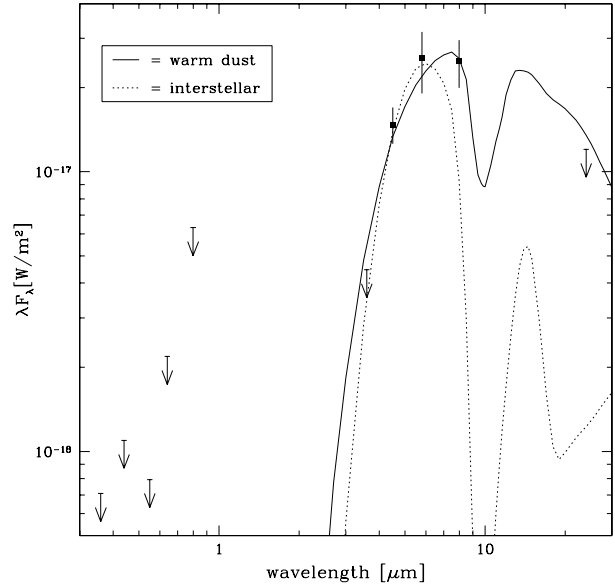}}
\caption{The MIR SED of the SN 2008S progenitor star and the two fits obtained 
using the radiative transfer model DUSTY. Model $A$ (thin line) consists of a warm circumstellar, spherical,
dust shell with T$_\mathrm{dust}=800$\,K (at the inner boundary), which has
an optical depth of $\tau_{V} = 150$. Model $B$ (dotted line) consists of a block 
of optically thick interstellar dust with $A_{V} = 140$\,mag. }
\label{dustymodel}
\end{figure}

To determine whether the precursor object detected in the Spitzer
images is a viable stellar progenitor we modelled the MIR emission
using the radiative transfer model DUSTY, in a manner similar
 to that applied to dusty red supergiants (RSG) and AGB stars in the
LMC by \cite{vanLoon2005}. 
In Fig.~\ref{dustymodel} we
show two fits.
In the first case (model $A$) we show a warm circumstellar, spherical,
dust shell with T$_\mathrm{dust}=800$\,K (at the inner boundary), which has
an optical depth of $\tau_{V} = 150$ ($\tau_\mathrm{8{\mu}m} \sim$ 2).  We assume that the  central exciting source is a blackbody of
3000\,K  (a cooler temperature does not make a major difference to the
MIR output). The outer radius of the shell is 454\,AU (2.2
milli-pc), but its thickness needs to be R$_\mathrm{outer}$/R$_\mathrm{inner} < 5$ to fit the SED.  The dust in this case consists of silicates
with a typical grain size distribution. The inferred
luminosity of the central star is $\log \mathrm{L}/ \mathrm{L}_{\odot} \simeq 4.6$
and with such a dense shell all of the stellar energy is absorbed by
the dust and hence all appears as reprocessed MIR flux. While this
may seem a plausible model and scenario for a dust shell around a
massive, embedded star, the extinction toward the central star would
be $A_{V} = 163$\,mag. Hence in this case of a spherical, dense (but
radially thin) circumstellar shell the optical and $K'$ bands offer no
meaningful constraints on the stellar SED. \cite{Prieto2008}
suggested a  thermally radiating sphere 
to account for the observed SED (with a blackbody temperature T$_\mathrm{dust}=440$\,K, $\log \mathrm{L}/ \mathrm{L}_{\odot} \simeq 4.5$, and R$_\mathrm{BB} =150$\,AU).

The second case (model $B$) shown in Fig.~\ref{dustymodel} is a block 
of optically thick interstellar dust with $A_{V} = 140$\,mag.
The extinction law from \cite{Cardelli1989}
is used, although this does 
not fit particularly well around the 8\,$\mu$m detection. The
extremely high value of extinction would imply densities similar
to those in the cores of molecular clouds and the central 
stellar source would need to be $\log \mathrm{L}/ \mathrm{L}_{\odot} \simeq 6.9$
(again we have assumed a blackbody of 3000\,K). 

The main problem with either of the above two scenarios is that the
extinction we see toward SN 2008S can be robustly estimated to be less
than $A_{V} < 4$\,mag and is more likely to be of order $A_{V} = 2.2$\,mag. Hence if a spherical dust shell surrounding an exciting
source is the explanation for the precursor SED, then the explosion
must have destroyed nearly all the dust within the 454\,AU
shell. \cite{Bode1980}, \cite{Wright1980} and \cite{Dwek1983} have suggested that the UV-optical luminosity
from SNe could destroy dust within a spherical
cavity around the progenitor star.  Additionally  UV flashes have been recently
observed from type II-P SNe
\citep{Gezari2008,Schawinski2008}. However,
 these calculations assume a small optical depth, and a more detailed
estimate of an optically thick CSM which is evaporated by UV-optical
photons within a gamma-ray burst beam has been undertaken by
\cite{WaxmanDraine2000}. They suggest that a GRB could clear very
high column densities of dust (e.g. $A_{V}\sim30$\,mag) within several
parsec of the explosion. Taking these calculations,  scaling the
destruction radius R$_\mathrm{d}$ by $\sqrt \mathrm{L}$ (where for a SN X-ray/UV
flash we take E$ \sim$\,$10^{46}$\,erg, as found in
\cite{Gezari2008} and \cite{Schawinski2008}) and 
 reducing the effective flux by the solid angle of the GRB beam in
comparison to the isotropic SN energy, a destruction radius of a few
milli-parsec is found. In the optically thick regime of
\cite{WaxmanDraine2000}   the destruction radius does not depend strongly 
on the density, or $A_{V}$, with R$_\mathrm{d}$ changing by a factor of
less than 2 when n$_\mathrm{H}$ changes by a factor $10^{5}$. Hence it may be feasible for a dust embedded SN to clear
a large enough cavity to become relatively unobscured, even at
relatively high dust densities. 

\cite{Prieto2008} proposed that  the properties of the precursor
source are very similar to  those of dust enshrouded RSG and AGB
stars in the LMC as studied by \cite{vanLoon2005}.
All of the carbon stars in \cite{vanLoon2005}   have
$\log \mathrm{L}/ \mathrm{L}_{\odot} < 4.09$ which is somewhat too low to result in
a CCSN. \cite{Smartt2008} show the lowest luminosity
progenitors to be around 4.3\,dex and stellar evolutionary models
would suggest progenitors of luminosity $<4.1$\,dex would have initial
masses of 5\,\msol or less. Hence these C type stars are not plausible
counterparts to the SN 2008S progenitor and are not viable progenitors
for a CCSN. More luminous and massive stars than these AGB
objects (which are above the core-collapse threshold) tend to have
less optically thick envelopes. For example,  IRAS 04516-6902 is likely
to have $A_{V} \sim 13$\,mag but would be too faint to match the precursor
of SN 2008S. In fact all of the RSGs in LMC with luminosities
above 4.3\,dex would not be bright enough at 8\,$\mu$m to account for
the observed flux before the explosion of SN 2008S. In addition, many
of them would be too bright in the $K$ or $I$ band to be consistent
with our observed upper limits (although adjusting the extinction could
help hide this shorter wavelength flux).

\begin{figure}
\resizebox{\hsize}{!}{\includegraphics{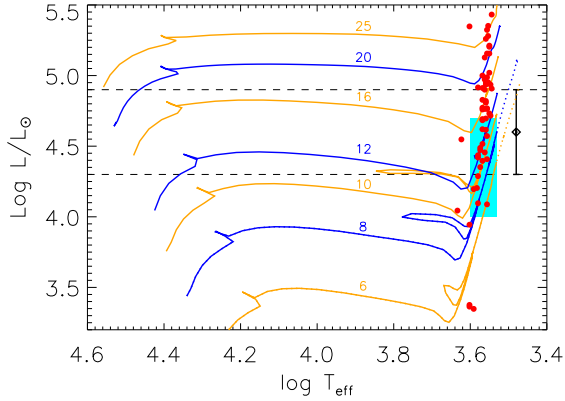}}
\caption{An HR diagram showing the positions of RSGs in the Galaxy from Levesque et al. (2005) (points) and the region (shaded region) in which red supergiant progenitors of normal II-P SNe have been seen (Smartt et al. 2008).  
The luminosity of the progenitor of SN 2008S is estimated as $\log \mathrm{L}/ \mathrm{L}_{\odot}=4.6\pm0.3$, and $T_\mathrm{eff}= 3000$\,K is consistent with the SED (black point with error bar). However the $T_\mathrm{eff}$ is unconstrained from the DUSTY model. The dotted lines limit the region of luminosities allowed for the progenitor. The tracks are the STARS models from Eldridge \& Tout (2004).}
\label{prog-HRD}
\end{figure}

IRAS 05280-6910 \citep{vanLoon2005a} is an extreme case where
the star is hardly detectable in the optical and might appear an
appealing source for comparison. However in this case the 24\,$\mu$m
flux of IRAS 05280-6910 would be much brighter than the limit set for
the pre-explosion source. Hence we agree that the suggestion of
\cite{Prieto2008} of a dust enshrouded RSG is
initially very appealing, and the dust destruction even at such high
column densities does not appear unrealistic. 
However,  we cannot easily
match SN 2008S progenitor  quantitatively with any of the known dusty RSG or AGB stars in the LMC and Galactic samples, in full
agreement across the optical and NIR non-detections.
Our conclusions are consistent with the work of
\citet{Thompson2008}, who show that stars with the same  MIR properties as the progenitor of SN 2008S
are very rare in the nearby spiral M 33.   They
find only $\sim$\,10 objects with similar magnitudes and colour which
they associate with the high luminosity tail of the AGB sequence. They
conclude that this phase is a short  period in the lives of a
reasonable fraction of massive stars, perhaps linked to large mass
ejections and subsequent dust formation episodes in the last
$\sim$\,0.1$\%$ of a stellar lifetime.  The fact that these types of stars
are not common in the LMC studied populations supports this
conclusion.

In Fig.~\ref{prog-HRD} we show an  HertzsprungÐRussell  diagram (HRD) with the STARS model tracks of
\cite{Eldridge2004}. The positions of Galactic RSGs are
shown \citep[from][]{Levesque2005}
along with the region in which RSG progenitors of
recent nearby type II-P SNe lie  \citep[$\log \mathrm{L}/ \mathrm{L}_{\odot} =4.3^{+0.5}_{-0.3}$] [] {Smartt2008}.  The luminosity of the
progenitor has been estimated at $\logl = 4.6\pm0.3$. Although we have estimated the temperature of
the exciting source as T$_\mathrm{eff}= 3000$\,K in the DUSTY model,
this number is not well constrained, as a hotter star can be placed
inside a denser, geometrically thinner envelope and produce a similar
SED. The dotted lines at the extrema of the luminosity ranges show
where the progenitor may lie. The tracks of 6-8\,\msol  stars which have
gone through 2nd dredge-up are shown as dotted line. As discussed by \cite{Eldridge2007}, many different 
stellar evolutionary 
models predict that super-AGB stars (in the 6-9\,\msol range) 
that have gone through 2nd dredge-up
can rise to  higher luminosities and lower effective temperatures
than their higher mass counterparts. These stars may be thermally 
unstable, pulsating, and prone to large mass ejection events. 
The luminosity of the SN 2008S progenitor star is, within the 
uncertainties of both the measurements and models,  consistent
with the position we would expect in the HRD for a 6-8\,\msol,
star which has gone through 2nd dredge-up and has developed an 
O-Ne-Mg core within which electron-capture collapse could occur. 
The luminosity of the progenitor has been 
interpreted as an indication of a mass 
of 10-20\,\msol  \citep{Prieto2008,Smith2008c,Bond2009,Berger2009}. However, as seen in 
 Fig.~\ref{prog-HRD}, the luminosities of 6-8\,\msol  progenitors
after 2nd-dredge up are consistent with the bolometric 
luminosity of the progenitor of  SN 2008S.

So far our analysis has assumed that the pre-explosion MIR source  is a single 
object and not, for example, an embedded cluster of stars within
which the progenitor arose. One cannot definitively rule out
the latter and we noted above that there is some evidence
to suggest that the 4.5\,$\mu$m image is extended. However, 
we would argue that it is unlikely to be a cluster for two reasons. 
Firstly, the total luminosity is not unusually high for a 
stellar source, and if it is a cluster that hosted a SN explosion
then the progenitor is likely the dominant source of flux.
Secondly, the cluster would still need to have an 
extinction similar to that found for the stellar source assumption
and the dust would, presumably, be extended over a 
few parsec (typical cluster size). However,  we estimated
the likely destruction
radius at only a few milli-parsec which seems too small
to have an embedded object within a cluster clear the 
line of sight of intervening dust. The CS dust shell
appears to be more plausible.

\section{The Nature of SN~2008S}
\label{nature}

\subsection{Pre-explosion CSM}
\begin{figure}
\resizebox{\hsize}{!}{\includegraphics{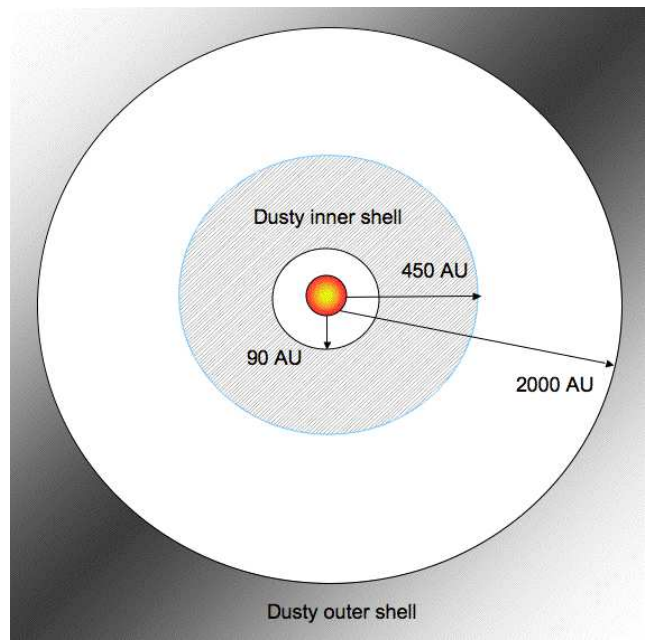}}
\caption{Schematic illustration of  the pre-explosion geometry of SN2008S.}
\label{schematic}
\end{figure}

Super-AGB stars lose mass at a high rate and their circumstellar
medium have a high density and complex geometry.  The circumstellar
environment of SN 2008S suggests such a progenitor star. The
progenitor analysis demonstrates the presence of a dusty optically
thick shell around SN 2008S with an inner radius of nearly 90\,AU ,
while the MIR echo analysis indicates a second, dusty outer shell of
inner radius $\sim$2000\,AU (Fig.~\ref{schematic}).  The two different
shells imply that the progenitor star experienced  a variation in the mass-loss rate.  The physical size of the inner
shell indicates that it was probably circumstellar and the result of
mass loss due to a steady wind rather than eruptive ejection (see also
\cite{Thompson2008}).
If we assume that the optical depth to the source scales as r$^{-1}$
in a freely expanding wind, $\tau_{V}$ to the progenitor was about a
factor of $\sim$ 10--20 larger before explosion i.e. the inner shell
was highly-obscuring.  The dust within this shell was probably
evaporated in the explosion. In contrast the much larger distance of
the dust in the outer shell meant that it was not destroyed by the
radiation from the SN photosphere. Instead, the enormous MIR flux
observed at 17.3 days can be explained as an IR echo from the outer
shell dust whose mass we estimate to be $\sim 10^{-3}$\, \msun.   We
also propose that, if we invoke a toroidal geometry in the outer
shell, then the asymmetric profile of the [CaII] and Fe lines can
also be explained.  This is addressed below.

\subsection{The photometric and spectroscopic evolution}\label{summaryev}
SN 2008S exhibited a very slow photometric evolution and almost no
 spectral variability during the first nine months, implying a long
 photo-diffusion time and a high density CSM \citep{Schlegel1990}.
 Nevertheless, photometric and spectroscopic observations suggest that
 there are three distinct phases:  the maximum light phase (0--50 days after
 the explosion), the flattening of H$\alpha$ intensity 
 (60--100 days) and the NIR excess phase (from about 120 days). 
 
  During the first
 two phases the SN 2008S optical-NIR SED can be fitted with a single blackbody with a radius of 
 about $2 \times 10^{14}$\,cm. The blackbody temperature and radius declined
 monotonically, with the temperature falling from $\sim 8000$\,K to 5000\,K.

During the first phase, all the spectra showed prominent emission
lines of H$\alpha$, the [Ca II] doublet, the Ca II NIR triplet, faint
Fe II emission lines and Na I D in absorption.  None of the H, Ca
or Fe lines exhibit P-Cygni profiles, providing additional evidence of
a high density CSM.  Some of the Na I D absorption probably has an
origin in the CSM, given the observed high density and rapid evolution
of the EW during the first month.  The only forbidden lines visible
in the early spectra are the intense [Ca II] and weak [O I] doublets,
probably produced in the low-density inter-shell zone.

At about 60 days after the explosion the decline  of SN 2008S light
curves steepened,  while the H$\alpha$ intensity remained constant
until about 100~days.  During this phase the H$\alpha$ narrow
component disappeared and  the [Ca II]
profile lost its earlier asymmetry.   
We suggest, as an explanation of the change in the [Ca II] profile,
that the progenitor CSM took the form of a toroid, viewed almost
edge-on.  At early times, obscuration of the [Ca II] emission zone by
the outer shell dust would affect predominantly the fraction of the
gas moving towards us, thus attenuating the blue wing. However, by
60~days, much of the ejecta would have reached the inner shell. (It
would take a velocity of only $\sim 2500$\,\kms.) This would be
expected to produce a strong burst of hard radiation arising from the
ejecta-shell impact. This, in turn, might evaporate the outer shell
dust resulting in the increasing symmetry of the [Ca II] line
profiles. However, there are several difficulties with this
scenario. (a) We surmise that the [O I] doublet formed in the same low
density region as did the [Ca II] feature, and yet no comparable
wavelength shift is seen in the [O I] feature. (b) The IR echo model
indicates an extinction through the outer shell dust of only about
$A=0.15$\,mag at 7300~\AA.  Even allowing for the toroidal geometry,
it is not clear that this would be sufficient to account for the
profile change. (c) If part of the ejecta/inner CSM impact radiation
flowed inward, it might destroy or inhibit the formation of new ejecta
dust.  However, it may be that the impact radiation would be severely
attenuated by intervening material before it could reach the
dust-formation zone. Further discussion of the [Ca II] profile
evolution is beyond the scope of this paper.  We note that the release
in 2009 August of the Spitzer observations of SN~2008S at
$\sim$180~days will provide a useful test of the outer shell dust
evaporation scenario.  We conclude that while the NIR excess provides
clear evidence for newly-formed dust, the evolution of the [CaII]
feature is somewhat more difficult to explain.

The last phase started after about 120 days when the appearance of the
NIR excess indicated the presence of an additional, warm
component. The temperature and radius (and therefore the luminosity)
of the hot component showed a slower decline during this phase. The
warm component cooled from 1400\,K at 160 days to 1200\,K at about 300
days, while its radius and luminosity increased.  An explanation for
the NIR excess is thermal emission from newly-formed dust in the
ejecta or in a cool dense shell formed by the ejecta-inner CSM impact.
At this epoch the decline rate of the SN 2008S light curves flattened
to 1.3 to 0 mag/100d, depending on the band.  The bolometric light
curve shows a decay rate of 0.88 mag/100d, very similar to that of
$^{56}$Co.  The velocity width of the broad component and the
intensity of H$\alpha$ declined slower while the velocity width of the
intermediate component decreased as in the first phases.  The [Ca
II]/H$\alpha$ intensity ratio halved by 260 days while the (Ca II
triplet)/H$\alpha$ declined only slightly. During this later phase the
Na I D and O I appeared in emission.

\subsection{Comparison with  NGC 300 OT2008-1 and M 85 OT2006-1}
\label{NGC300-M85-comp}

Several recent papers  have suggested an analogy between
SN 2008S and two other transients: M 85 OT2006-1 \citep{Kulkarni2007,Pastorello2007} and NGC 300 OT2008-1 \citep{Thompson2008,Berger2009,Bond2009}.
The progenitors of the SN 2008S and NGC 300 OT2008-1 transients have
been detected only on Spitzer pre-explosion images
\citep{Prieto2008,Thompson2008,Berger2009}  which indicates
very similar properties  and geometry of the obscuring CSM.
\cite{Thompson2008} has proposed that these transients are members of a 
new class given the similar pre-explosion properties  
(see Sect.~\ref{proganalysis}). The extensive multi-wavelength 
monitoring campaigns for both these transients ( 
\cite{Smith2008c}  and this paper for SN 2008S;  \cite{Bond2009} and \cite{Berger2009} for NGC 300 OT2008-1) show that they are indeed
very similar in their kinetic and radiative energies. 
M 85 OT2006-1 had a peak $R$ band
absolute magnitude of $\sim -12$ , a peak luminosity L$_\mathrm{p} \sim 2
\times 10^{40}$\,erg\,s$^{-1}$ and a total radiated energy in the first
two months E$_\mathrm{ph} \sim 6 \times 10^{46}$\,erg \citep{Kulkarni2007}, 
while NGC 300 OT2008-1 had a $V$ band peak of $-13$, a luminosity of $1.6\times 10^{40}$\,erg\,s$^{-1}$ and
a total energy of $\sim$\,2$\times 10^{47}$\,erg and no radio or X-ray
emission as SN 2008S \citep{Bond2009,Berger2009}.

\begin{figure}
\resizebox{\hsize}{!}{\includegraphics{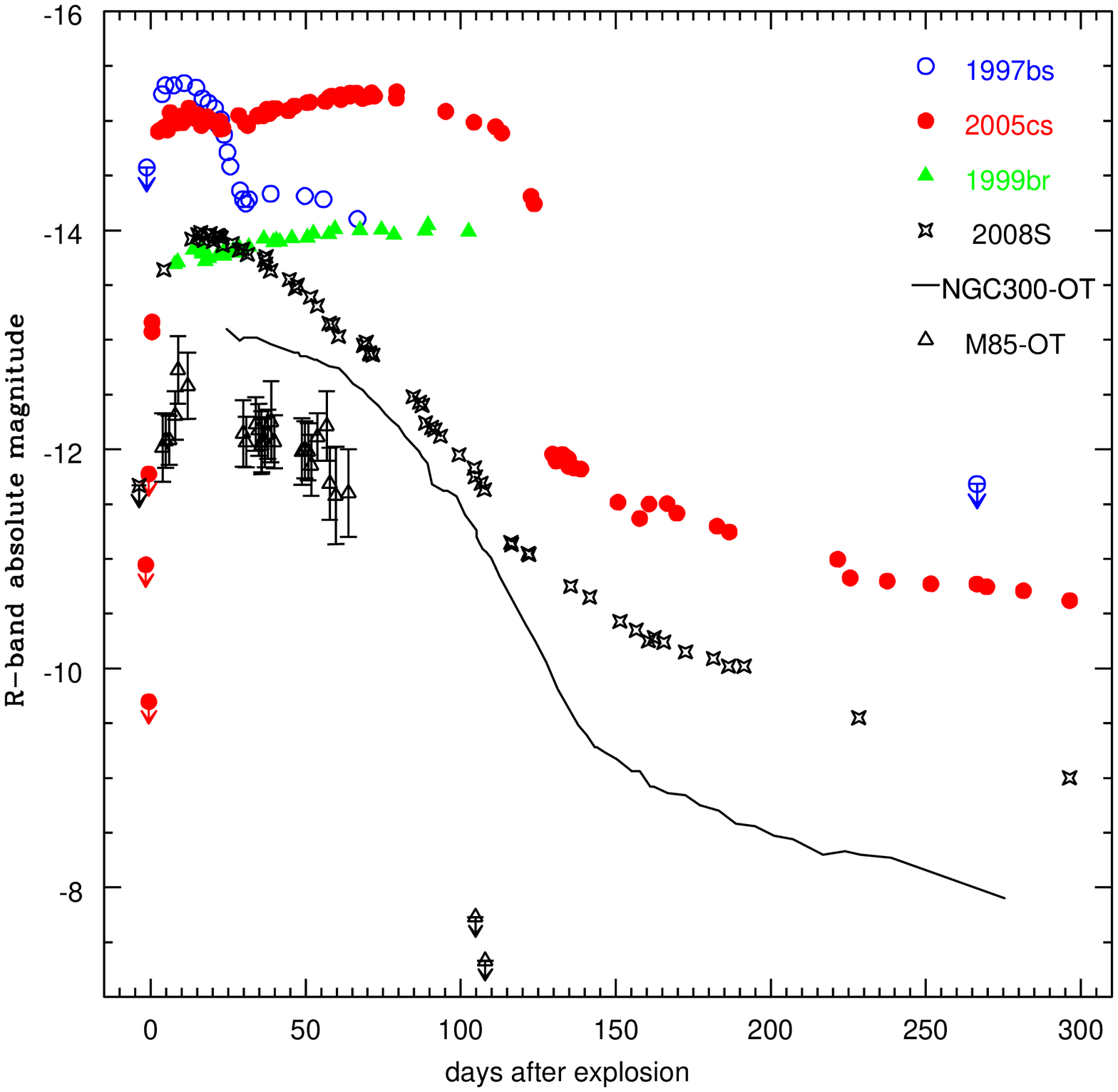}}
\caption{Comparison between light curves of the  transients SN 2008S, M85 OT2006-1 \citep{Kulkarni2007}, NGC 300 OT2008-1 \citep{Bond2009}, the subluminous type II-P SN 1999br  \citep{Pastorello2004} and SN 2005cs  \citep{Pastorello2006,Tsvetkov2006,Pastorello2009}, and the SN impostor SN 1997bs  \citep{VanDyk2000}.}
\label{compRlc}
\end{figure}

The light curves of both transients are very similar to that of
SN 2008S although in the latest phases M 85 OT2006-1
 faded quickly  \citep{Kulkarni2007}, as shown in Fig.~\ref{compRlc}.
 In the first phases of its evolution M 85 OT2006-1 showed a warm
infrared-bright component likely due to an IR echo by circumstellar
dust surviving the explosion \citep{PrietoAtel2}.  The spectra of both
M 85 OT2006-1 and NGC 300 OT2008-1 showed Balmer lines,  the [Ca II]  doublet and  the Ca II NIR
triplet   \citep{Kulkarni2007,Pastorello2007,Bond2009,Berger2009}. Moreover,  in NGC 300 OT2008-1 He I (in emission),  and Ca H$\&$K and OI
$\lambda8446$  (in absorption) are visible while M 85 OT2006-1 showed also
prominent K I lines.  The  velocity width of  H$\alpha$  is
very similar in SN 2008S and NGC 300 OT2008-1,  while in M 85 OT2006-1 is
narrower.

\cite{Smith2008c} have 
interpreted SN 2008S as a SN impostor analogous to the eruptions of
LBVs. They proposed that SN 2008S was a super Eddington outburst of a
star of about 20\,\msol, highly obscured because an outburst which had 
occurred shortly after the recent blue loop transition.
However,  a 20\,\msol  star has a luminosity of $\logl\simeq5.0-5.3$
which is not consistent with the total MIR 
luminosity of the progenitor of $\logl \simeq 4.6$
derived from the  analysis of the pre-explosion images. 
\cite{Smith2008c} also suggested that the spectral similarity of SN 2008S 
in outburst with 
the hypergiant IRC+10420 might indicate that
SN 2008S was also a star of similar evolutionary state. 
However, the similarity between the SN 2008S and IRC+10420 spectra only
points to similar
physical conditions of the regions where the emission lines form. The fact that the 
T$_\mathrm{eff}$ of the continuum of SN 2008S decreases dramatically, while
the emission-line spectrum does not evolve, suggests that the lines are not formed 
in a stellar like expanding photosphere. IRC+10420, in its 
quiescent phase, has a luminosity 20 times higher than that of the 
MIR progenitor but 50 times lower than that of 
SN 2008S  at maximum. 
Finally IRC+10420 is not enshrouded by a
dusty shell and likely has a different mass loss history with respect
to that of progenitors of these transients. Overall,  a physical  or 
evolutionary link between possible LBV-like outbursts and SN 2008S  does not seem convincing to us.

In contrast  \cite{Thompson2008} compared the 
MIR properties of known LBVs in M 33 to the progenitors of SN 2008S and 
NGC 300 OT2008-1 and claimed that a LBV
explanation is unlikely for these transients since the LBV luminosity
is higher and the LBV MIR colours are much bluer than those of the
two transients. Moreover,  they stressed that the time scale of the
LBV variability is not consistent with the lack of variability of the
MIR progenitors. We would agree that this is an evidence against
the stellar eruption scenario, at least in any LBV or LBV-like
event. 

As an explanation for NGC 300 OT2008-1, 
\cite{Berger2009} and \cite{Bond2009} also favour a stellar eruption
which is not unlike that proposed by \cite{Smith2008c}. 
But nevertheless, the issues discussed
        above still argue against this interpretation for SN~2008S.
        
        \cite{Bond2009} proposed as the SN 2008S-like transient progenitors
heavily dust enshrouded luminous stars of about 10--15\,M$_{\odot}$,
likely an OH/IR sources, which have begun to evolve on a blue loop toward warmer temperatures.  During this transition progenitor stars
reached a state in which they exceeded the Eddington limit for their
luminosities and masses and suddenly initiated outflows.  The reason of these eruptions nevertheless remains uncertain.
\cite{Berger2009} explained the nature of NGC 300 OT2008-1 and SN 2008S as an
eruption of a blue supergiant or pre-WR star with a mass of
10--20\,M$_{\odot}$ likely in a binary system.  

\subsection{Comparison with SN 1998S and SN 1979C}\label{comp98S79C}
\begin{figure}
\resizebox{\hsize}{!}{\includegraphics{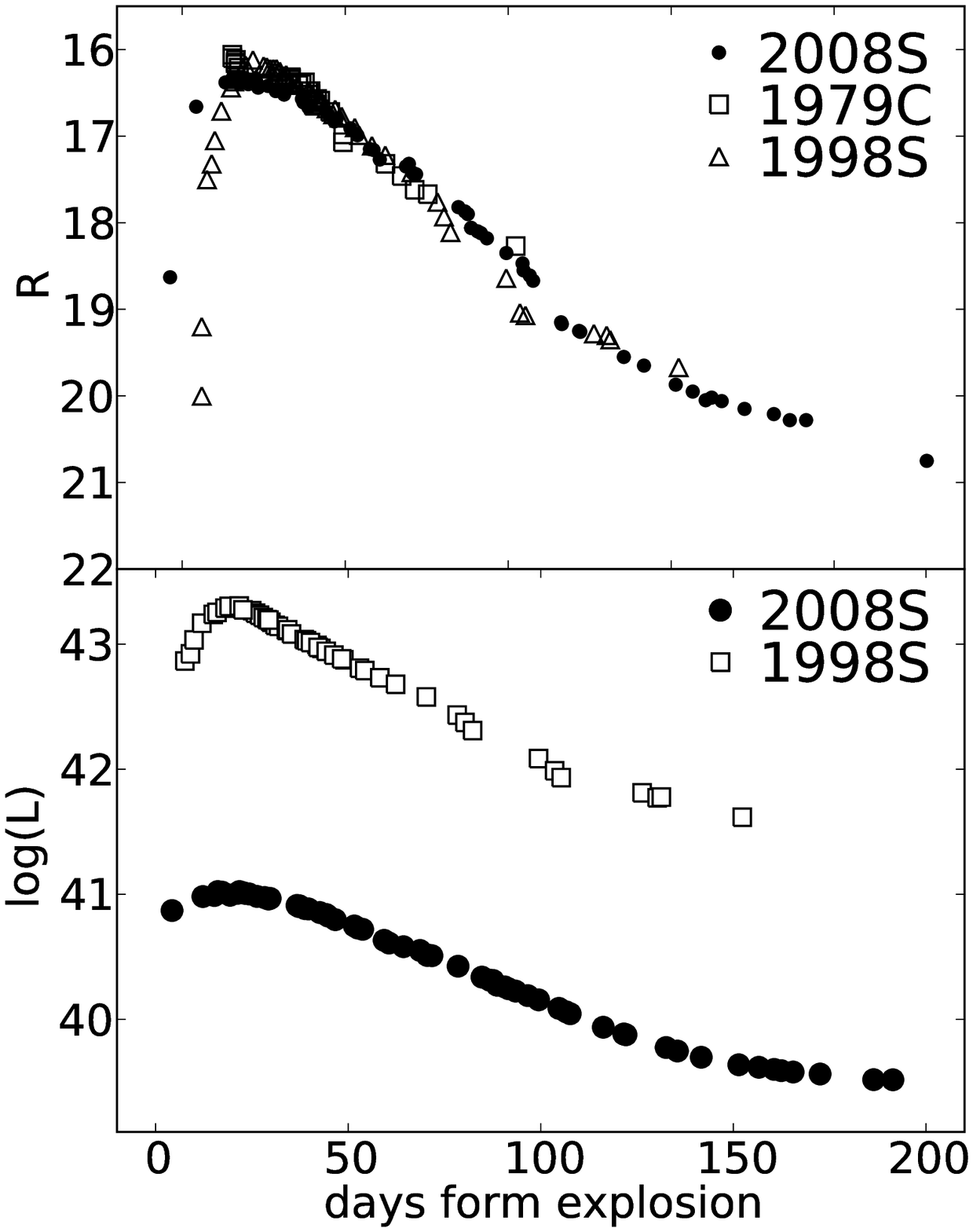}} 
\caption{Comparison between light curves of SN 2008S (filled symbols),  SN 1998S (empty triangles) \citep{Liu2000,Fassia2001} and SN 1979C (empty squares) \citep{Balinskaia1980, deVaucouleurs1981, Barbon1982}:  in the top panel the $R$ band light curves (the SN 1998S and SN 1979C light curves  are  both shifted by 4 mag) and tin the bottom panel the quasi bolometric light curves of SN 2008S and SN 1998S. Phase is in days after  the explosion epoch (JD $2\,450\,869$ for SN 1998S, JD $2\,443\,979$ for SN 1979C and JD $2\,454\,486$ for SN 2008S).}
\label{bolconf}
\end{figure}

\begin{figure}
\resizebox{\hsize}{!}{\includegraphics{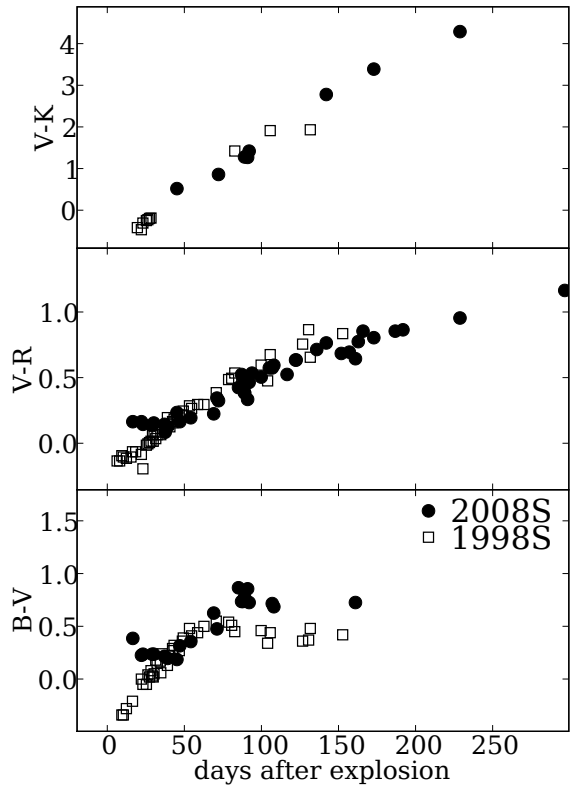}} 
\caption{Comparison  between  $V-K, V-R$ and $B-V$ colours of SN 2008S  (filled symbols) and SN 1998S (empty symbols). The colours of SN 1998S are corrected for $A_{B,Gal}= 0.86$ and $A_{B,int}=0.09$. Phase is in days after the explosion epoch (JD $2\,450\,869$ for SN 1998S, JD $2\,443\,979$ for SN 1979C and JD $2\,454\,486$ for SN 2008S).}
\label{colconf}
\end{figure}

The optical light curves of SN 2008S are surprisingly similar to those
of type IIn SN 1998S and  type II-L SN 1979C (Fig.~ \ref{bolconf}), although these
 were much more luminous (M$_B= -19.6$ from \cite{Fassia2000} and M$_B= -19.4$  from \cite{Panagia1980}, respectively).  In the NIR bands the decline rates of SN 2008S
at $\sim$\,170 days after the explosion are similar to those of slowly
declining CC SNe templates of \cite{Mattila2001}  based on SN 1979C and SN 1998S at the same epoch ($J \sim 0.9$\,mag/100d, $H \sim 0.3$\,mag/100d and $K \sim 0$\,mag/100d).
In Fig.~\ref{bolconf} the $R$ band and  $UBVRIJHK$ quasi-bolometric light curves of
SN 2008S are compared with those of SN 1998S,
while in Fig.~\ref{colconf} the evolution of the $B-V$,  $V-R$ and $V-K$ colours
is shown for SN 2008S and SN 1998S.  The result of this comparison is
intriguing: the overall photometric evolution of these  events is
very similar,  the only two differences being  the absolute luminosity and 
a broader peak for SN 2008S.

Hydrodynamical modelling of type II-L  SNe has suggested three quite
different evolutionary scenarios and explosion mechanisms. 
\cite{Swartz1991} modeled the collapse of a O-Ne-Mg star,
suggesting the electron capture as a mechanism for  II-L SNe. 
\cite{Blinnikov1993} proposed a large supergiant progenitor, while
\cite{Young2005} presented a two component model of a GRB
afterglow with underlying SN ejecta.  
However,  there is still not a clear consensus on the bright  type 
IIn and II-L  events such as SNe 1998S and 1979C. 

The spectral evolution of SN 1979C and SN 1998S
is quite different with respect to
that of SN 2008S but there are some common characteristics.  Both
SN 1979C and SN 1998S showed strong emission lines and absence of P-Cygni
profiles during the first months after the discovery
\citep{Fassia2000,Branch1981}. 
The high
density of the CS shell in the case of SN 1998S implies that a cool
dense shell (CDS) forms at the interface of SN ejecta and the CSM.
The absence of broad P-Cygni profiles was explained with obscuration
by the CDS \citep{Chugai2001}. 
\cite{Fassia2001} showed that absorption troughs  appeared in the spectra
of SN 1998S after only 12 days,  and suggested that the SN ejecta had
over-run the inner CS by that point.


SN 1998S and SN 1979C have been suggested to be the results of
explosions of a red supergiant with an extended envelope,   R$\sim$
(1--10)$\times 10^{3}$\,R$_{\odot}$, and a moderate mass of ejecta
$\sim 5$\,M$_{\odot}$. \cite{Chugai2001} suggested also that the origin
of the CS shell around SN 1998S may be a violent mass loss during the
Ne and O burning in cores of $\sim$\,11\,M$_{\odot}$.
The mass-loss rate of the progenitor by optical, radio and X ray
estimates, is about $1-5 \times 10^{-4}$\,M$_{\odot}$\,yr$^{-1}$.  Likely
the wind of the SN 1998S and SN 1979C progenitor underwent significant
changes in the rate of mass outflow resulting in several distinct CSM
shells.





\subsection{SN 2008S: an electron capture SN}
An alternative explanation of SN 2008S  is the explosion of a massive star in a
low, or moderate, energetic SN.  We show that the observed
characteristics of SN 2008S and its progenitor share characteristics
that are common to  models of electron-capture supernovae
(ECSNe). 
Such
models  have been developed extensively in the
last three decades. 
\cite{Thompson2008} suggested that the ECSN was a viable 
explanation for these SN 2008S-like objects based on the progenitor
properties.  The energetics of the events and their
evolution now add weight to this argument. 
The strongest evidence for this
interpretation is based on the observational discovery that
the light curve tail of SN 2008S follows the decay rate of
$^{56}$Co. The observation of $^{56}$Ni and $^{56}$Co decay can be used
to distinguish between a SN explosion and multiple shell interaction
scenarios \citep{Dessart2008}.  The tail phase luminosity 
decay rate is a strong argument that  $^{56}$Ni was
produced by SN 2008S,  and there is no  physical channel
to produce this radioactive isotope other than explosive
burning of oxygen and silicon at temperatures in excess of $\sim 10^9$\,K. 

A star in the mass range of $\sim$\, 8--10\,M$_{\odot}$ can form an electron
degenerate O-Ne-Mg core during the final stage, as it becomes 
a super-AGB star \citep{Ritossa1999,Siess2007,Eldridge2004}. 
Super-AGB stars
may end their lives either as
massive O-Ne-Mg white dwarfs or as ECSN before Ne ignition
\citep{Miyaji1980,Nomoto1984,Hillebrandt1984,Baron1987,Mayle1988,
Poelarends2008,Wanajo2003,Wanajo2008}.  The final
fate of super-AGB stars depends on the competing effects of 
core growth and mass loss during the late evolutionary stages, 
particularly during carbon burning.  If the
core mass reaches the Chandrasekhar mass, high pressure and density
lead to electron capture onto $^{24}$Mg and $^{20}$Ne, the electron
degeneracy pressure decreases and the core may collapse before an iron
core is  formed \citep{Miyaji1980,Miyaji1987,Hashimoto1993,Poelarends2008,Wanajo2008}.  If
the mass-loss rate is high enough the envelope is lost before the core
reaches the Chandrasekhar mass and the star ends its life as an O-Ne-Mg white dwarf.  

\cite{Nomoto1984,Nomoto1987} estimated that stars with a mass of 8--10\,M$_{\odot}$ can become ECSNe but recent work has suggested a more narrow  
mass range  \citep{Siess2007,Poelarends2008} 
and a lower mass limit of $\sim$\,9\, M$_{\odot}$.
However, \cite{Podsiadlowski2004} suggested that the initial mass range may be
wider if one considers binary systems, 
 about 8--11\,M$_{\odot}$.
ECSNe and their progenitors are predicted to
show three properties that might allow us to distinguish
them from ordinary CCSNe: they might produce low energy explosions;
the enormous mass-loss rate of the progenitor star in the super-AGB phase
may produce signatures of CSM interaction in SN the light curve;  the ECSN
progenitors have luminosities of the order of $10^{5}$\,L$_{\odot}$
and cool effective temperatures \citep{Eldridge2007}. 
All these characteristics appear consistent with
SN 2008S and its progenitor star.
An important implication of the low explosion energy is a small
$^{56}$Ni mass produced by the explosion.
In ECSN explosion models, \cite{Kitaura2006}
estimated an explosion energy of  1--$2\times 10^{50}$\,erg, 
a Ni mass of about $10^{-2}$\,M$_{\odot}$, and an ejecta velocity after the
shock breakout from the stellar surface of $\le 3000$\,\kms. 
The estimated values of  energy and $^{56}$Ni
mass in SN 2008S are somewhat lower than the prediction from
\cite{Kitaura2006},  but more recent models of 
ECSNe by  \cite{Wanajo2008}
are closer to the observed properties of SN 2008S. 
These weak explosions  give an ejected $^{56}$Ni mass of 
0.002--0.004\,\msol.

The progenitor of SN 2008S is too faint to be a luminous RSG, and too bright to be an intermediate-mass AGB 
star. It could thus be a super-AGB star or 
a "low-mass" (close to $\sim$\,8\,\msol   also) RSG.
One argument in favour of a super-AGB star progenitor, is that low-mass 
RSGs generally do not lose mass fast enough to avoid exploding 
as a RSG, by far. So they still have massive mantles, but 
fairly diluted circumstellar envelopes. AGB stars and super-AGB stars lose 
mass at high rate for longer prior to their ends, which in case of AGB 
stars means they lose their mantles before core-collapse could occur 
whereas in super-AGB stars it's a race between mass loss and core growth 
that determines who wins. The stellar mantle prior to explosion would be 
thin anyway, one would expect, and the circumstellar medium would be 
dense.     This strong mass loss is an obvious way to
 account for the
optically thick shell around SN 2008S and NGC 300
OT2008-1.  
    \cite{Weaver1979} 
suggested that the super-AGB stars might produce strong flashes in the
semi-degenerate core a few years prior to a SN explosion, and the
strongest flash could eject most of the H envelope with velocities of 
$\sim$ 100\,\kms.  
The wind speed inferred from the narrow component of H$\alpha$ (about 200\,\kms), which has likely CSM as origin,  has a
         larger velocity width  than is seen normally in AGB stars,
         although such velocities are not unprecedented in post-AGB stars. 
Super-AGB stars, or  ECSN progenitors, may be distinguished from very massive
progenitors of similar luminosity by their much cooler effective
temperatures ( $\le 3000$\,K for ECSN vs. $\sim$ 3400\,K for CCSN
progenitors \citep{Wanajo2008}).  The exciting star in the DUSTY model described above
has a temperature of 3000\,K, hence consistent with cool super-AGB stars
\citep{Eldridge2007}. 
A lack of $\alpha$ elements  in the case of ECSN is
also a key to distinguish between the ECSN and CCSNe \citep{Kitaura2006,Wanajo2008}.
It is interesting to note that 
the Crab remnant has also been suggested
to have originated in 
an ECSN given the low kinetic energy and a small amount of $\alpha$
elements \citep{Nomoto1982,Chevalier1984}.
 Moreover 
\cite{Swartz1991}, as already discussed in the previous section,  suggested that the ECSN progenitor  may lose a
 large fraction of their envelope and become a type II-L SNe.
  
 Finally the rate of these types of transients would be a 
useful guide to their physical origins. 
\cite{Wanajo2008} suggest, from nucleosynthesis arguments, that 
ECSNe must be $\le 30\%$ of all
CCSN events, while \cite{Poelarends2008} suggest that they should
be somewhat rarer ($\sim7$--8$\%$) if they come from a
         narrow mass range. \cite{Thompson2008} propose that there are possibly 4 known 
events that are similar, and that their faint peak magnitudes
mean that many more may be undiscovered in the Local Universe. 
The local SN sample compiled by \cite{Smartt2008} shows 92 CCSNe within a local  28\,Mpc volume, within a 10.5\,yr period. 
At least there are 4 candidates  which are SN2008S-like, 
as discussed in \cite{Thompson2008}. Hence a lower limit 
to the rate of these events of $\geq5$\%  can be inferred, which is 
not inconsistent with the arguments from theory.


\section{Conclusions}

In summary, the combination of our monitoring data and the evidence
from the progenitor studies suggests that a weak 
SN explosion, due to core-collapse
through the electron-capture SN mechanism,  
of a massive star with an initial mass around 6--10\,\msol\ 
in the super-AGB stage 
is a plausible explanation for SN 2008S. 
By implication
NGC 300 OT2008-1 and M 85 OT2006-1 are  likely  of similar 
origin. 
The progenitor star was not a normal RSG which
should  produce the standard type II-P SNe
\citep{Smartt2008}. However,  if  the pre-explosion MIR flux reflects the
stellar photospheric flux, the object has a luminosity similar to 
models of super-AGB stars. Extremely high mass-loss rates could 
create circumstellar shells which are optically thick, 
cool, dense and dusty  \citep{Thompson2008}.  Such high mass-loss through thermal pulses
has been predicted by theory  and obscured massive oxygen-rich
AGB stars are certainly observed by \cite{vanLoon2005}

The evidence that SN 2008S was indeed  a SN explosion comes from two 
observations. Firstly, SN 2008S is similar in the total radiated energy to other faint SNe, and it shows  moderate velocities of about
3000\,\kms.  The latter are hard to reconcile with either an LBV-like
or non-destructive stellar eruption.  Secondly, we detect a tail phase
which has a decay slope matching that of radioactive $^{56}$Co. We
estimate the mass of $^{56}$Ni ejected to be $0.0014\pm 0.0003$\,M$_{\odot}$, 
which  is marginally lower than that observed in  the faintest known type II-P
SNe ($\sim$0.002--0.008\,\msol from \cite{Turatto1998,Benetti2001,Pastorello2004})  but is close enough that SN 2008S could be an extension
of these low-energy explosions. 

\cite{Wanajo2008} have presented models of
electron capture supernovae of progenitor AGB stars with an O-Ne-Mg core
 and an initial mass of 8.8\,\msol that are in 
plausible agreement with  SN 2008S and similar transients. 
These weak explosions, as also discussed by \cite{Kitaura2006}, 
give an ejected $^{56}$Ni mass of 0.002--0.004\,\msol. 
SN~2008S is also significant in that, apart from SN~1987A,  it
         has allowed the earliest ever MIR observation of a SN,
         and the remarkably large flux detected confirms the presence
         of substantial circumstellar material around the progenitor
         star. In addition, SN~2008S developed a NIR excess at later
         times. This can be interpreted as optically thin thermal
         emission from $\sim$\,$10^{-6}$\,M$_{\odot}$ of amorphous carbon
         grains or $\sim$\,$10^{-5}$\,M$_{\odot}$ of silicate
         grains. However, larger masses of grains, which would be
         optically thick, are not ruled out. The grain location and
         heating mechanisms are uncertain. While some of the NIR
         emission may originate in new, radioactively-heated ejecta
         grains, the magnitude of the later NIR flux requires that at
         least a proportion must come from an additional source.  This
         might be reverse shock heating of new ejecta
         grains. Alternatively it might be due to shock heating of new
         (CDS) dust or old dust in the circumstellar region.

Our suggested scenario can be tested in the future. If the 
star has cataclysmically exploded as a SN,  then,  once  the 
remnant fades, there should be no 
source left with a luminosity similar to the progenitor
($10^{4.6}$\,L$_{\odot}$). 
This needs to be tested over
         the entire optical to MIR wavelength range to ensure that no
         progenitor star remains, whether exposed or concealed by
         newly-formed dust. In addition, if the SN ejecta become visible, 
one might expect to see broad
lines of forbidden oxygen (e.g. [O I] $\lambda\lambda$ 6300,6364) and
other intermediate mass elements. This may be difficult, as the
events are intrinsically faint, but it could be possible  in particular for 
NGC 300 OT2008-1 as its distance is only 2\,Mpc
and hence it can be monitored for  a long time. The 
tail phase, following  the $^{56}$Co decay,  should continue  at least for 
another $\sim$\,300 days, and the slope of this decay should
be monitored closely in both SN 2008S and NGC 300 OT2008-1.  
Finally the true rate of these transients
will give further insights into their nature, and deeper 
searches by future sky surveys may discover more of these events
\citep{2008A&A...489..359Y}.




\section*{Acknowledgments}
We would like to thank the referee,  P. Bouchet,  and  J. Danziger  for helpful discussions  and   F. Sabbadin for  his suggestions on the evolution of  line profiles.
This work, conducted as part of the award "Understanding the lives of
massive stars from birth to supernovae" (S.J. Smartt) made under the
European Heads of Research Councils and European Science Foundation
EURYI (European Young Investigator) Awards scheme, was supported by
funds from the Participating Organisations of EURYI and the EC Sixth
Framework Programme.
S. M. acknowledges financial support from Academy of Finland (project 8120503),  D. T. acknowledges financial support from the Program of Support for Leading Scientific Schools of Russian Federation (project
NSh.433.2008.2),  V.S. acknowledges financial support from the Funda\c{c}\~{a}o para a Ci\^{e}ncia e a Tecnologia and  I.M. V. acknowledges financial support from SAI scholarship.
F.P. K. is grateful to AWE Aldermaston for the award of a William Penney Fellowship.
This work is based on observations collected at WHT (La Palma), NOT (La Palma), INT (La Palma), TNG (La Palma), Copernico 1.82 m telescope, the 2.2 m Telescope (Calar Alto).
The WHT and INT are operated  by the Isaac Newton Group,  the NOT is operated jointly by Denmark, Finland, Iceland, Norway and Sweden, the Liverpool Telescope is operated by the Astrophysics Research Institute of Liverpool John Moores University and the TNG is operated by the Fundazione Galileo Galilei - National Institute for Astrophysics (INAF), Fundacion Canaria on the island of La Palma in the Spanish Observatorio del Roque de los Muchachos of the Instituto de Astrofisica de Canarias. 
 The 1.82 m telescope is operated by the Osservatorio di Padova INAF.  The 2.2m telescope is operated jointly by the Max-Planck-Institut fur Astronomie (MPIA) in Heidelberg, Germany, and the Instituto de Astrofisica de Andalucia (CSIC) in Granada/Spain in the Centro Astronomico Hispano Aleman at Calar Alto. 
We are grateful to the support astronomers at these telescopes for performing the follow up observations of SN 2008S in particular to P. Rodriguez Gil at Isaac Newton Group of Telescopes, V.P.Goranskij at SAO and T.R.Irsmambetova at SAI Crimean laboratory. 
Moreover, we are grateful to U. Hopp for arranging the observations at the Wendelstein Observatory, to M. Dolci and E. Di Carlo for arraging observations at Osservatorio di Campo Imperatore and to S. D. Van Dyk, and J. Mauerhan for  observations at  Palomar 5m telescope.  
We thank the members of the LBT partnership who contributed to the Science Demonstration Time observation and J. Knapen for the images of the Bok telescope.
This work is based in part on archival data obtained with the Spitzer Space Telescope and made use of the NASA/IPAC Extragalactic Database (NED),
 which are operated by the Jet Propulsion Laboratory, California Institute of Technology under a contract with National Aeronautics and Space Administration.
 Support for this work was provided by an award issued by JPL/Caltech.
 We also exploited data products from the Two Micron All Sky Survey (2MASS), which is a joint project of the University of Massachusetts and the Infrared Processing and Analysis center/California Institute of Technology, funded by the National Aeronautics and Space Administration and the National Science Foundation.We acknowledge the usage of the HyperLeda database (http://leda.univ-lyon1.fr).


\begin{thebibliography}{}


\bibitem[\protect\citeauthoryear{{Alard}}{{Alard}}{2000}]{Alard2000}
{Alard} C.,  2000, \aaps, 144, 363

\bibitem[\protect\citeauthoryear{{Arbour}}{{Arbour}}{2008}]{Arbour2008CBET2}
{Arbour} R.,  2008, Central Bureau Electronic Telegrams, 1235, 2

\bibitem[\protect\citeauthoryear{{Arbour} \& {Boles}}{{Arbour} \&
  {Boles}}{2008}]{Arbour2008CBET1}
{Arbour} R.,  {Boles} T.,  2008, Central Bureau Electronic Telegrams, 1234, 1

\bibitem[\protect\citeauthoryear{{Arnett} \& {Fu}}{{Arnett} \&
  {Fu}}{1989}]{Arnett1989}
{Arnett} W.~D.,  {Fu} A.,  1989, \apj, 340, 396

\bibitem[\protect\citeauthoryear{{Balinskaia}, {Bychkov} \&
  {Neizvestnyi}}{{Balinskaia} et~al.}{1980}]{Balinskaia1980}
{Balinskaia} I.~S.,  {Bychkov} K.~V.,    {Neizvestnyi} S.~I.,  1980, \aap, 85,
  L19+

\bibitem[\protect\citeauthoryear{{Barbon}, {Ciatti}, {Rosino}, {Ortolani} \&
  {Rafanelli}}{{Barbon} et~al.}{1982}]{Barbon1982}
{Barbon} R.,  {Ciatti} F.,  {Rosino} L.,  {Ortolani} S.,    {Rafanelli} P.,
  1982, \aap, 116, 43

\bibitem[\protect\citeauthoryear{{Baron}, {Cooperstein} \& {Kahana}}{{Baron}
  et~al.}{1987}]{Baron1987}
{Baron} E.,  {Cooperstein} J.,    {Kahana} S.,  1987, \apj, 320, 300

\bibitem[\protect\citeauthoryear{{Benetti}, {Turatto}, {Balberg}, {Zampieri},
  {Shapiro}, {Cappellaro}, {Nomoto}, {Nakamura}, {Mazzali} \&
  {Patat}}{{Benetti} et~al.}{2001}]{Benetti2001}
{Benetti} S.,  {Turatto} M.,  {Balberg} S.,  {Zampieri} L.,  {Shapiro} S.~L.,
  {Cappellaro} E.,  {Nomoto} K.,  {Nakamura} T.,  {Mazzali} P.~A.,    {Patat}
  F.,  2001, \mnras, 322, 361

\bibitem[\protect\citeauthoryear{{Berger}, {Soderberg}, {Chevalier},
  {Fransson}, {Foley}, {Leonard}, {Debes}, {Diamond-Stanic}, {Dupree}, {Ivans},
  {Simmerer}, {Thompson} \& {Tremonti}}{{Berger} et~al.}{2009}]{Berger2009}
{Berger} E.,  {Soderberg} A.~M.,  {Chevalier} R.~A.,  {Fransson} C.,  {Foley}
  R.~J.,  {Leonard} D.~C.,  {Debes} J.~H.,  {Diamond-Stanic} A.~M.,  {Dupree}
  A.~K.,  {Ivans} I.~I.,  {Simmerer} J.,  {Thompson} I.~B.,    {Tremonti}
  C.~A.,  2009, ArXiv e-prints

\bibitem[\protect\citeauthoryear{{Blinnikov} \& {Bartunov}}{{Blinnikov} \&
  {Bartunov}}{1993}]{Blinnikov1993}
{Blinnikov} S.~I.,  {Bartunov} O.~S.,  1993, \aap, 273, 106

\bibitem[\protect\citeauthoryear{{Blondin}, {Prieto}, {Patat}, {Challis},
  {Hicken}, {Kirshner}, {Matheson} \& {Modjaz}}{{Blondin}
  et~al.}{2008}]{Blondin2008}
{Blondin} S.,  {Prieto} J.~L.,  {Patat} F.,  {Challis} P.,  {Hicken} M.,
  {Kirshner} R.~P.,  {Matheson} T.,    {Modjaz} M.,  2008, ArXiv e-prints

\bibitem[\protect\citeauthoryear{{Bode} \& {Evans}}{{Bode} \&
  {Evans}}{1980}]{Bode1980}
{Bode} M.~F.,  {Evans} A.,  1980, \mnras, 193, 21P

\bibitem[\protect\citeauthoryear{{Bond}, {Bonanos}, {Humphreys}, {Berto
  Monard}, {Prieto} \& {Walter}}{{Bond} et~al.}{2009}]{Bond2009}
{Bond} H.~E.,  {Bonanos} A.~Z.,  {Humphreys} R.~M.,  {Berto Monard} L.~A.~G.,
  {Prieto} J.~L.,    {Walter} F.~M.,  2009, ArXiv e-prints

\bibitem[\protect\citeauthoryear{{Boomsma}, {Oosterloo}, {Fraternali}, {van der
  Hulst} \& {Sancisi}}{{Boomsma} et~al.}{2008}]{Boomsma2008}
{Boomsma} R.,  {Oosterloo} T.~A.,  {Fraternali} F.,  {van der Hulst} J.~M.,
  {Sancisi} R.,  2008, \aap, 490, 555

\bibitem[\protect\citeauthoryear{{Bouchet}, {Danziger} \& {Lucy}}{{Bouchet}
  et~al.}{1991}]{Bouchet1991}
{Bouchet} P.,  {Danziger} I.~J.,    {Lucy} L.~B.,  1991, \aj, 102, 1135

\bibitem[\protect\citeauthoryear{{Boulanger} \& {Viallefond}}{{Boulanger} \&
  {Viallefond}}{1992}]{Boulanger1992}
{Boulanger} F.,  {Viallefond} F.,  1992, \aap, 266, 37

\bibitem[\protect\citeauthoryear{{Bowen}, {Roth}, {Meyer} \& {Blades}}{{Bowen}
  et~al.}{2000}]{Bowen2000}
{Bowen} D.~V.,  {Roth} K.~C.,  {Meyer} D.~M.,    {Blades} J.~C.,  2000, \apj,
  536, 225

\bibitem[\protect\citeauthoryear{{Branch}, {Falk}, {Uomoto}, {Wills}, {McCall}
  \& {Rybski}}{{Branch} et~al.}{1981}]{Branch1981}
{Branch} D.,  {Falk} S.~W.,  {Uomoto} A.~K.,  {Wills} B.~J.,  {McCall} M.~L.,
   {Rybski} P.,  1981, \apj, 244, 780

\bibitem[\protect\citeauthoryear{{Cardelli}, {Clayton} \& {Mathis}}{{Cardelli}
  et~al.}{1989}]{Cardelli1989}
{Cardelli} J.~A.,  {Clayton} G.~C.,    {Mathis} J.~S.,  1989, \apj, 345, 245

\bibitem[\protect\citeauthoryear{{Carignan}, {Charbonneau}, {Boulanger} \&
  {Viallefond}}{{Carignan} et~al.}{1990}]{Carignan1990}
{Carignan} C.,  {Charbonneau} P.,  {Boulanger} F.,    {Viallefond} F.,  1990,
  \aap, 234, 43

\bibitem[\protect\citeauthoryear{{Chandra} \& {Soderberg}}{{Chandra} \&
  {Soderberg}}{2008}]{Chandra2008ATel}
{Chandra} P.,  {Soderberg} A.,  2008, The Astronomer's Telegram, 1382, 1

\bibitem[\protect\citeauthoryear{{Chevalier}}{{Chevalier}}{1984}]{Chevalier1984}
{Chevalier} R.~A.,  1984, \apj, 280, 797

\bibitem[\protect\citeauthoryear{{Chevalier} \& {Fransson}}{{Chevalier} \&
  {Fransson}}{1994}]{ChevalierFransson1994}
{Chevalier} R.~A.,  {Fransson} C.,  1994, \apj, 420, 268

\bibitem[\protect\citeauthoryear{{Chugai}}{{Chugai}}{2001}]{Chugai2001}
{Chugai} N.~N.,  2001, \mnras, 326, 1448

\bibitem[\protect\citeauthoryear{{Chugai}}{{Chugai}}{2008}]{Chugai2008a}
{Chugai} N.~N.,  2008, Astronomy Letters, 34, 389

\bibitem[\protect\citeauthoryear{{Chugai} \& {Utrobin}}{{Chugai} \&
  {Utrobin}}{2008}]{Chugai2008}
{Chugai} N.~N.,  {Utrobin} V.~P.,  2008, Astronomy Letters, 34, 589

\bibitem[\protect\citeauthoryear{{Danziger}, {Lucy}, {Bouchet} \&
  {Gouiffes}}{{Danziger} et~al.}{1991}]{Danziger1991}
{Danziger} I.~J.,  {Lucy} L.~B.,  {Bouchet} P.,    {Gouiffes} C.,  1991, in
  {Woosley} S.~E.,  ed., Supernovae {Molecules Dust and Ionic Abundances in
  Supernova 1987A}.
pp 69--+

\bibitem[\protect\citeauthoryear{{de Vaucouleurs}, {de Vaucouleurs}, {Buta},
  {Ables} \& {Hewitt}}{{de Vaucouleurs} et~al.}{1981}]{deVaucouleurs1981}
{de Vaucouleurs} G.,  {de Vaucouleurs} A.,  {Buta} R.,  {Ables} H.~D.,
  {Hewitt} A.~V.,  1981, \pasp, 93, 36

\bibitem[\protect\citeauthoryear{{Degioia-Eastwood}, {Grasdalen}, {Strom} \&
  {Strom}}{{Degioia-Eastwood} et~al.}{1984}]{Degioia1984}
{Degioia-Eastwood} K.,  {Grasdalen} G.~L.,  {Strom} S.~E.,    {Strom} K.~M.,
  1984, \apj, 278, 564

\bibitem[\protect\citeauthoryear{{Dessart}, {Hillier}, {Gezari}, {Basa} \&
  {Matheson}}{{Dessart} et~al.}{2008}]{Dessart2008}
{Dessart} L.,  {Hillier} D.~J.,  {Gezari} S.,  {Basa} S.,    {Matheson} T.,
  2008, ArXiv e-prints

\bibitem[\protect\citeauthoryear{{Di Carlo}, {Corsi}, {Arkharov}, {Massi},
  {Larionov}, {Efimova}, {Dolci}, {Napoleone} \& {Di Paola}}{{Di Carlo}
  et~al.}{2008}]{DiCarlo2008}
{Di Carlo} E.,  {Corsi} C.,  {Arkharov} A.~A.,  {Massi} F.,  {Larionov} V.~M.,
  {Efimova} N.~V.,  {Dolci} M.,  {Napoleone} N.,    {Di Paola} A.,  2008, \apj,
  684, 471

\bibitem[\protect\citeauthoryear{{Drake} \& {Ulrich}}{{Drake} \&
  {Ulrich}}{1980}]{Drake1980}
{Drake} S.~A.,  {Ulrich} R.~K.,  1980, \apjs, 42, 351

\bibitem[\protect\citeauthoryear{{Draper}}{{Draper}}{2000}]{Draper2000}
{Draper} P.~W.,  2000, in {Manset} N.,  {Veillet} C.,   {Crabtree} D.,  eds,
  Astronomical Data Analysis Software and Systems IX Vol.~216 of Astronomical
  Society of the Pacific Conference Series, {GAIA: Recent Developments}.
pp 615--+

\bibitem[\protect\citeauthoryear{{Dwek}}{{Dwek}}{1983}]{Dwek1983}
{Dwek} E.,  1983, \apj, 274, 175

\bibitem[\protect\citeauthoryear{{Eldridge}, {Mattila} \& {Smartt}}{{Eldridge}
  et~al.}{2007}]{Eldridge2007}
{Eldridge} J.~J.,  {Mattila} S.,    {Smartt} S.~J.,  2007, \mnras, 376, L52

\bibitem[\protect\citeauthoryear{{Eldridge} \& {Tout}}{{Eldridge} \&
  {Tout}}{2004}]{Eldridge2004}
{Eldridge} J.~J.,  {Tout} C.~A.,  2004, \mnras, 353, 87

\bibitem[\protect\citeauthoryear{{Elmhamdi}, {Danziger}, {Chugai},
  {Pastorello}, {Turatto}, {Cappellaro}, {Altavilla}, {Benetti}, {Patat} \&
  {Salvo}}{{Elmhamdi} et~al.}{2003}]{Elmhamdi2003}
{Elmhamdi} A.,  {Danziger} I.~J.,  {Chugai} N.,  {Pastorello} A.,  {Turatto}
  M.,  {Cappellaro} E.,  {Altavilla} G.,  {Benetti} S.,  {Patat} F.,    {Salvo}
  M.,  2003, \mnras, 338, 939

\bibitem[\protect\citeauthoryear{{Engargiola}}{{Engargiola}}{1991}]{Engargiola%
1991}
{Engargiola} G.,  1991, \apjs, 76, 875

\bibitem[\protect\citeauthoryear{{Fassia}, {Meikle}, {Chugai}, {Geballe},
  {Lundqvist}, {Walton}, {Pollacco}, {Veilleux}, {Wright}, {Pettini}, {Kerr},
  {Puchnarewicz}, {Puxley}, {Irwin}, {Packham}, {Smartt} \& {Harmer}}{{Fassia}
  et~al.}{2001}]{Fassia2001}
{Fassia} A.,  {Meikle} W.~P.~S.,  {Chugai} N.,  {Geballe} T.~R.,  {Lundqvist}
  P.,  {Walton} N.~A.,  {Pollacco} D.,  {Veilleux} S.,  {Wright} G.~S.,
  {Pettini} M.,  {Kerr} T.,  {Puchnarewicz} E.,  {Puxley} P.,  {Irwin} M.,
  {Packham} C.,  {Smartt} S.~J.,    {Harmer} D.,  2001, \mnras, 325, 907

\bibitem[\protect\citeauthoryear{{Fassia}, {Meikle}, {Vacca}, {Kemp}, {Walton},
  {Pollacco}, {Smartt} \& et al}{{Fassia} et~al.}{2000}]{Fassia2000}
{Fassia} A.,  {Meikle} W.~P.~S.,  {Vacca} W.~D.,  {Kemp} S.~N.,  {Walton}
  N.~A.,  {Pollacco} D.~L.,  {Smartt} S.,    et al 2000, \mnras, 318, 1093

\bibitem[\protect\citeauthoryear{{Gerardy}, {Fesen}, {Nomoto}, {Garnavich},
  {Jha}, {Challis}, {Kirshner}, {H{\"o}flich} \& {Wheeler}}{{Gerardy}
  et~al.}{2002}]{Gerardy2002}
{Gerardy} C.~L.,  {Fesen} R.~A.,  {Nomoto} K.,  {Garnavich} P.~M.,  {Jha} S.,
  {Challis} P.~M.,  {Kirshner} R.~P.,  {H{\"o}flich} P.,    {Wheeler} J.~C.,
  2002, \apj, 575, 1007

\bibitem[\protect\citeauthoryear{{Gezari}, {Dessart}, {Basa}, {Martin},
  {Neill}, {Woosley}, {Hillier}, {Bazin}, {Forster}, {Friedman}, {Le Du},
  {Mazure}, {Morrissey}, {Neff}, {Schiminovich} \& {Wyder}}{{Gezari}
  et~al.}{2008}]{Gezari2008}
{Gezari} S.,  {Dessart} L.,  {Basa} S.,  {Martin} D.~C.,  {Neill} J.~D.,
  {Woosley} S.~E.,  {Hillier} D.~J.,  {Bazin} G.,  {Forster} K.,  {Friedman}
  P.~G.,  {Le Du} J.,  {Mazure} A.,  {Morrissey} P.,  {Neff} S.~G.,
  {Schiminovich} D.,    {Wyder} T.~K.,  2008, \apjl, 683, L131

\bibitem[\protect\citeauthoryear{{Hamuy}}{{Hamuy}}{2003}]{Hamuy2003}
{Hamuy} M.,  2003, \apj, 582, 905

\bibitem[\protect\citeauthoryear{{Hamuy} \& {Pinto}}{{Hamuy} \&
  {Pinto}}{2002}]{Hamuy2002}
{Hamuy} M.,  {Pinto} P.~A.,  2002, \apjl, 566, L63

\bibitem[\protect\citeauthoryear{{Hashimoto}, {Iwamoto} \&
  {Nomoto}}{{Hashimoto} et~al.}{1993}]{Hashimoto1993}
{Hashimoto} M.,  {Iwamoto} K.,    {Nomoto} K.,  1993, \apjl, 414, L105

\bibitem[\protect\citeauthoryear{{Herrmann}, {Ciardullo}, {Feldmeier} \&
  {Vinciguerra}}{{Herrmann} et~al.}{2008}]{Herrmann2008}
{Herrmann} K.~A.,  {Ciardullo} R.,  {Feldmeier} J.~J.,    {Vinciguerra} M.,
  2008, \apj, 683, 630

\bibitem[\protect\citeauthoryear{{Hillebrandt}, {Nomoto} \&
  {Wolff}}{{Hillebrandt} et~al.}{1984}]{Hillebrandt1984}
{Hillebrandt} W.,  {Nomoto} K.,    {Wolff} R.~G.,  1984, \aap, 133, 175

\bibitem[\protect\citeauthoryear{{Huo} et al.}{{Huo} 
  et~al.}{1987}]{Huo1987}
{Huo} J.,  et al. 1987, Nuclear Data Sheets, 51, 1

\bibitem[\protect\citeauthoryear{{Kamphuis} \& {Sancisi}}{{Kamphuis} \&
  {Sancisi}}{1993}]{Kamphuis1993}
{Kamphuis} J.,  {Sancisi} R.,  1993, \aap, 273, L31+

\bibitem[\protect\citeauthoryear{{Karachentsev}, {Sharina} \&
  {Huchtmeier}}{{Karachentsev} et~al.}{2000}]{Karachentsev2000}
{Karachentsev} I.~D.,  {Sharina} M.~E.,    {Huchtmeier} W.~K.,  2000, \aap,
  362, 544

\bibitem[\protect\citeauthoryear{{Kitaura}, {Janka} \& {Hillebrandt}}{{Kitaura}
  et~al.}{2006}]{Kitaura2006}
{Kitaura} F.~S.,  {Janka} H.-T.,    {Hillebrandt} W.,  2006, \aap, 450, 345

\bibitem[\protect\citeauthoryear{{Knapen}, {de Jong}, {Stedman} \&
  {Bramich}}{{Knapen} et~al.}{2003}]{Knapen2003}
{Knapen} J.~H.,  {de Jong} R.~S.,  {Stedman} S.,    {Bramich} D.~M.,  2003,
  \mnras, 344, 527


\bibitem[\protect\citeauthoryear{{Kotak}, {Meikle},  {Farrah}}{{Kotak} et~al.}{2009}]{Kotak2009}
{Kotak} R.,  {Meikle} P.,  {Farrah} D. et al., 2009, ArXiv e-prints
 

\bibitem[\protect\citeauthoryear{{Kulkarni}, {Ofek}, {Rau}, {Cenko},
  {Soderberg}, {Fox}, {Gal-Yam}, {Capak}, {Moon}, {Li}, {Filippenko}, {Egami},
  {Kartaltepe} \& {Sanders}}{{Kulkarni} et~al.}{2007}]{Kulkarni2007}
{Kulkarni} S.~R.,  {Ofek} E.~O.,  {Rau} A.,  {Cenko} S.~B.,  {Soderberg} A.~M.,
   {Fox} D.~B.,  {Gal-Yam} A.,  {Capak} P.~L.,  {Moon} D.~S.,  {Li} W.,
  {Filippenko} A.~V.,  {Egami} E.,  {Kartaltepe} J.,    {Sanders} D.~B.,  2007,
  \nat, 447, 458

\bibitem[\protect\citeauthoryear{{Lacey}, {Duric} \& {Goss}}{{Lacey}
  et~al.}{1997}]{Lacey1997}
{Lacey} C.,  {Duric} N.,    {Goss} W.~M.,  1997, \apjs, 109, 417

\bibitem[\protect\citeauthoryear{{Langer}, {Norman}, {de Koter}, {Vink},
  {Cantiello} \& {Yoon}}{{Langer} et~al.}{2007}]{Langer2007}
{Langer} N.,  {Norman} C.~A.,  {de Koter} A.,  {Vink} J.~S.,  {Cantiello} M.,
   {Yoon} S.-C.,  2007, \aap, 475, L19

\bibitem[\protect\citeauthoryear{{Levesque}, {Massey}, {Olsen}, {Plez},
  {Josselin}, {Maeder} \& {Meynet}}{{Levesque} et~al.}{2005}]{Levesque2005}
{Levesque} E.~M.,  {Massey} P.,  {Olsen} K.~A.~G.,  {Plez} B.,  {Josselin} E.,
  {Maeder} A.,    {Meynet} G.,  2005, \apj, 628, 973

\bibitem[\protect\citeauthoryear{{Li}, {McCray} \& {Sunyaev}}{{Li}
  et~al.}{1993}]{Li1993}
{Li} H.,  {McCray} R.,    {Sunyaev} R.~A.,  1993, \apj, 419, 824

\bibitem[\protect\citeauthoryear{{Liddle}}{{Liddle}}{2004}]{lid04}
{Liddle} A.~R.,  2004, \mnras, 351, L49

\bibitem[\protect\citeauthoryear{{Liu}, {Hu}, {Hang}, {Qiu}, {Zhu} \&
  {Qiao}}{{Liu} et~al.}{2000}]{Liu2000}
{Liu} Q.-Z.,  {Hu} J.-Y.,  {Hang} H.-R.,  {Qiu} Y.-L.,  {Zhu} Z.-X.,    {Qiao}
  Q.-Y.,  2000, \aaps, 144, 219

\bibitem[\protect\citeauthoryear{{Lucy}, {Danziger}, {Gouiffes} \&
  {Bouchet}}{{Lucy} et~al.}{1991}]{Lucy1991}
{Lucy} L.~B.,  {Danziger} I.~J.,  {Gouiffes} C.,    {Bouchet} P.,  1991, in
  {Woosley} S.~E.,  ed., Supernovae {Dust Condensation in the Ejecta of
  Supernova 1987A - Part Two}.


\bibitem[\protect\citeauthoryear{{Matonick} \& {Fesen}}{{Matonick} \&
  {Fesen}}{1997}]{Matonick1997}
{Matonick} D.~M.,  {Fesen} R.~A.,  1997, \apjs, 112, 49

\bibitem[\protect\citeauthoryear{{Mattila} \& {Meikle}}{{Mattila} \&
  {Meikle}}{2001}]{Mattila2001}
{Mattila} S.,  {Meikle} W.~P.~S.,  2001, \mnras, 324, 325

\bibitem[\protect\citeauthoryear{{Mattila}, {Meikle}, {Lundqvist},
  {Pastorello}, {Kotak}, {Eldridge}, {Smartt}, {Adamson}, {Gerardy}, {Rizzi},
  {Stephens} \& {van Dyk}}{{Mattila} et~al.}{2008}]{Mattila2008}
{Mattila} S.,  {Meikle} W.~P.~S.,  {Lundqvist} P.,  {Pastorello} A.,  {Kotak}
  R.,  {Eldridge} J.,  {Smartt} S.,  {Adamson} A.,  {Gerardy} C.~L.,  {Rizzi}
  L.,  {Stephens} A.~W.,    {van Dyk} S.~D.,  2008, \mnras, 389, 141

\bibitem[\protect\citeauthoryear{{Mayle} \& {Wilson}}{{Mayle} \&
  {Wilson}}{1988}]{Mayle1988}
{Mayle} R.,  {Wilson} J.~R.,  1988, \apj, 334, 909

\bibitem[\protect\citeauthoryear{{Meikle}, {Mattila}, {Gerardy}, {Kotak},
  {Pozzo}, {van Dyk}, {Farrah}, {Fesen}, {Filippenko}, {Fransson}, {Lundqvist},
  {Sollerman} \& {Wheeler}}{{Meikle} et~al.}{2006}]{Meikle2006}
{Meikle} W.~P.~S.,  {Mattila} S.,  {Gerardy} C.~L.,  {Kotak} R.,  {Pozzo} M.,
  {van Dyk} S.~D.,  {Farrah} D.,  {Fesen} R.~A.,  {Filippenko} A.~V.,
  {Fransson} C.,  {Lundqvist} P.,  {Sollerman} J.,    {Wheeler} J.~C.,  2006,
  \apj, 649, 332

\bibitem[\protect\citeauthoryear{{Meikle}, {Mattila}, {Pastorello}, {Gerardy},
  {Kotak}, {Sollerman}, {Van Dyk}, {Farrah}, {Filippenko}, {H{\"o}flich},
  {Lundqvist}, {Pozzo} \& {Wheeler}}{{Meikle} et~al.}{2007}]{Meikle2007}
{Meikle} W.~P.~S.,  {Mattila} S.,  {Pastorello} A.,  {Gerardy} C.~L.,  {Kotak}
  R.,  {Sollerman} J.,  {Van Dyk} S.~D.,  {Farrah} D.,  {Filippenko} A.~V.,
  {H{\"o}flich} P.,  {Lundqvist} P.,  {Pozzo} M.,    {Wheeler} J.~C.,  2007,
  \apj, 665, 608

\bibitem[\protect\citeauthoryear{{Meikle}, {Spyromilio}, {Allen}, {Varani} \&
  {Cumming}}{{Meikle} et~al.}{1993}]{Meikle1993}
{Meikle} W.~P.~S.,  {Spyromilio} J.,  {Allen} D.~A.,  {Varani} G.-F.,
  {Cumming} R.~J.,  1993, \mnras, 261, 535

\bibitem[\protect\citeauthoryear{{Miyaji} \& {Nomoto}}{{Miyaji} \&
  {Nomoto}}{1987}]{Miyaji1987}
{Miyaji} S.,  {Nomoto} K.,  1987, \apj, 318, 307

\bibitem[\protect\citeauthoryear{{Miyaji}, {Nomoto}, {Yokoi} \&
  {Sugimoto}}{{Miyaji} et~al.}{1980}]{Miyaji1980}
{Miyaji} S.,  {Nomoto} K.,  {Yokoi} K.,    {Sugimoto} D.,  1980, \pasj, 32, 303

\bibitem[\protect\citeauthoryear{{Moseley}, {Dwek}, {Silverberg}, {Glaccum},
  {Graham} \& {Loewenstein}}{{Moseley} et~al.}{1989}]{Moseley1989}
{Moseley} S.~H.,  {Dwek} E.,  {Silverberg} R.~F.,  {Glaccum} W.,  {Graham}
  J.~R.,    {Loewenstein} R.~F.,  1989, \apj, 347, 1119

\bibitem[\protect\citeauthoryear{{Mould}, {Huchra}, {Freedman}, {Kennicutt}
  Jr., {Ferrarese}, {Ford}, {Gibson}, {Graham}, {Hughes}, {Illingworth},
  {Kelson}, {Macri}, {Madore}, {Sakai}, {Sebo}, {Silbermann} \&
  {Stetson}}{{Mould} et~al.}{2000}]{Mould2000}
{Mould} J.~R.,  {Huchra} J.~P.,  {Freedman} W.~L.,  {Kennicutt} Jr. R.~C.,
  {Ferrarese} L.,  {Ford} H.~C.,  {Gibson} B.~K.,  {Graham} J.~A.,  {Hughes}
  S.~M.~G.,  {Illingworth} G.~D.,  {Kelson} D.~D.,  {Macri} L.~M.,  {Madore}
  B.~F.,  {Sakai} S.,  {Sebo} K.~M.,  {Silbermann} N.~A.,    {Stetson} P.~B.,
  2000, \apj, 529, 786

\bibitem[\protect\citeauthoryear{{Munari} \& {Zwitter}}{{Munari} \&
  {Zwitter}}{1997}]{Munari1997}
{Munari} U.,  {Zwitter} T.,  1997, \aap, 318, 269

\bibitem[\protect\citeauthoryear{{Nomoto}}{{Nomoto}}{1984}]{Nomoto1984}
{Nomoto} K.,  1984, \apj, 277, 791

\bibitem[\protect\citeauthoryear{{Nomoto}}{{Nomoto}}{1987}]{Nomoto1987}
{Nomoto} K.,  1987, \apj, 322, 206

\bibitem[\protect\citeauthoryear{{Nomoto}, {Sugimoto}, {Sparks}, {Fesen},
  {Gull} \& {Miyaji}}{{Nomoto} et~al.}{1982}]{Nomoto1982}
{Nomoto} K.,  {Sugimoto} D.,  {Sparks} W.~M.,  {Fesen} R.~A.,  {Gull} T.~R.,
  {Miyaji} S.,  1982, \nat, 299, 803

\bibitem[\protect\citeauthoryear{{Nugent}, {Sullivan}, {Ellis}, {Gal-Yam},
  {Leonard}, {Howell}, {Astier}, {Carlberg}, {Conley}, {Fabbro}, {Fouchez},
  {Neill}, {Pain}, {Perrett}, {Pritchet} \& {Regnault}}{{Nugent}
  et~al.}{2006}]{Nugent2006}
{Nugent} P.,  {Sullivan} M.,  {Ellis} R.,  {Gal-Yam} A.,  {Leonard} D.~C.,
  {Howell} D.~A.,  {Astier} P.,  {Carlberg} R.~G.,  {Conley} A.,  {Fabbro} S.,
  {Fouchez} D.,  {Neill} J.~D.,  {Pain} R.,  {Perrett} K.,  {Pritchet} C.~J.,
   {Regnault} N.,  2006, \apj, 645, 841

\bibitem[\protect\citeauthoryear{{Panagia}, {Vettolani}, {Boksenberg},
  {Ciatti}, {Ortolani}, {Rafanelli} \& et al.}{{Panagia}
  et~al.}{1980}]{Panagia1980}
{Panagia} N.,  {Vettolani} G.,  {Boksenberg} A.,  {Ciatti} F.,  {Ortolani} S.,
  {Rafanelli} P.,    et al. 1980, \mnras, 192, 861

\bibitem[\protect\citeauthoryear{{Pannuti}, {Schlegel} \& {Lacey}}{{Pannuti}
  et~al.}{2007}]{Pannuti2007}
{Pannuti} T.~G.,  {Schlegel} E.~M.,    {Lacey} C.~K.,  2007, \aj, 133, 1361

\bibitem[\protect\citeauthoryear{{Pastorello}, {Della Valle}, {Smartt},
  {Zampieri}, {Benetti}, {Cappellaro}, {Mazzali}, {Patat}, {Spiro}, {Turatto}
  \& {Valenti}}{{Pastorello} et~al.}{2007}]{Pastorello2007}
{Pastorello} A.,  {Della Valle} M.,  {Smartt} S.~J.,  {Zampieri} L.,  {Benetti}
  S.,  {Cappellaro} E.,  {Mazzali} P.~A.,  {Patat} F.,  {Spiro} S.,  {Turatto}
  M.,    {Valenti} S.,  2007, \nat, 449, 1

\bibitem[\protect\citeauthoryear{{Pastorello}, {Sauer}, {Taubenberger} \& et
  al.}{{Pastorello} et~al.}{2006}]{Pastorello2006}
{Pastorello} A.,  {Sauer} D.,  {Taubenberger} S.,    et al. 2006, \mnras, 370,
  1752

\bibitem[\protect\citeauthoryear{{Pastorello}, {Valenti}, {Zampieri} \& et
  al.}{{Pastorello} et~al.}{2009}]{Pastorello2009}
{Pastorello} A.,  {Valenti} S.,  {Zampieri} L.,    et al. 2009, ArXiv e-prints

\bibitem[\protect\citeauthoryear{{Pastorello}, {Zampieri}, {Turatto},
  {Cappellaro}, {Meikle}, {Benetti}, {Branch}, {Baron}, {Patat}, {Armstrong},
  {Altavilla}, {Salvo} \& {Riello}}{{Pastorello} et~al.}{2004}]{Pastorello2004}
{Pastorello} A.,  {Zampieri} L.,  {Turatto} M.,  {Cappellaro} E.,  {Meikle}
  W.~P.~S.,  {Benetti} S.,  {Branch} D.,  {Baron} E.,  {Patat} F.,  {Armstrong}
  M.,  {Altavilla} G.,  {Salvo} M.,    {Riello} M.,  2004, \mnras, 347, 74

\bibitem[\protect\citeauthoryear{{Patat}, {Benetti}, {Justham}, {Mazzali},
  {Pasquini}, {Cappellaro}, {Della Valle}, {-Podsiadlowski}, {Turatto},
  {Gal-Yam} \& {Simon}}{{Patat} et~al.}{2007}]{Patat2007}
{Patat} F.,  {Benetti} S.,  {Justham} S.,  {Mazzali} P.~A.,  {Pasquini} L.,
  {Cappellaro} E.,  {Della Valle} M.,  {-Podsiadlowski} P.,  {Turatto} M.,
  {Gal-Yam} A.,    {Simon} J.~D.,  2007, \aap, 474, 931

\bibitem[\protect\citeauthoryear{{Paturel}, {Theureau}, {Bottinelli},
  {Gouguenheim}, {Coudreau-Durand}, {Hallet} \& {Petit}}{{Paturel}
  et~al.}{2003}]{Paturel2003}
{Paturel} G.,  {Theureau} G.,  {Bottinelli} L.,  {Gouguenheim} L.,
  {Coudreau-Durand} N.,  {Hallet} N.,    {Petit} C.,  2003, \aap, 412, 57

\bibitem[\protect\citeauthoryear{{Pierce}}{{Pierce}}{1994}]{Pierce1994}
{Pierce} M.~J.,  1994, \apj, 430, 53

\bibitem[\protect\citeauthoryear{{Pilyugin}, {V{\'{\i}}lchez} \&
  {Contini}}{{Pilyugin} et~al.}{2004}]{Pilyugin2004}
{Pilyugin} L.~S.,  {V{\'{\i}}lchez} J.~M.,    {Contini} T.,  2004, \aap, 425,
  849

\bibitem[\protect\citeauthoryear{{Podsiadlowski}, {Langer}, {Poelarends},
  {Rappaport}, {Heger} \& {Pfahl}}{{Podsiadlowski}
  et~al.}{2004}]{Podsiadlowski2004}
{Podsiadlowski} P.,  {Langer} N.,  {Poelarends} A.~J.~T.,  {Rappaport} S.,
  {Heger} A.,    {Pfahl} E.,  2004, \apj, 612, 1044

\bibitem[\protect\citeauthoryear{{Poelarends}, {Herwig}, {Langer} \&
  {Heger}}{{Poelarends} et~al.}{2008}]{Poelarends2008}
{Poelarends} A.~J.~T.,  {Herwig} F.,  {Langer} N.,    {Heger} A.,  2008, \apj,
  675, 614

\bibitem[\protect\citeauthoryear{{Pozzo}, {Meikle}, {Fassia}, {Geballe},
  {Lundqvist}, {Chugai} \& {Sollerman}}{{Pozzo} et~al.}{2004}]{Pozzo2004}
{Pozzo} M.,  {Meikle} W.~P.~S.,  {Fassia} A.,  {Geballe} T.,  {Lundqvist} P.,
  {Chugai} N.~N.,    {Sollerman} J.,  2004, \mnras, 352, 457

\bibitem[\protect\citeauthoryear{{Pozzo}, {Meikle}, {Rayner}, {Joseph},
  {Filippenko}, {Foley}, {Li}, {Mattila} \& {Sollerman}}{{Pozzo}
  et~al.}{2006}]{Pozzo2006}
{Pozzo} M.,  {Meikle} W.~P.~S.,  {Rayner} J.~T.,  {Joseph} R.~D.,  {Filippenko}
  A.~V.,  {Foley} R.~J.,  {Li} W.,  {Mattila} S.,    {Sollerman} J.,  2006,
  \mnras, 368, 1169

\bibitem[\protect\citeauthoryear{{Prieto}, {Kistler}, {Stanek}, {Thompson},
  {Kochanek} \& {Beacom}}{{Prieto} et~al.}{2008}]{PrietoAtel2}
{Prieto} J.~L.,  {Kistler} M.~D.,  {Stanek} K.~Z.,  {Thompson} T.~A.,
  {Kochanek} C.~S.,    {Beacom} J.~F.,  2008, The Astronomer's Telegram, 1596,
  1

\bibitem[\protect\citeauthoryear{{Prieto}, {Kistler}, {Thompson}, {Y{\"u}ksel},
  {Kochanek}, {Stanek}, {Beacom}, {Martini}, {Pasquali} \& {Bechtold}}{{Prieto}
  et~al.}{2008}]{Prieto2008}
{Prieto} J.~L.,  {Kistler} M.~D.,  {Thompson} T.~A.,  {Y{\"u}ksel} H.,
  {Kochanek} C.~S.,  {Stanek} K.~Z.,  {Beacom} J.~F.,  {Martini} P.,
  {Pasquali} A.,    {Bechtold} J.,  2008, \apjl, 681, L9

\bibitem[\protect\citeauthoryear{{Quimby}, {Aldering}, {Wheeler},
  {H{\"o}flich}, {Akerlof} \& {Rykoff}}{{Quimby} et~al.}{2007}]{Quimby2007}
{Quimby} R.~M.,  {Aldering} G.,  {Wheeler} J.~C.,  {H{\"o}flich} P.,  {Akerlof}
  C.~W.,    {Rykoff} E.~S.,  2007, \apjl, 668, L99

\bibitem[\protect\citeauthoryear{{Ritossa}, {Garc{\'{\i}}a-Berro} \&
  {Iben}}{{Ritossa} et~al.}{1999}]{Ritossa1999}
{Ritossa} C.,  {Garc{\'{\i}}a-Berro} E.,    {Iben} I.~J.,  1999, \apj, 515, 381

\bibitem[\protect\citeauthoryear{{Roche}, {Aitken} \& {Smith}}{{Roche}
  et~al.}{1993}]{Roche1993}
{Roche} P.~F.,  {Aitken} D.~K.,    {Smith} C.~H.,  1993, \mnras, 261, 522

\bibitem[\protect\citeauthoryear{{Rouleau} \& {Martin}}{{Rouleau} \&
  {Martin}}{1991}]{Rouleau91}
{Rouleau} F.,  {Martin} P.~G.,  1991, \apj, 377, 526

\bibitem[\protect\citeauthoryear{{Sahu}, {Anupama}, {Srividya} \&
  {Muneer}}{{Sahu} et~al.}{2006}]{Sahu2006}
{Sahu} D.~K.,  {Anupama} G.~C.,  {Srividya} S.,    {Muneer} S.,  2006, \mnras,
  372, 1315

\bibitem[\protect\citeauthoryear{{Schawinski}, {Justham}, {Wolf} \& et
  al.}{{Schawinski} et~al.}{2008}]{Schawinski2008}
{Schawinski} K.,  {Justham} S.,  {Wolf} C.,    et al. 2008, Science, 321, 223

\bibitem[\protect\citeauthoryear{{Schlegel}, {Finkbeiner} \&
  {Davis}}{{Schlegel} et~al.}{1998}]{Schlegel1998}
{Schlegel} D.~J.,  {Finkbeiner} D.~P.,    {Davis} M.,  1998, \apj, 500, 525

\bibitem[\protect\citeauthoryear{{Schlegel}}{{Schlegel}}{1990}]{Schlegel1990}
{Schlegel} E.~M.,  1990, \mnras, 244, 269

\bibitem[\protect\citeauthoryear{{Schlegel}}{{Schlegel}}{1994}]{Schlegel1994}
{Schlegel} E.~M.,  1994, \apj, 434, 523

\bibitem[\protect\citeauthoryear{{Schlegel}, {Blair} \& {Fesen}}{{Schlegel}
  et~al.}{2000}]{Schlegel2000}
{Schlegel} E.~M.,  {Blair} W.~P.,    {Fesen} R.~A.,  2000, \aj, 120, 791

\bibitem[\protect\citeauthoryear{{Schmeer}}{{Schmeer}}{2008}]{Schmeer2008CBET}
{Schmeer} P.,  2008, Central Bureau Electronic Telegrams, 1236, A260000+

\bibitem[\protect\citeauthoryear{{Schmidt}, {Kirshner}, {Eastman}, {Phillips},
  {Suntzeff}, {Hamuy}, {Maza} \& {Aviles}}{{Schmidt}
  et~al.}{1994}]{Schmidt1994}
{Schmidt} B.~P.,  {Kirshner} R.~P.,  {Eastman} R.~G.,  {Phillips} M.~M.,
  {Suntzeff} N.~B.,  {Hamuy} M.,  {Maza} J.,    {Aviles} R.,  1994, \apj, 432,
  42

\bibitem[\protect\citeauthoryear{{Schoniger} \& {Sofue}}{{Schoniger} \&
  {Sofue}}{1994}]{Schoniger1994}
{Schoniger} F.,  {Sofue} Y.,  1994, \aap, 283, 21

\bibitem[\protect\citeauthoryear{{Sharina}, {Karachentsev} \&
  {Tikhonov}}{{Sharina} et~al.}{1997}]{Sharina1997}
{Sharina} M.~E.,  {Karachentsev} I.~D.,    {Tikhonov} N.~A.,  1997, Astronomy
  Letters, 23, 373

\bibitem[\protect\citeauthoryear{{Siess}}{{Siess}}{2007}]{Siess2007}
{Siess} L.,  2007, \aap, 476, 893

\bibitem[\protect\citeauthoryear{{Smartt}, {Eldridge}, {Crockett} \&
  {Maund}}{{Smartt} et~al.}{2009}]{Smartt2008}
{Smartt} S.~J.,  {Eldridge} J.~J.,  {Crockett} R.~M.,    {Maund} J.~R.,  2009,
  ArXiv e-prints

\bibitem[\protect\citeauthoryear{{Smith}, {Foley} \& {Filippenko}}{{Smith}
  et~al.}{2008}]{Smith2008d}
{Smith} N.,  {Foley} R.~J.,    {Filippenko} A.~V.,  2008, \apj, 680, 568

\bibitem[\protect\citeauthoryear{{Smith}, {Ganeshalingam}, {Li}, {Chornock},
  {Steele}, {Silverman}, {Filippenko} \& {Mobberley}}{{Smith}
  et~al.}{2008}]{Smith2008c}
{Smith} N.,  {Ganeshalingam} M.,  {Li} W.,  {Chornock} R.,  {Steele} T.~N.,
  {Silverman} J.~M.,  {Filippenko} A.~V.,    {Mobberley} M.~P.,  2008, ArXiv
  e-prints

\bibitem[\protect\citeauthoryear{{Smith}, {Li}, {Foley}, {Wheeler}, {Pooley},
  {Chornock}, {Filippenko}, {Silverman}, {Quimby}, {Bloom} \& {Hansen}}{{Smith}
  et~al.}{2007}]{Smith2007}
{Smith} N.,  {Li} W.,  {Foley} R.~J.,  {Wheeler} J.~C.,  {Pooley} D.,
  {Chornock} R.,  {Filippenko} A.~V.,  {Silverman} J.~M.,  {Quimby} R.,
  {Bloom} J.~S.,    {Hansen} C.,  2007, \apj, 666, 1116

\bibitem[\protect\citeauthoryear{{Stanishev}, {Pastorello} \&
  {Pursimo}}{{Stanishev} et~al.}{2008}]{Stanishev2008}
{Stanishev} V.,  {Pastorello} A.,    {Pursimo} T.,  2008, Central Bureau
  Electronic Telegrams, 1236, 2

\bibitem[\protect\citeauthoryear{{Steele}, {Silverman}, {Ganeshalingam}, {Lee},
  {Li} \& {Filippenko}}{{Steele} et~al.}{2008}]{Steele2008CBET}
{Steele} T.~N.,  {Silverman} J.~M.,  {Ganeshalingam} M.,  {Lee} N.,  {Li} W.,
   {Filippenko} A.~V.,  2008, Central Bureau Electronic Telegrams, 1275,
  A260000+

\bibitem[\protect\citeauthoryear{{Sugerman}, {Ercolano}, {Barlow}, {Tielens},
  {Clayton}, {Zijlstra}, {Meixner}, {Speck}, {Gledhill}, {Panagia}, {Cohen},
  {Gordon}, {Meyer}, {Fabbri}, {Bowey}, {Welch}, {Regan} \&
  {Kennicutt}}{{Sugerman} et~al.}{2006}]{Sugerman2006}
{Sugerman} B.~E.~K.,  {Ercolano} B.,  {Barlow} M.~J.,  {Tielens} A.~G.~G.~M.,
  {Clayton} G.~C.,  {Zijlstra} A.~A.,  {Meixner} M.,  {Speck} A.,  {Gledhill}
  T.~M.,  {Panagia} N.,  {Cohen} M.,  {Gordon} K.~D.,  {Meyer} M.,  {Fabbri}
  J.,  {Bowey} J.~E.,  {Welch} D.~L.,  {Regan} M.~W.,    {Kennicutt} R.~C.,
  2006, Science, 313, 196

\bibitem[\protect\citeauthoryear{{Suntzeff} \& {Bouchet}}{{Suntzeff} \&
  {Bouchet}}{1990}]{Suntzeff1990}
{Suntzeff} N.~B.,  {Bouchet} P.,  1990, \aj, 99, 650

\bibitem[\protect\citeauthoryear{{Swartz}, {Wheeler} \& {Harkness}}{{Swartz}
  et~al.}{1991}]{Swartz1991}
{Swartz} D.~A.,  {Wheeler} J.~C.,    {Harkness} R.~P.,  1991, \apj, 374, 266

\bibitem[\protect\citeauthoryear{{Tacconi} \& {Young}}{{Tacconi} \&
  {Young}}{1990}]{Tacconi1990}
{Tacconi} L.~J.,  {Young} J.~S.,  1990, \apj, 352, 595

\bibitem[\protect\citeauthoryear{{Terry}, {Paturel} \& {Ekholm}}{{Terry}
  et~al.}{2002}]{Terry2002}
{Terry} J.~N.,  {Paturel} G.,    {Ekholm} T.,  2002, \aap, 393, 57

\bibitem[\protect\citeauthoryear{{Thompson}, {Prieto}, {Stanek}, {Kistler},
  {Beacom} \& {Kochanek}}{{Thompson} et~al.}{2008}]{Thompson2008}
{Thompson} T.~A.,  {Prieto} J.~L.,  {Stanek} K.~Z.,  {Kistler} M.~D.,  {Beacom}
  J.~F.,    {Kochanek} C.~S.,  2008, ArXiv e-prints

\bibitem[\protect\citeauthoryear{{Tsvetkov}, {Volnova}, {Shulga}, {Korotkiy},
  {Elmhamdi}, {Danziger} \& {Ereshko}}{{Tsvetkov} et~al.}{2006}]{Tsvetkov2006}
{Tsvetkov} D.~Y.,  {Volnova} A.~A.,  {Shulga} A.~P.,  {Korotkiy} S.~A.,
  {Elmhamdi} A.,  {Danziger} I.~J.,    {Ereshko} M.~V.,  2006, \aap, 460, 769

\bibitem[\protect\citeauthoryear{{Turatto}, {Mazzali}, {Young} \& et
  al}{{Turatto} et~al.}{1998}]{Turatto1998}
{Turatto} M.,  {Mazzali} P.~A.,  {Young} T.~R.,    et al 1998, \apjl, 498,
  L129+

\bibitem[\protect\citeauthoryear{{Van Dyk}, {Peng}, {King}, {Filippenko},
  {Treffers}, {Li} \& {Richmond}}{{Van Dyk} et~al.}{2000}]{VanDyk2000}
{Van Dyk} S.~D.,  {Peng} C.~Y.,  {King} J.~Y.,  {Filippenko} A.~V.,  {Treffers}
  R.~R.,  {Li} W.,    {Richmond} M.~W.,  2000, \pasp, 112, 1532

\bibitem[\protect\citeauthoryear{{van Loon}, {Cioni}, {Zijlstra} \&
  {Loup}}{{van Loon} et~al.}{2005}]{vanLoon2005}
{van Loon} J.~T.,  {Cioni} M.-R.~L.,  {Zijlstra} A.~A.,    {Loup} C.,  2005,
  \aap, 438, 273

\bibitem[\protect\citeauthoryear{{van Loon}, {Marshall} \& {Zijlstra}}{{van
  Loon} et~al.}{2005}]{vanLoon2005a}
{van Loon} J.~T.,  {Marshall} J.~R.,    {Zijlstra} A.~A.,  2005, \aap, 442, 597

\bibitem[\protect\citeauthoryear{{Wanajo}, {Nomoto}, {Janka}, {Kitaura} \&
  {Mueller}}{{Wanajo} et~al.}{2008}]{Wanajo2008}
{Wanajo} S.,  {Nomoto} K.,  {Janka} H.~.,  {Kitaura} F.~S.,    {Mueller} B.,
  2008, ArXiv e-prints

\bibitem[\protect\citeauthoryear{{Wanajo}, {Tamamura}, {Itoh}, {Nomoto},
  {Ishimaru}, {Beers} \& {Nozawa}}{{Wanajo} et~al.}{2003}]{Wanajo2003}
{Wanajo} S.,  {Tamamura} M.,  {Itoh} N.,  {Nomoto} K.,  {Ishimaru} Y.,  {Beers}
  T.~C.,    {Nozawa} S.,  2003, \apj, 593, 968

\bibitem[\protect\citeauthoryear{{Waxman} \& {Draine}}{{Waxman} \&
  {Draine}}{2000}]{WaxmanDraine2000}
{Waxman} E.,  {Draine} B.~T.,  2000, \apj, 537, 796

\bibitem[\protect\citeauthoryear{{Weaver} \& {Woosley}}{{Weaver} \&
  {Woosley}}{1979}]{Weaver1979}
{Weaver} T.~A.,  {Woosley} S.~E.,  1979, in Bulletin of the American
  Astronomical Society Vol.~11 of Bulletin of the American Astronomical
  Society, {Evolution and Final Fate of 10M Stars}.
pp 724

\bibitem[\protect\citeauthoryear{{Weiler}, {van Dyk}, {Montes}, {Panagia} \&
  {Sramek}}{{Weiler} et~al.}{1998}]{Weiler1998}
{Weiler} K.~W.,  {van Dyk} S.~D.,  {Montes} M.~J.,  {Panagia} N.,    {Sramek}
  R.~A.,  1998, \apj, 500, 51

\bibitem[\protect\citeauthoryear{{Wesson}, {Fabbri}, {Barlow} \&
  {Meixner}}{{Wesson} et~al.}{2008}]{Wesson2008}
{Wesson} R.,  {Fabbri} J.,  {Barlow} M.,    {Meixner} M.,  2008, Central Bureau
  Electronic Telegrams, 1381, 1

\bibitem[\protect\citeauthoryear{{Woosley}, {Blinnikov} \& {Heger}}{{Woosley}
  et~al.}{2007}]{Woosley2007}
{Woosley} S.~E.,  {Blinnikov} S.,    {Heger} A.,  2007, \nat, 450, 390

\bibitem[\protect\citeauthoryear{{Wright}}{{Wright}}{1980}]{Wright1980}
{Wright} E.~L.,  1980, \apjl, 242, L23+

\bibitem[\protect\citeauthoryear{{Young}, {Smartt}, {Mattila}, {Tanvir},
  {Bersier}, {Chambers}, {Kaiser} \& {Tonry}}{{Young}
  et~al.}{2008}]{2008A&A...489..359Y}
{Young} D.~R.,  {Smartt} S.~J.,  {Mattila} S.,  {Tanvir} N.~R.,  {Bersier} D.,
  {Chambers} K.~C.,  {Kaiser} N.,    {Tonry} J.~L.,  2008, \aap, 489, 359

\bibitem[\protect\citeauthoryear{{Young}, {Smith} \& {Johnson}}{{Young}
  et~al.}{2005}]{Young2005}
{Young} T.~R.,  {Smith} D.,    {Johnson} T.~A.,  2005, \apjl, 625, L87

\bibitem[\protect\citeauthoryear{{Zwitter}, {Munari} \& {Moretti}}{{Zwitter}
  et~al.}{2004}]{Zwitter2004}
{Zwitter} T.,  {Munari} U.,    {Moretti} S.,  2004, IAU circular, 8413, 1

\end{thebibliography}


\newpage
\appendix
\section{Data Tables}
\begin{table*}
\caption{Magnitudes of the local sequence stars in the field of SN2008S.\label{seqstar}}
\begin{tabular}{cccccc}
\hline
ID & U & B & V & R & I  \\
\hline
1  &  16.41 $\pm$ 0.04     &  16.29 $\pm$ 0.04     &  15.45 $\pm$ 0.02   & 14.97 $\pm$ 0.02     &  14.46 $\pm$ 0.05  \\
2  &  14.56 $\pm$ 0.02     &  14.29 $\pm$ 0.02     &  13.55 $\pm$ 0.02   & 13.13 $\pm$ 0.01     &  12.77 $\pm$ 0.02  \\
3  &  18.35 $\pm$ 0.04     &  17.85 $\pm$ 0.03     &  16.85 $\pm$ 0.02   & 16.30 $\pm$ 0.03     &  15.75 $\pm$ 0.03  \\
4  &  14.77 $\pm$ 0.02     &  14.50 $\pm$ 0.02     &  13.76 $\pm$ 0.01   & 13.34 $\pm$ 0.02     &  12.95 $\pm$ 0.03  \\
6  &  16.83 $\pm$ 0.03     &  16.61 $\pm$ 0.03     &  15.80 $\pm$ 0.02   & 15.32 $\pm$ 0.03     &  14.83 $\pm$ 0.02  \\
7  &  17.94 $\pm$ 0.04     &  17.62 $\pm$ 0.04     &  16.68 $\pm$ 0.03   & 16.16 $\pm$ 0.02     &  15.58 $\pm$ 0.02  \\
12 &  16.79 $\pm$ 0.03     &  16.60 $\pm$ 0.02     &  15.83 $\pm$ 0.02   & 15.37 $\pm$ 0.02     &  14.89 $\pm$ 0.01 \\
13 &  18.41 $\pm$ 0.05     &  17.57 $\pm$ 0.04     &  16.43 $\pm$ 0.03   & 15.75 $\pm$ 0.02     &  15.14 $\pm$ 0.02  \\
16 &  15.87 $\pm$ 0.02     &  15.59 $\pm$ 0.02     &  14.74 $\pm$ 0.02   & 14.27 $\pm$ 0.03     &  13.81 $\pm$ 0.03 \\
17 &  17.88 $\pm$ 0.03     &  17.71 $\pm$ 0.04     &  16.91 $\pm$ 0.04   & 16.45 $\pm$ 0.03     &  15.97 $\pm$ 0.03 \\
19 &  16.13 $\pm$ 0.02     &  15.78 $\pm$ 0.02     &  14.85 $\pm$ 0.02   & 14.34 $\pm$ 0.02     &  13.88 $\pm$ 0.02  \\
20 &  19.89 $\pm$ 0.06     &  18.12 $\pm$ 0.05     &  16.86 $\pm$ 0.03   & 15.99 $\pm$ 0.03     &  15.32 $\pm$ 0.03  \\
21 &  16.75 $\pm$ 0.04     &  16.37 $\pm$ 0.03     &  15.49 $\pm$ 0.02   & 14.99 $\pm$ 0.02     &  14.53 $\pm$ 0.02  \\
22 &  16.55 $\pm$ 0.03     &  16.44 $\pm$ 0.03     &  15.66 $\pm$ 0.02   & 15.21 $\pm$ 0.02     &  14.74 $\pm$ 0.02   \\
23 &                       &  15.56 $\pm$ 0.02     &  13.82 $\pm$ 0.03   & 13.35 $\pm$ 0.03     &  12.91 $\pm$ 0.03  \\
26 &                       &  14.55 $\pm$ 0.03     &  14.38 $\pm$ 0.02   & 13.70 $\pm$ 0.02     &  13.09 $\pm$ 0.02  \\
\hline
\end{tabular}
\end{table*}

\begin{table*}
\caption{Emission features observed on SN 2008S spectra.\label{lineid}}
\begin{tabular}{llll}
\hline
 Identification & Rest Wavelength & Observed Wavelength & Notes\\  
  & (\AA) &(\AA) & \\  
\hline
$H_{\delta}$          & 4101                     &  4103    & blend           \\
FeII (28)              & 4178.85                 &  4182  & blend \\
FeII (27)              &4385.38                 &  4387   &            \\
FeII (37)              &4472.92                  & 4475    &            \\
FeII (37)              &4491                     & 4493    &            \\
FeII (38)              &4508.28                 & 4511  & blend           \\
FeII (37)              &4520                     & 4522    &            \\
FeII (38)              &4541.52                 & 4543   & blend            \\
FeII (38)              &4549.47                & 4553   &             \\
FeII (38)              &4576.331                 & 4578   &            \\
FeII (38)              &4583.829                 &  4586   &            \\
FeII (38)              &4620.5                   &  4622  &            \\
FeII (186)             &4635.3                   &4637   &            \\ 
$H_{\beta}$          & 4861                    &  4867    &  blend     \\
FeII (42)            & 4923.92                 &  4926   &        \\
FeII (42)            & 5018.43                   & 5020    &        \\
FeII (42)            & 5169.0                   & 5172    &        \\
FeII (49)            & 5197.56                   &  5199   &        \\
FeII (49)            & 5234                   &  5237  &        \\
FeII (49)            & 5254.9                 & 5257  &        \\         
FeII (48)            & 5264.8                 &  5267 &        \\       
FeII (49)            & 5275.99                &  5277  &        \\
FeII (49)            & 5316.6                 &  5318 &        \\
FeII (49)             & 5325.5                 &  5328&        \\
FeII (48)             & 5337.71                &  5340 &        \\                 
FeII (49)             & 5425.3                 &  5428&        \\
NaI                  & 5890  5896              &  5892     &          \\ 
FeII (46)              &  5991.38                & 5995       &          \\
FeII (46)              &  6084.11               & 6089       &          \\ 
FeII (46)              &  6113.33               & 6116       &          \\ 
FeII (74)            & 6238.38-6239.9          &  6243   &  blend     \\
FeII (74) + FeII (92)& 6247.55 -6248.89        &  6250   &  blend \\ 
$[OI]$                & 6300.23                 &  6304    & \\
$[OI]$                & 6363.88                 &  6368    & \\ 
FeII (40)+ FeII (93) & 6369.5-6371             &  6372     &       \\
FeII (74)            & 6416.9                  &  6418    &       \\
FeII (40)+ FeII (199)& 6432.7- 6433.9          &  6433-6440& blend       \\
FeII (74)            &  6456.38                &  6456-6461& blend      \\
FeII (92)+ FeII (92)  & 6491.2 + 6493.035       &6492-6496&   blend         \\
FeII (40)+ FeII (92)&   6516.1 + 6517.0       &  6519     &  blend      \\
$H_{\alpha}$         &  6563.5                   &  6568     &       \\
$[FeII]$(14)         &7155.14                 & 7159     &  ?    \\
FeII (73)            &  7222.39                  & 7226    &        \\
FeII (73)            &  7224.51                  & 7228    &         \\
$[CaII]$               & 7291                    &  7295   &       \\
FeII (73)            & 7310.24                   & 7313    &        \\
$[CaII]$               &  7324                   &  7327     &      \\
MnII (4)             &7353.52                 & 7358     &  ?    \\ 
MnII (4)             &7415.80                 & 7420     &  ?    \\
MnII (4)             &7432.27                 & 7438     &  ?    \\
FeII (73)             & 7449.3               &  7453     &  blend     \\
FeII (73)             & 7462.38                 &  7465     &       \\
FeII (73)             & 7515.9                 &  7520     &       \\
FeII (72)             &7533.42                 & 7538      &       \\
FeII (73)              &7711.71                  &  7716    &       \\
CaII (2)           & 8498                    &  8505   &       \\
CaII (2)           &  8542                   &  8547     &      \\
CaII (2)           & 8662.14                    &  8669     &      \\
OI (4)           & 8446.35                 &           & ph 256     \\
\hline
\end{tabular}
\end{table*}


\end{document}